%% file: theostat.tex
\documentclass[11pt,a4paper]{article}
\usepackage{pgf}
\usepackage{amsmath}
\usepackage{amssymb}
\usepackage{appendix}
\usepackage{multirow}
\usepackage{geometry}
\usepackage{float}
\usepackage{comment}

\pgfdeclareimage[width=15cm]{Signif1}{Signif1}
\pgfdeclareimage[width=15cm]{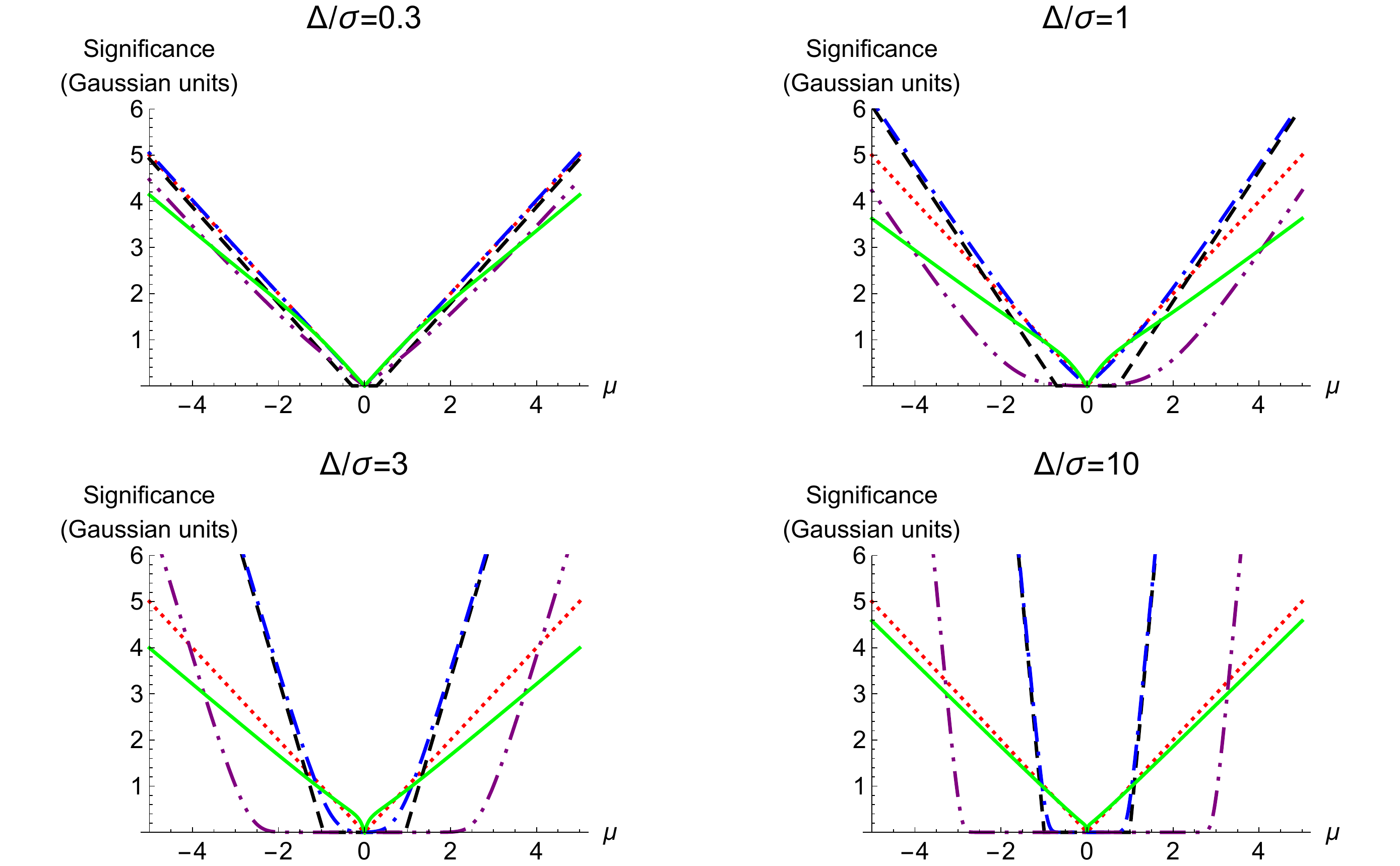}{Signif2}

\pgfdeclareimage[width=10cm]{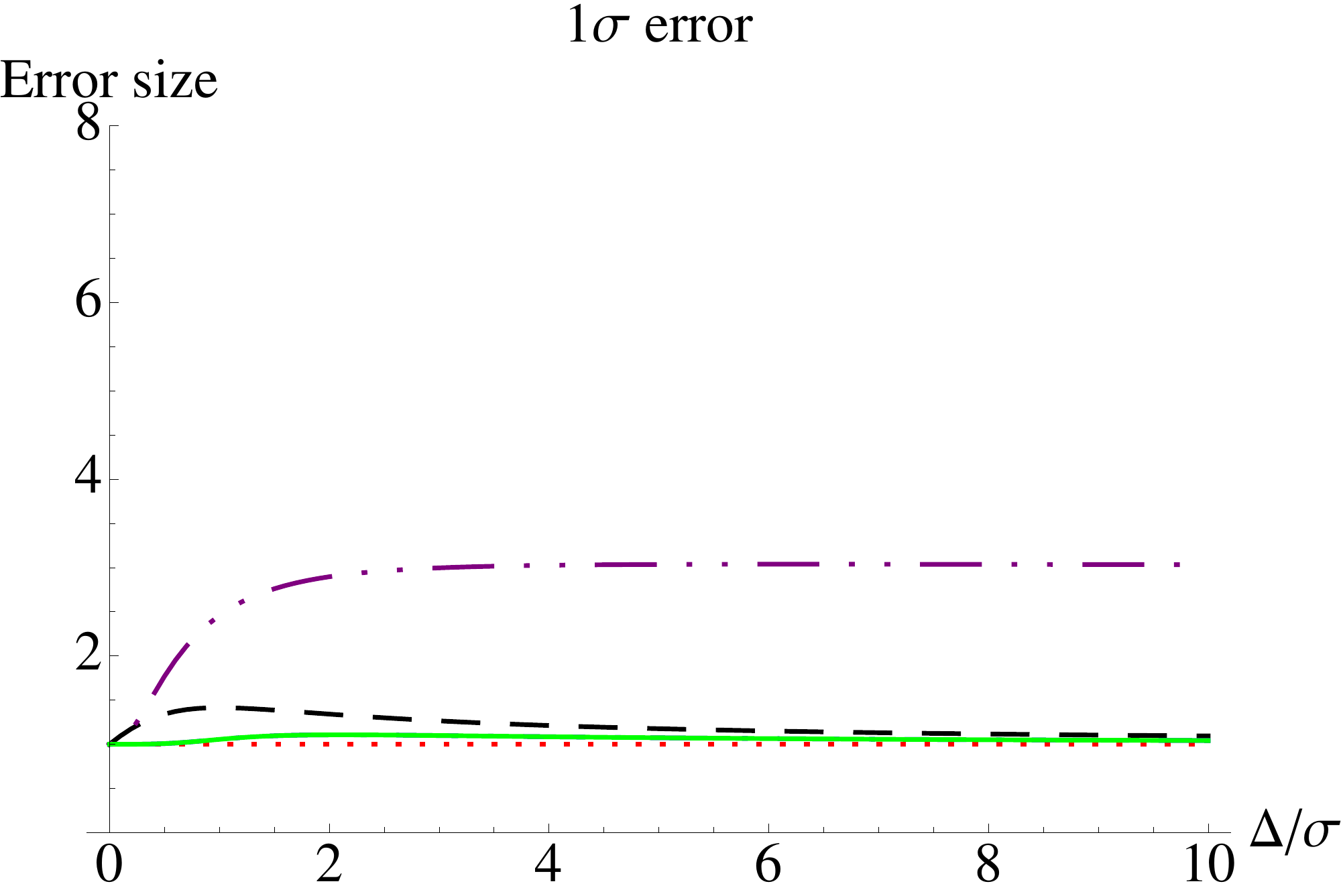}{CI1}
\pgfdeclareimage[width=10cm]{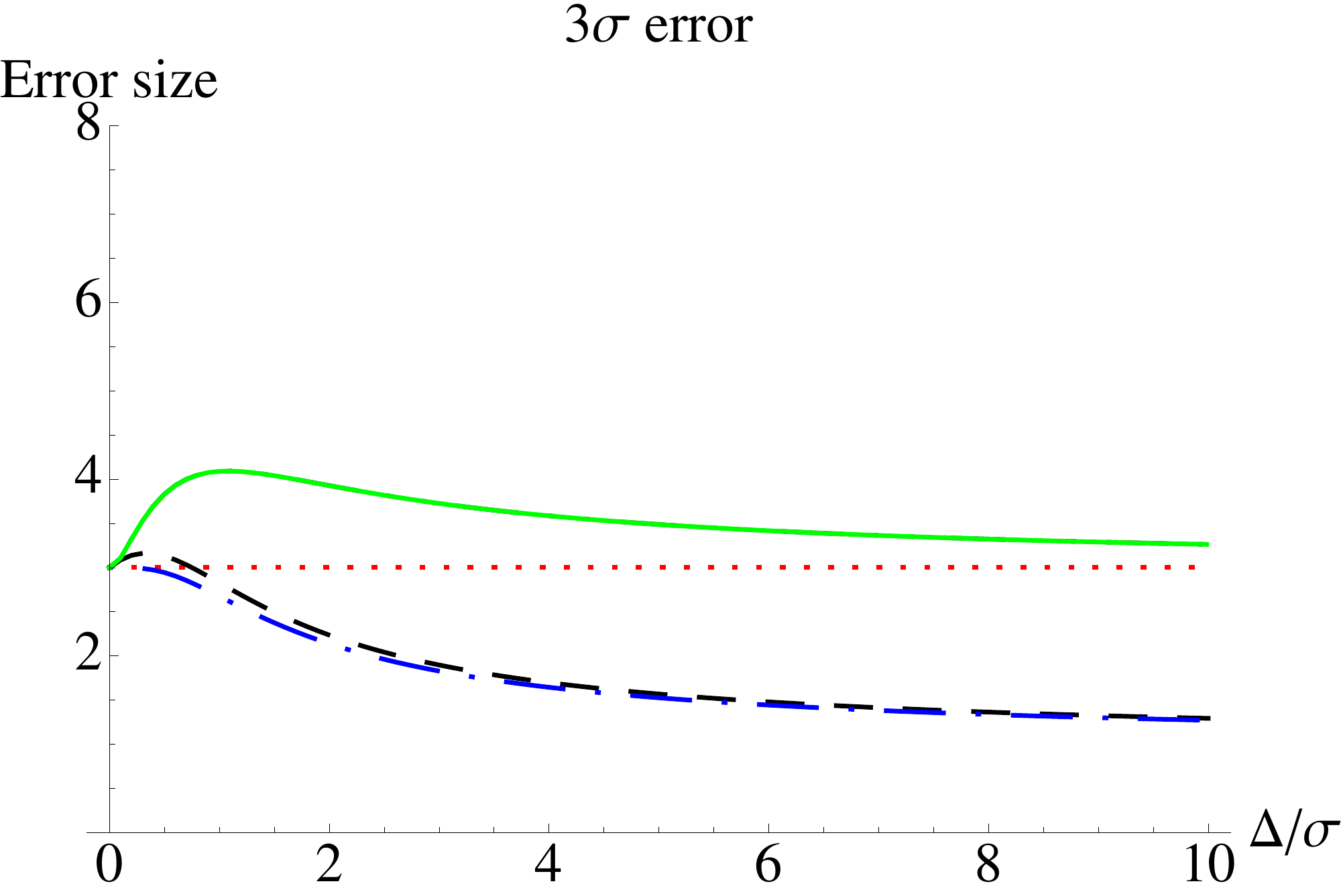}{CI2}
\pgfdeclareimage[width=10cm]{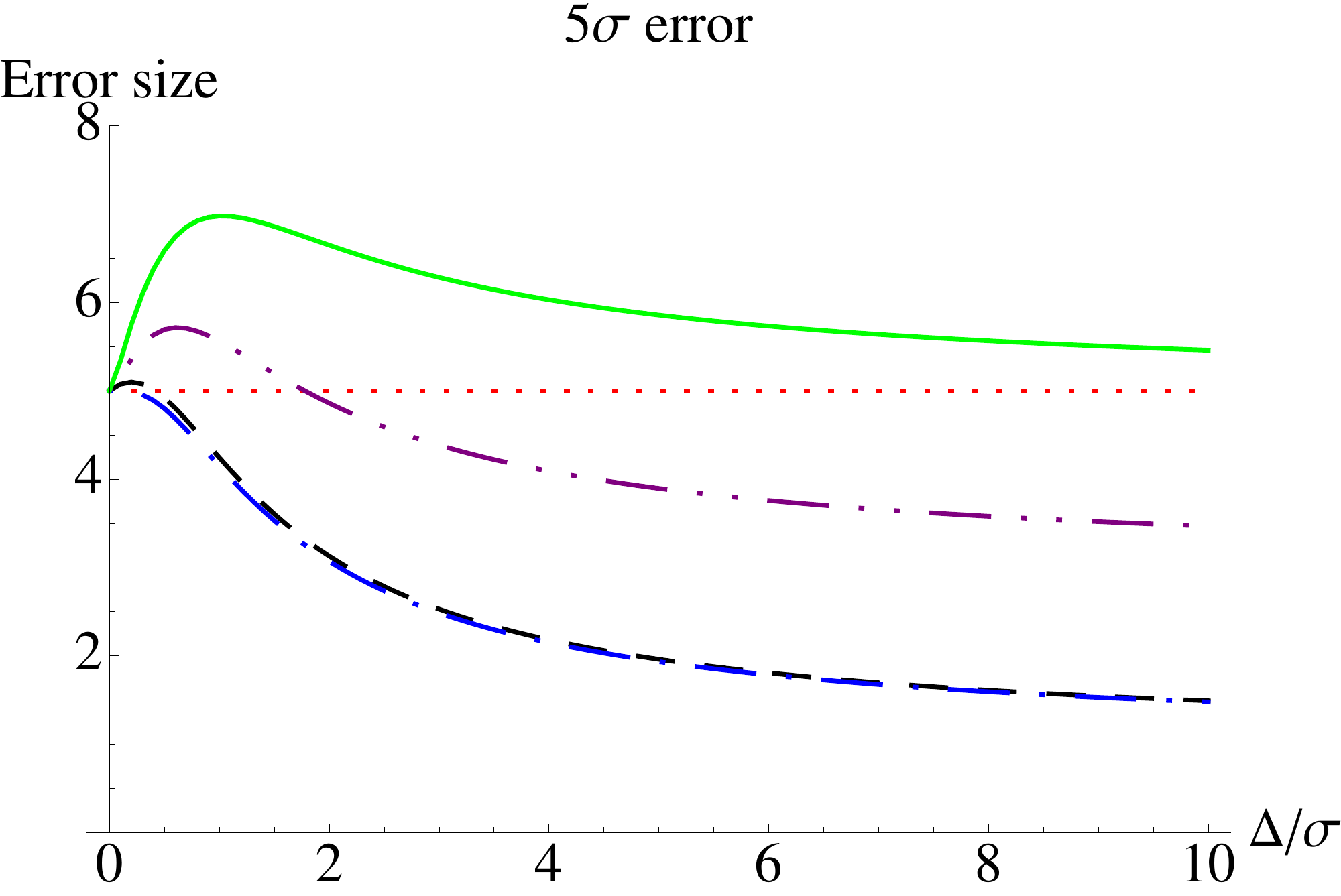}{CI3}

\pgfdeclareimage[height=5.5cm]{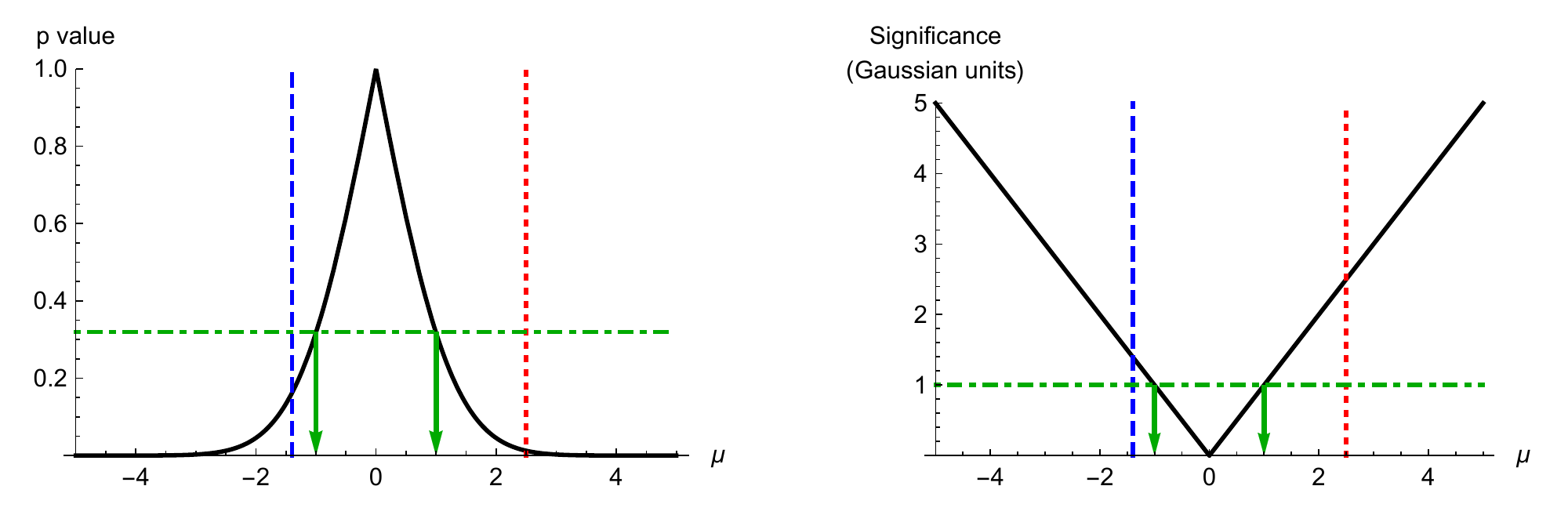}{CLs}
\pgfdeclareimage[height=6cm]{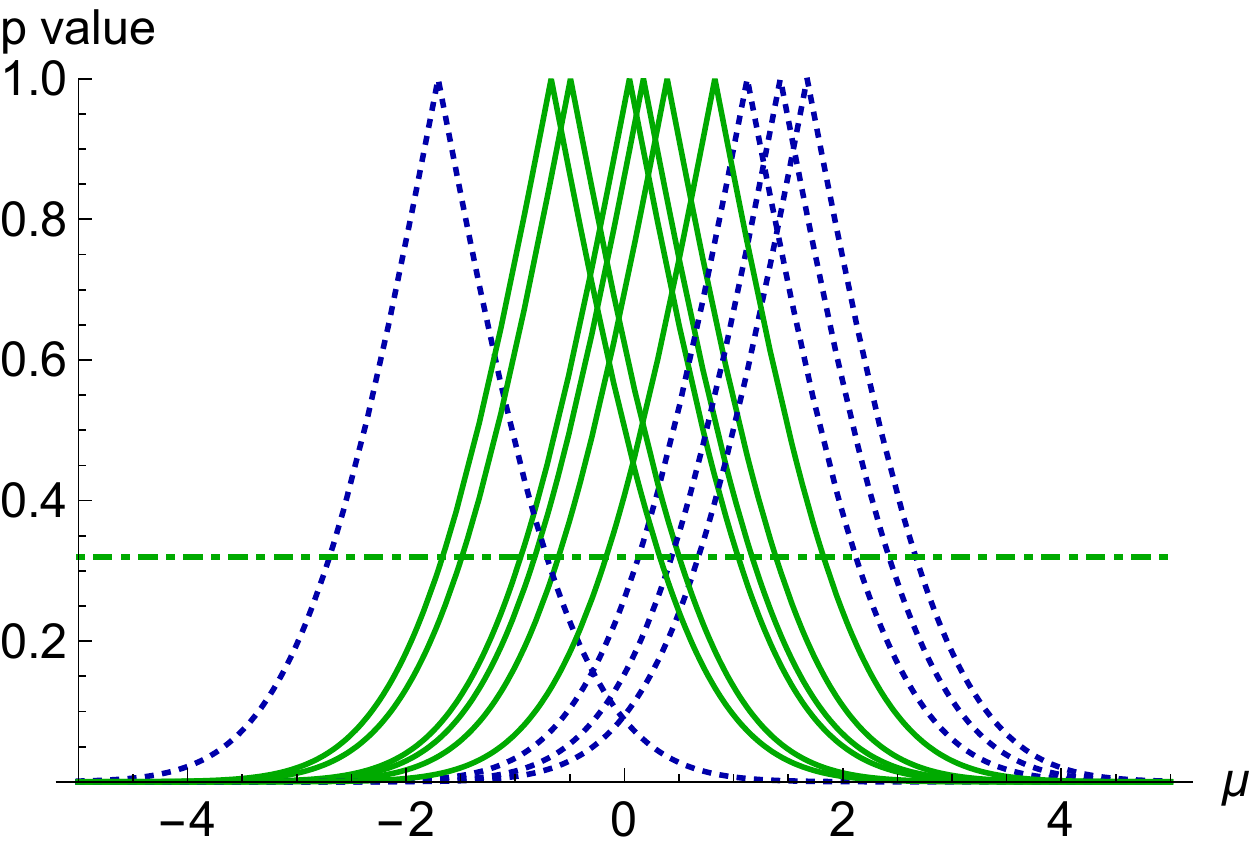}{coverage}
\pgfdeclareimage[height=6cm]{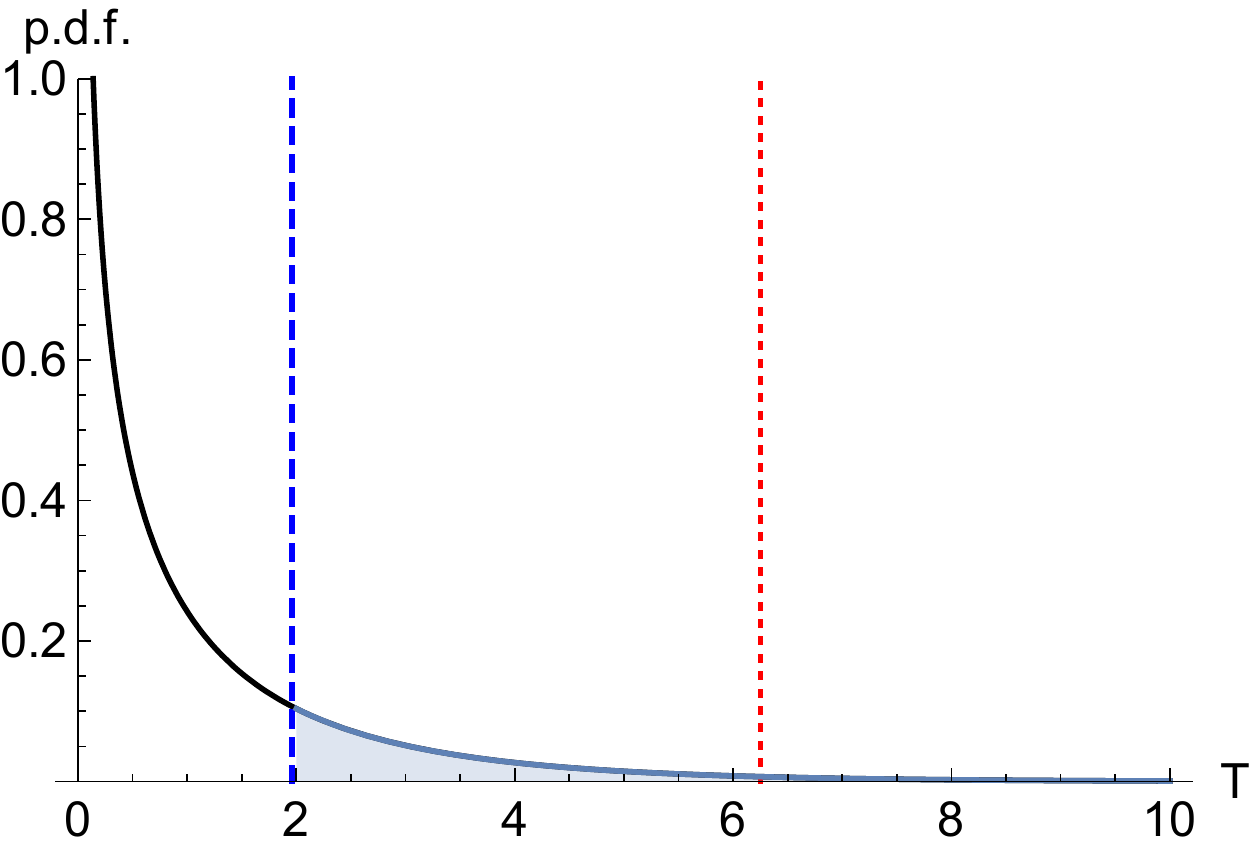}{Pdfchi2}

\pgfdeclareimage[width=15cm]{CombinRfit2}{CombinRfit2}

\pgfdeclareimage[height=5cm]{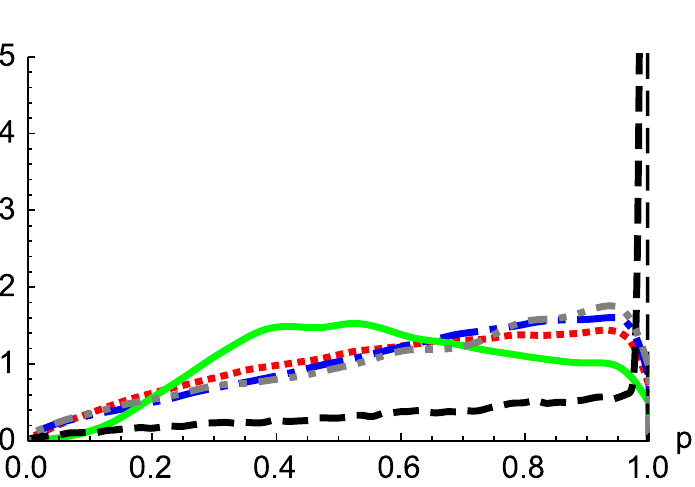}{coverage10}
\pgfdeclareimage[height=5cm]{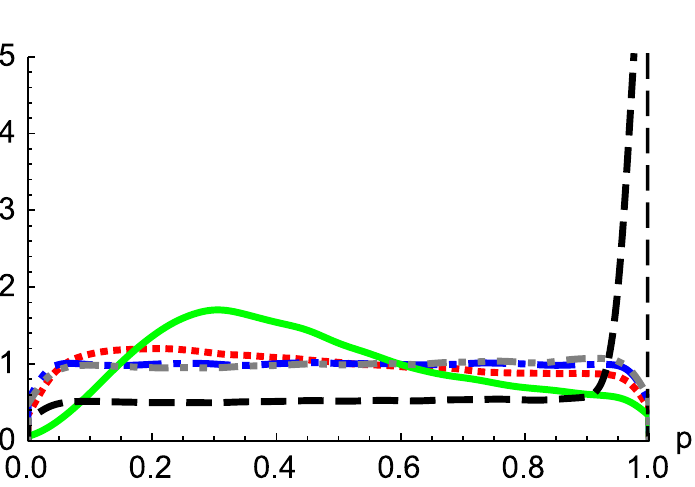}{coverage11}
\pgfdeclareimage[height=5cm]{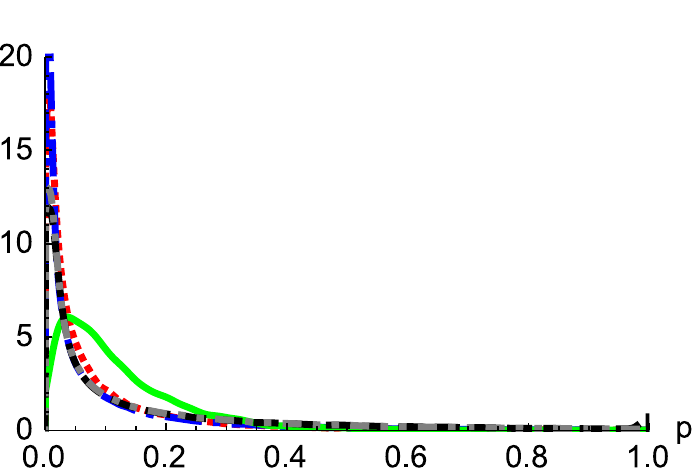}{coverage13}
\pgfdeclareimage[height=5cm]{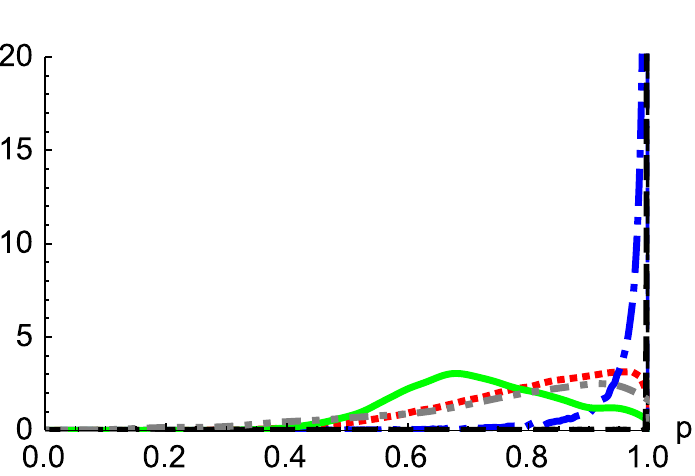}{coverage30}
\pgfdeclareimage[height=5cm]{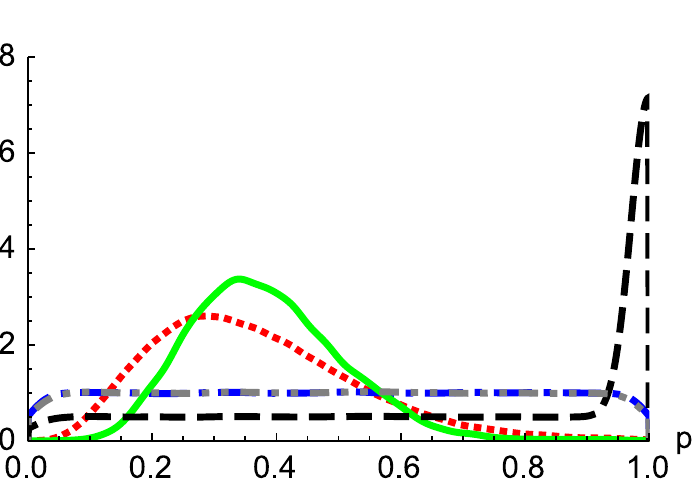}{coverage31}
\pgfdeclareimage[height=5cm]{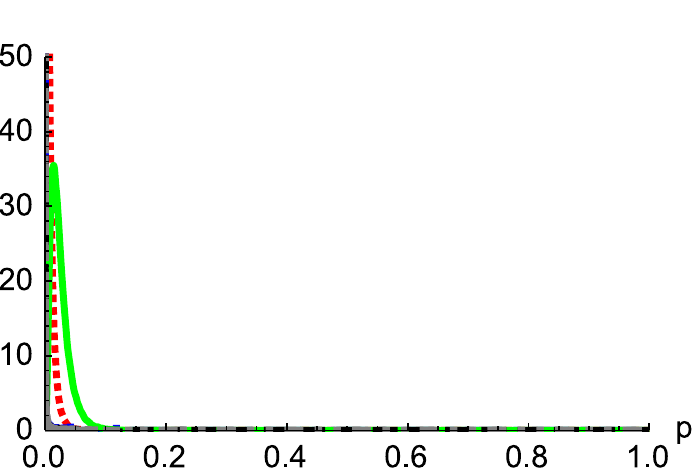}{coverage33}

\pgfdeclareimage[height=7cm]{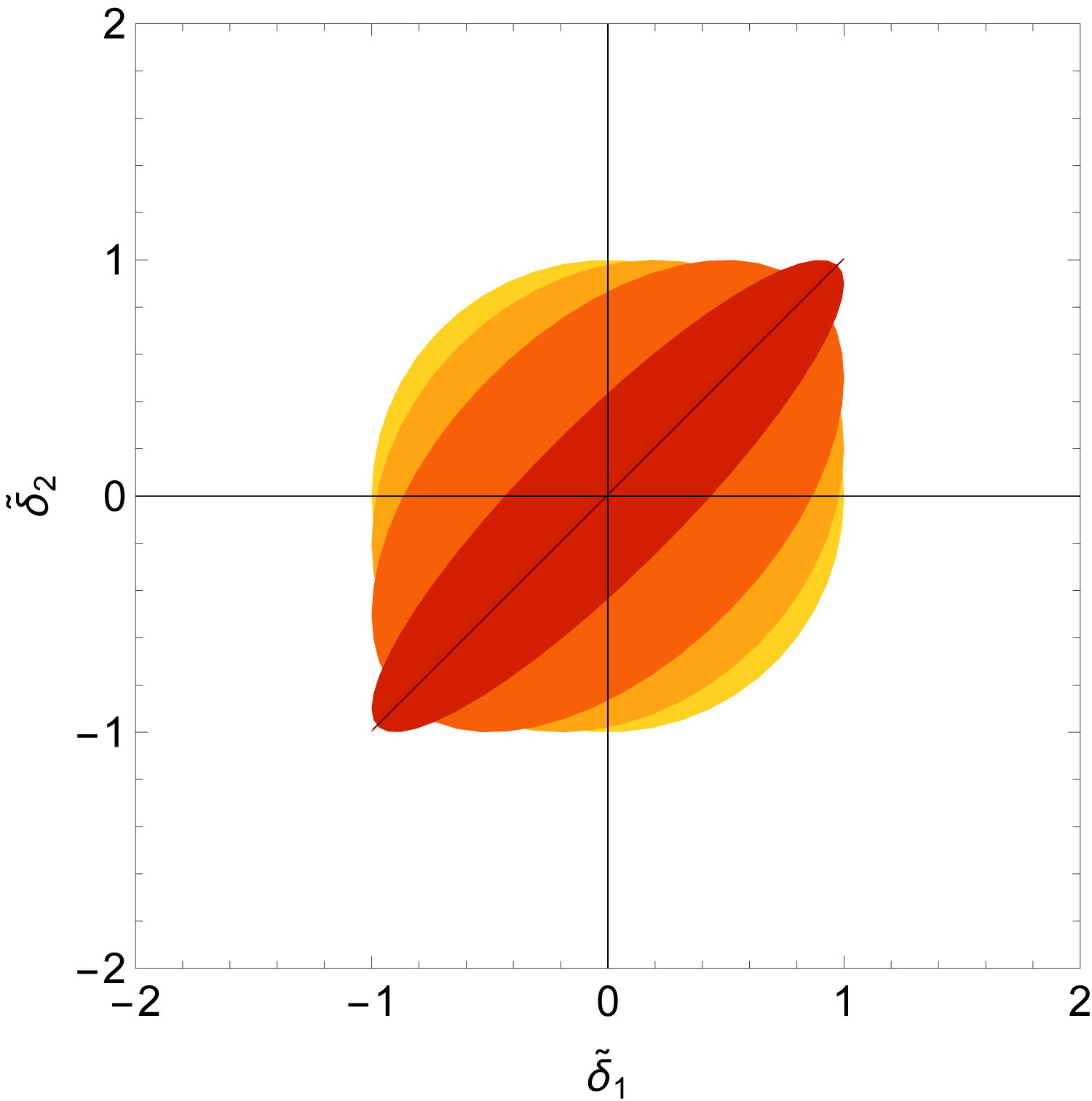}{theocorr2ball}
\pgfdeclareimage[height=7cm]{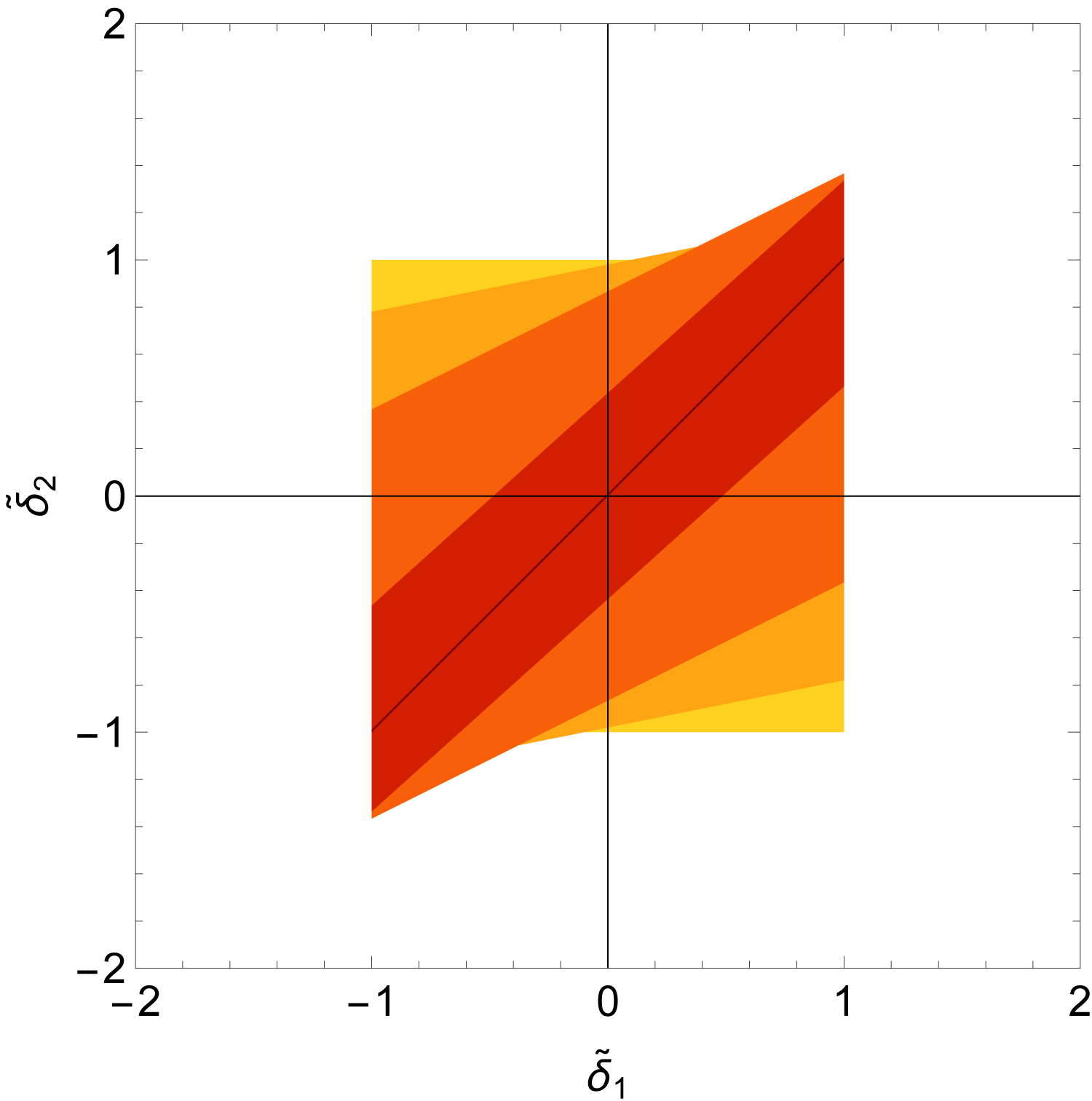}{theocorr2cube}
\pgfdeclareimage[height=3.5cm]{theocorr2rho}{theocorr2rho}
\pgfdeclareimage[height=5cm]{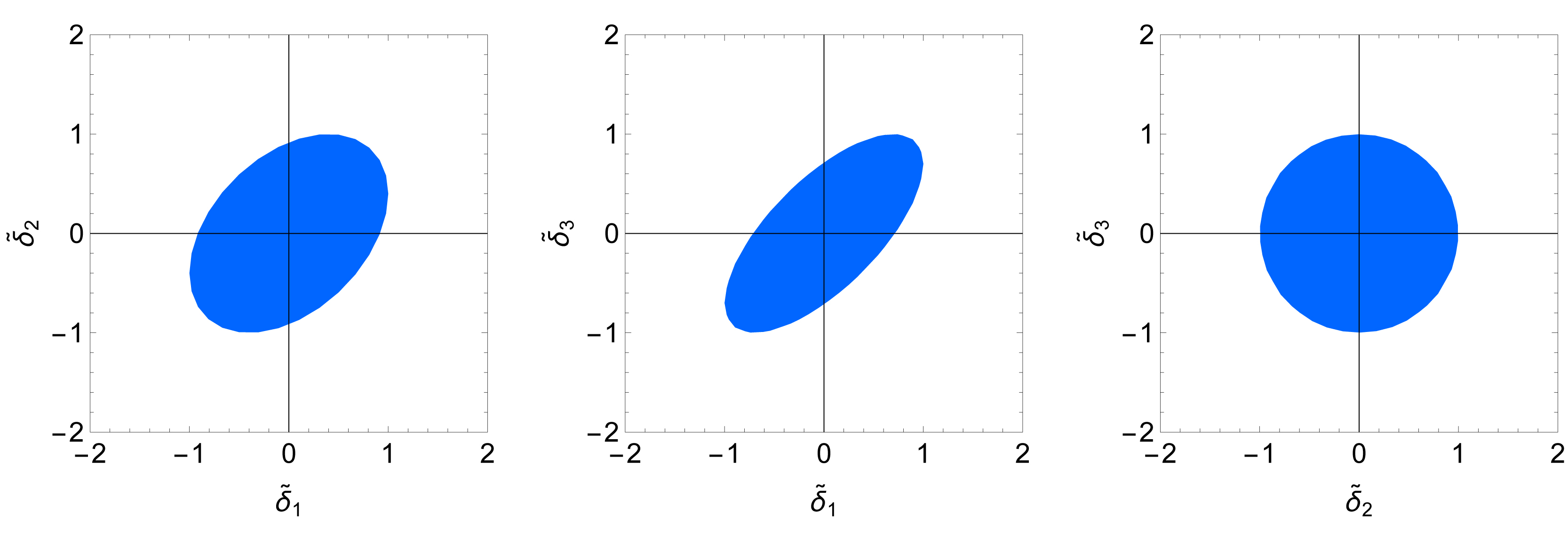}{theocorr3ball}
\pgfdeclareimage[height=5cm]{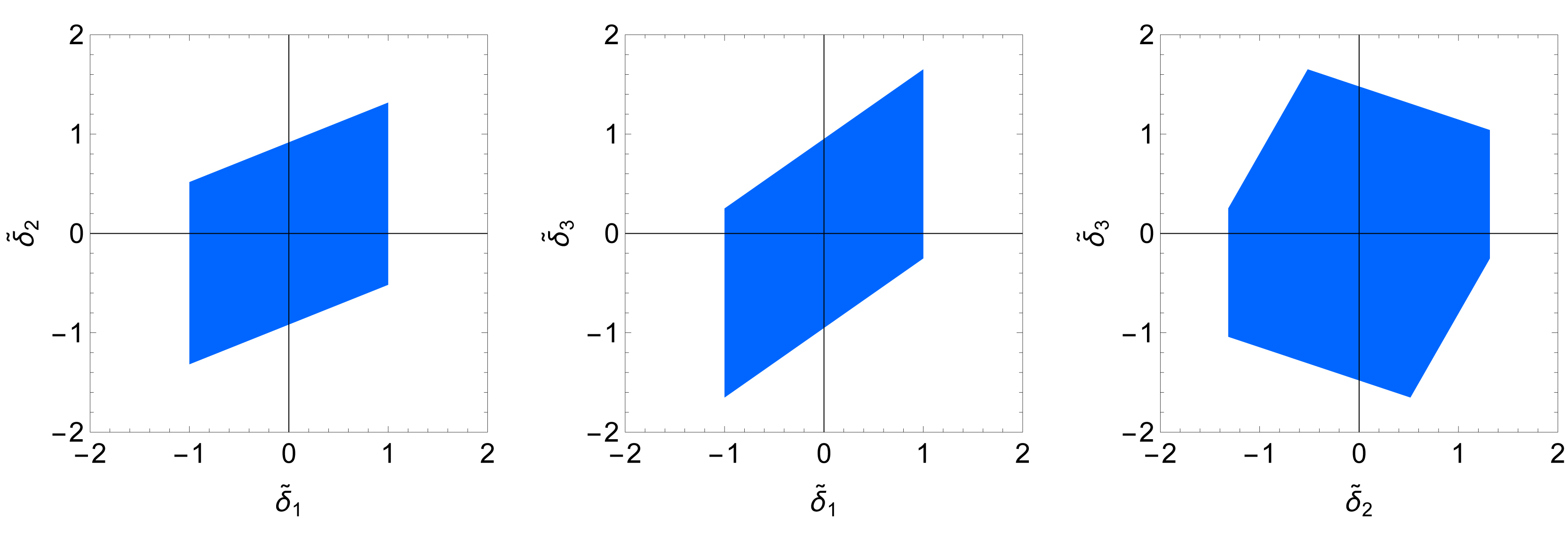}{theocorr3cube}

\pgfdeclareimage[height=6.3cm]{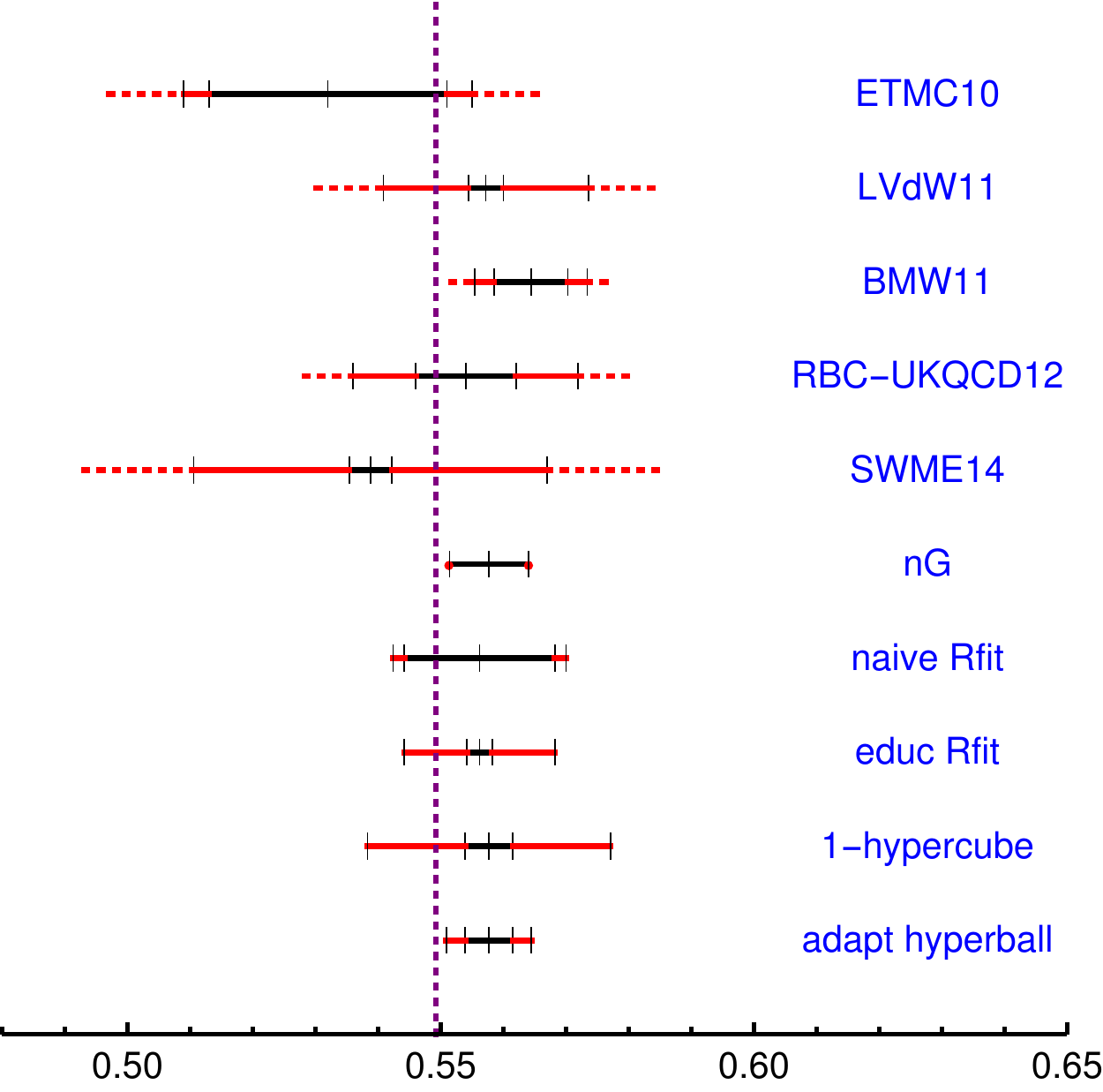}{Image_BK}
\pgfdeclareimage[height=6.3cm]{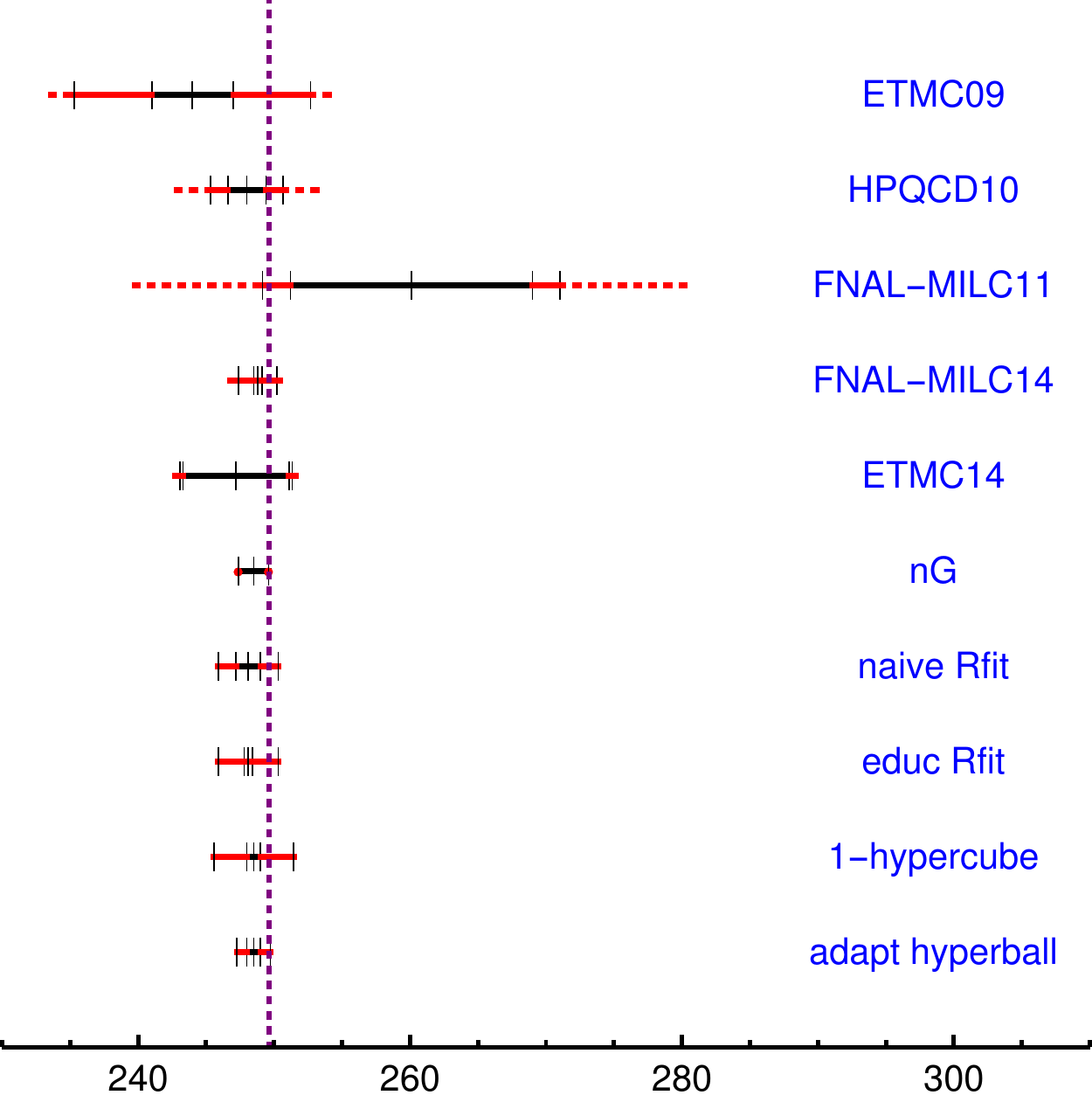}{Image_Ds}
\pgfdeclareimage[height=5cm]{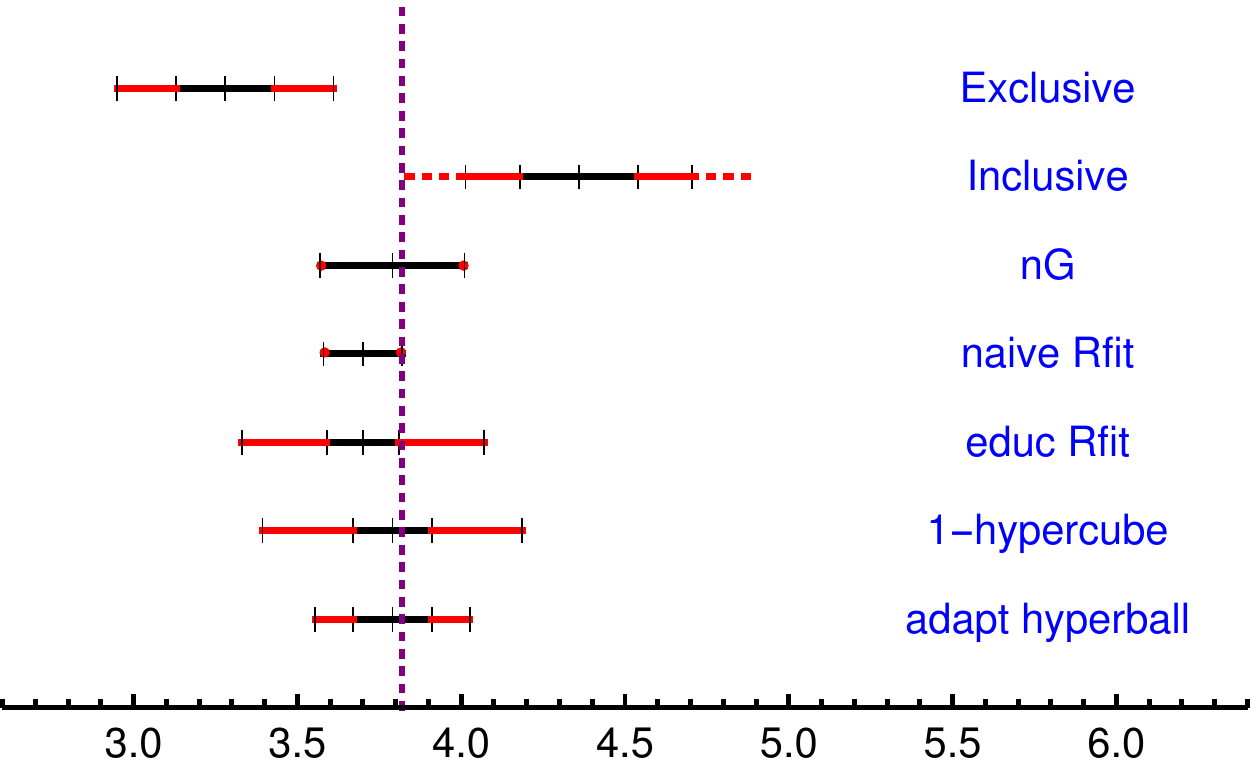}{Image_Vub}
\pgfdeclareimage[height=5cm]{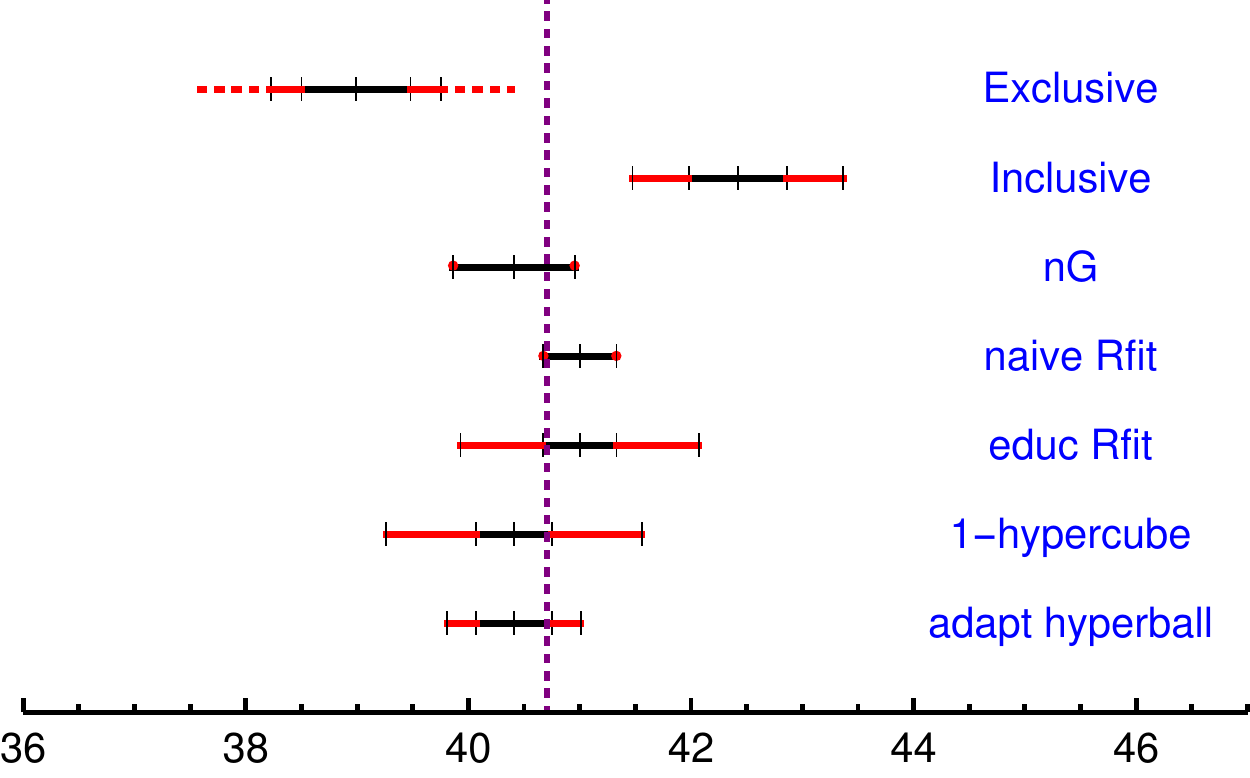}{Image_Vcb}
\pgfdeclareimage[height=10cm]{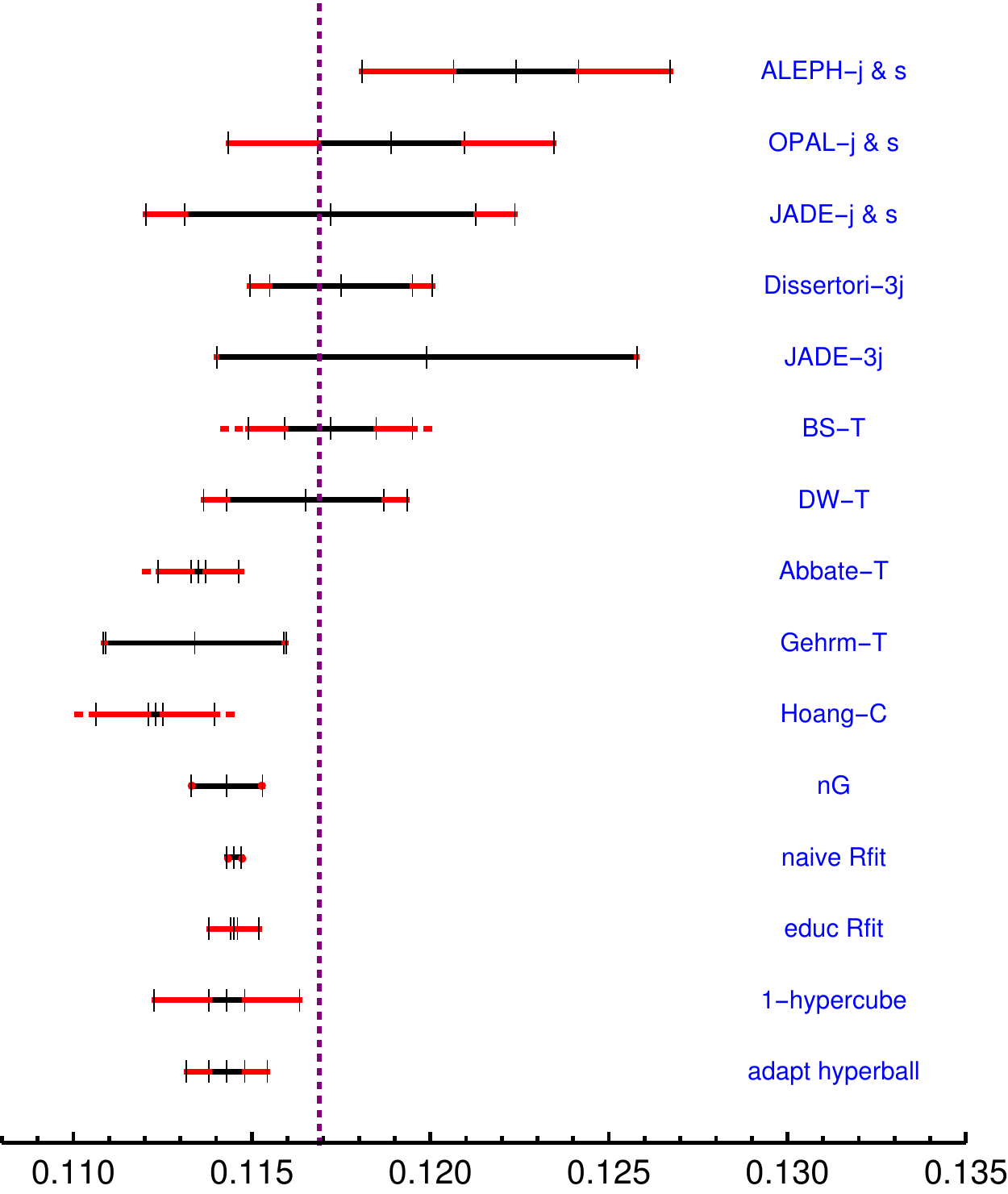}{Image_alphaSMZ}

\pgfdeclareimage[height=6.3cm]{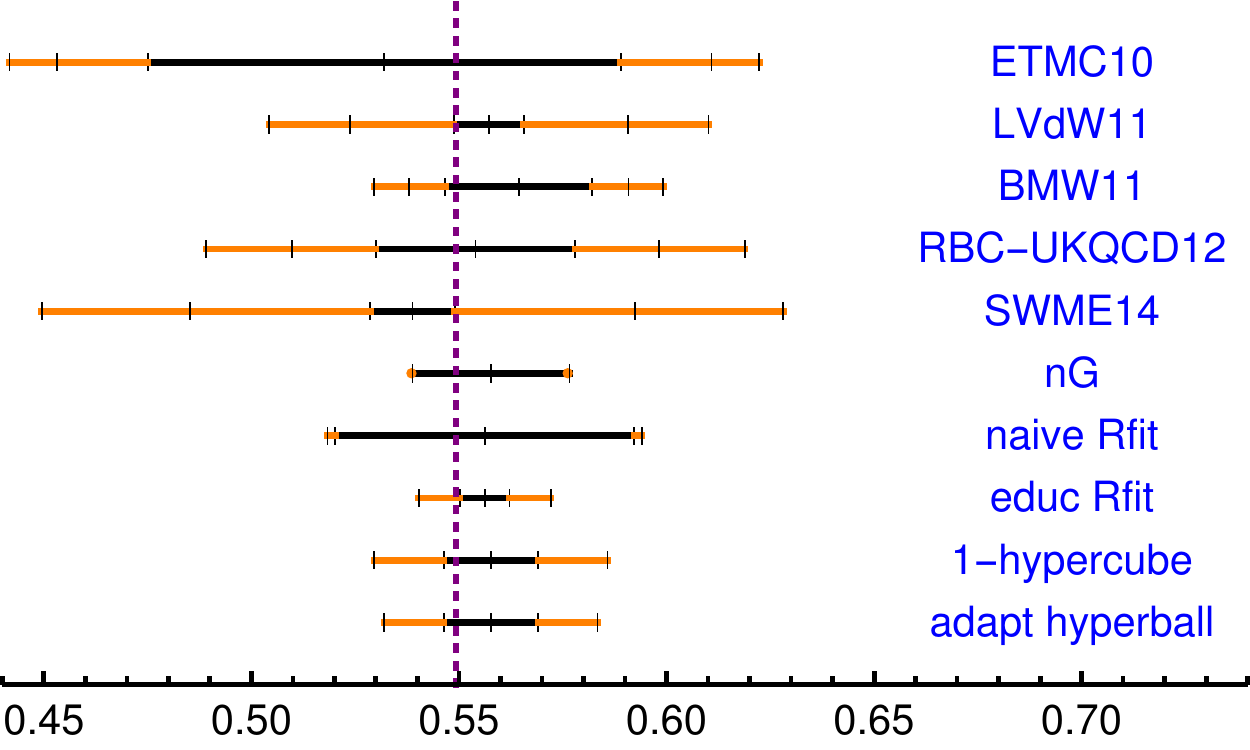}{3sigma_Image_BK}
\pgfdeclareimage[height=6.3cm]{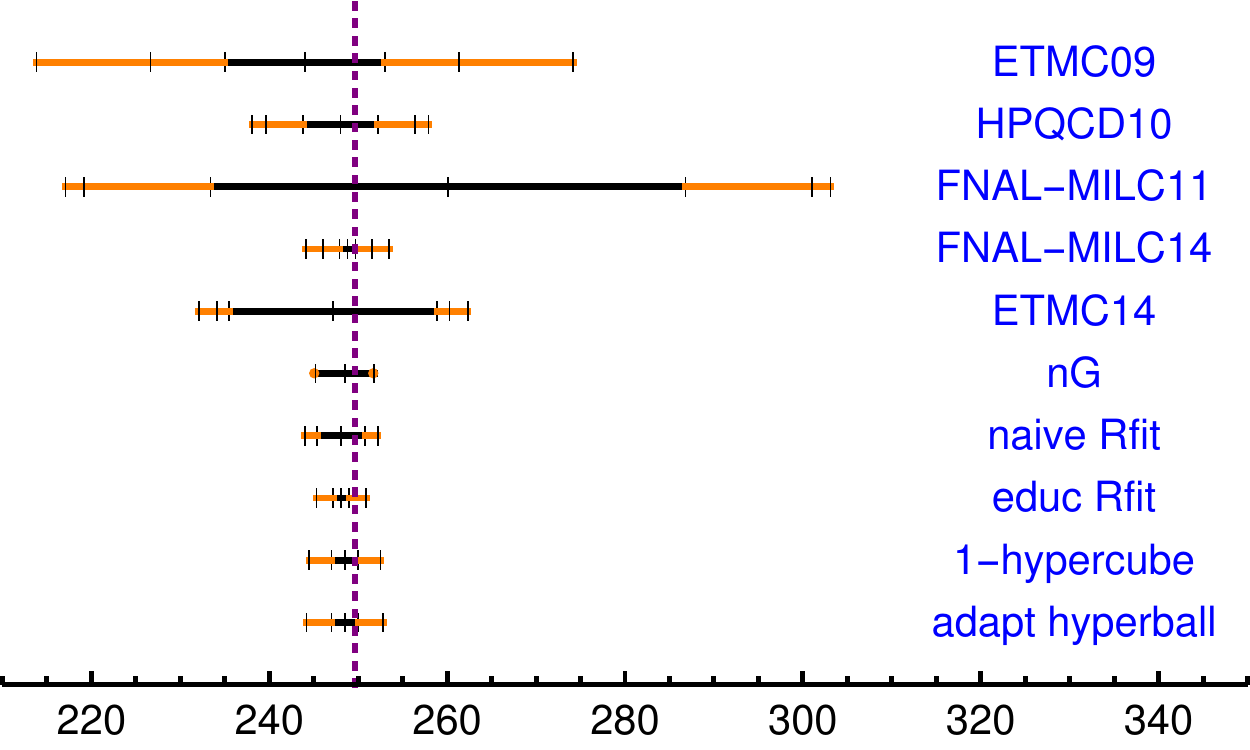}{3sigma_Image_Ds}
\pgfdeclareimage[height=5.0cm]{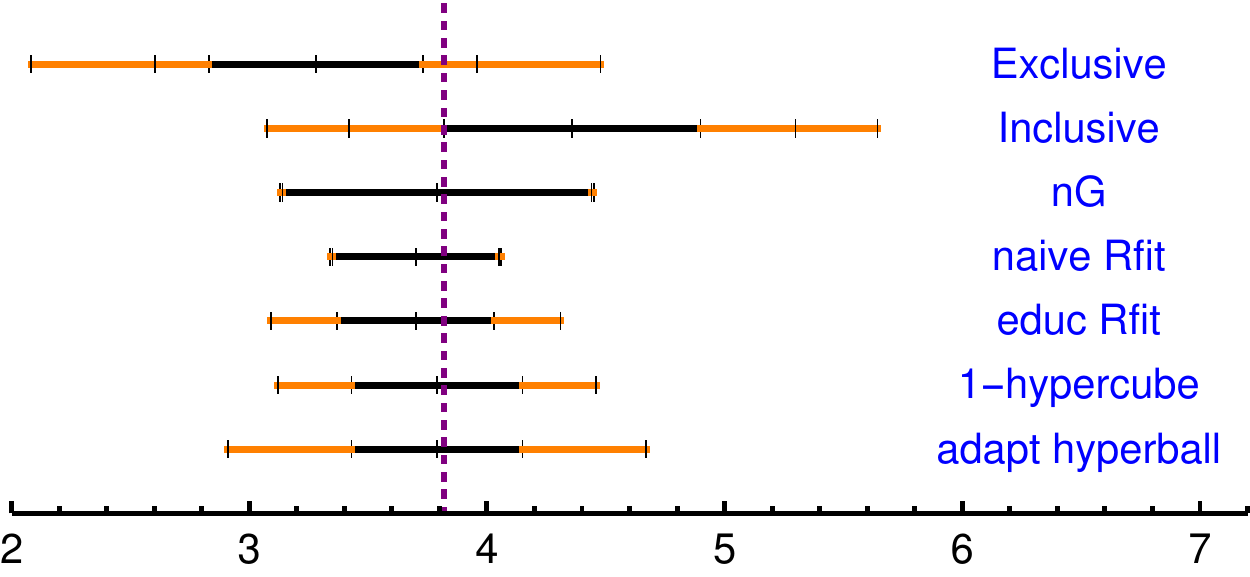}{3sigma_Image_Vub}
\pgfdeclareimage[height=5.0cm]{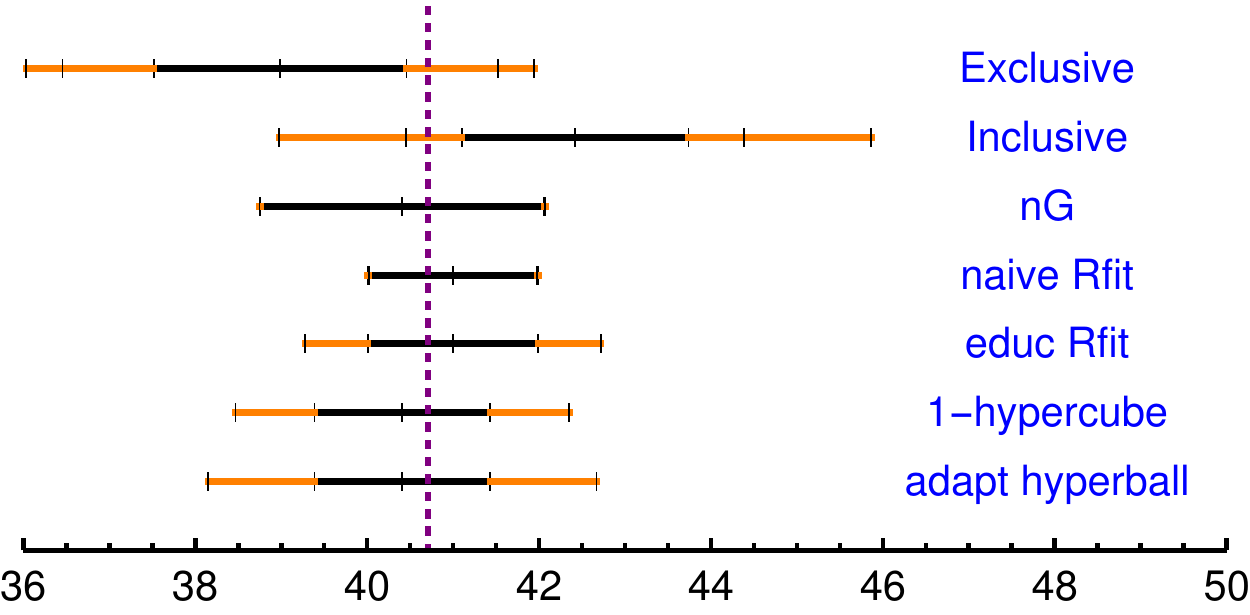}{3sigma_Image_Vcb}
\pgfdeclareimage[height=9cm]{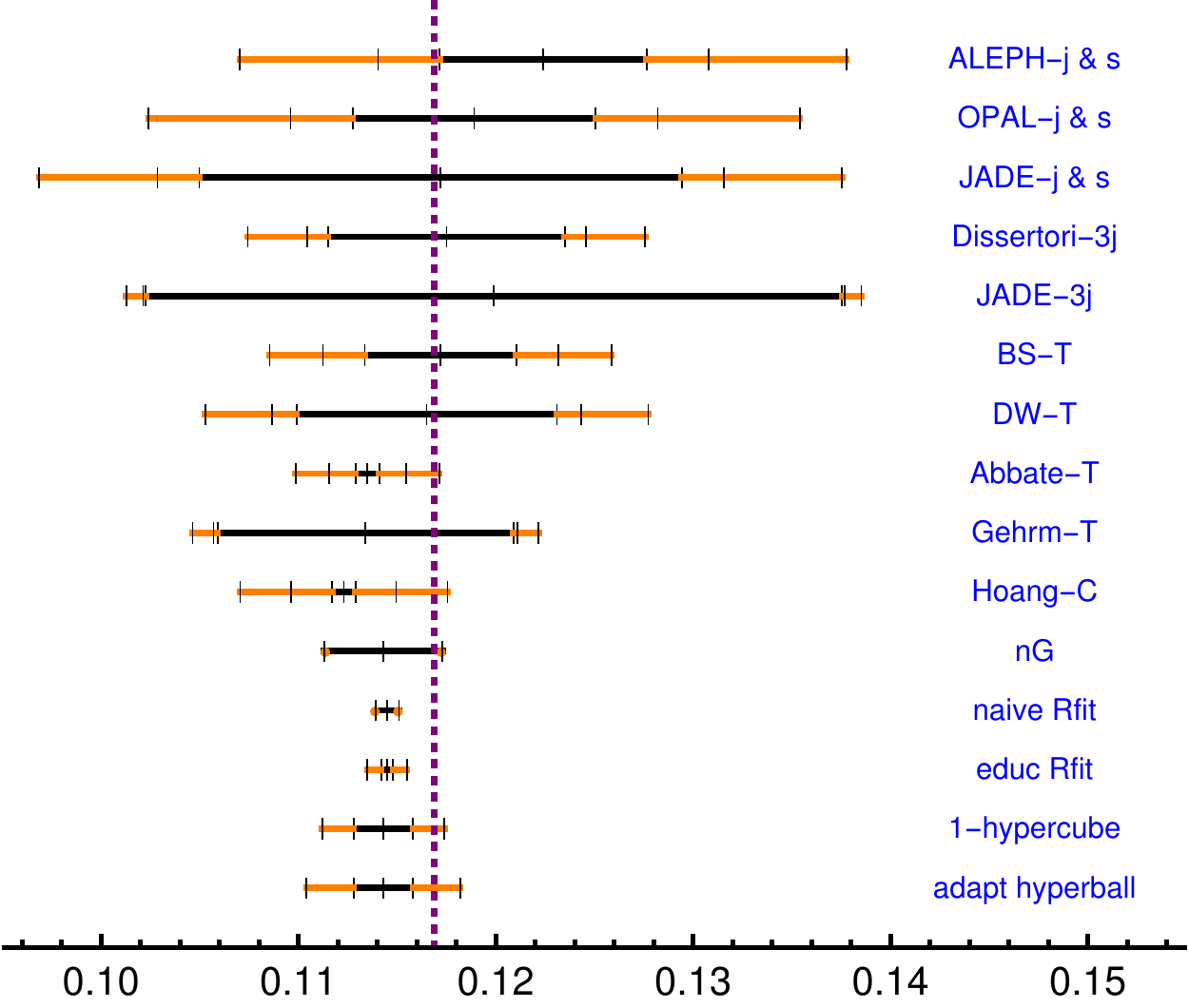}{3sigma_Image_alphaSMZ}

\pgfdeclareimage[height=5cm]{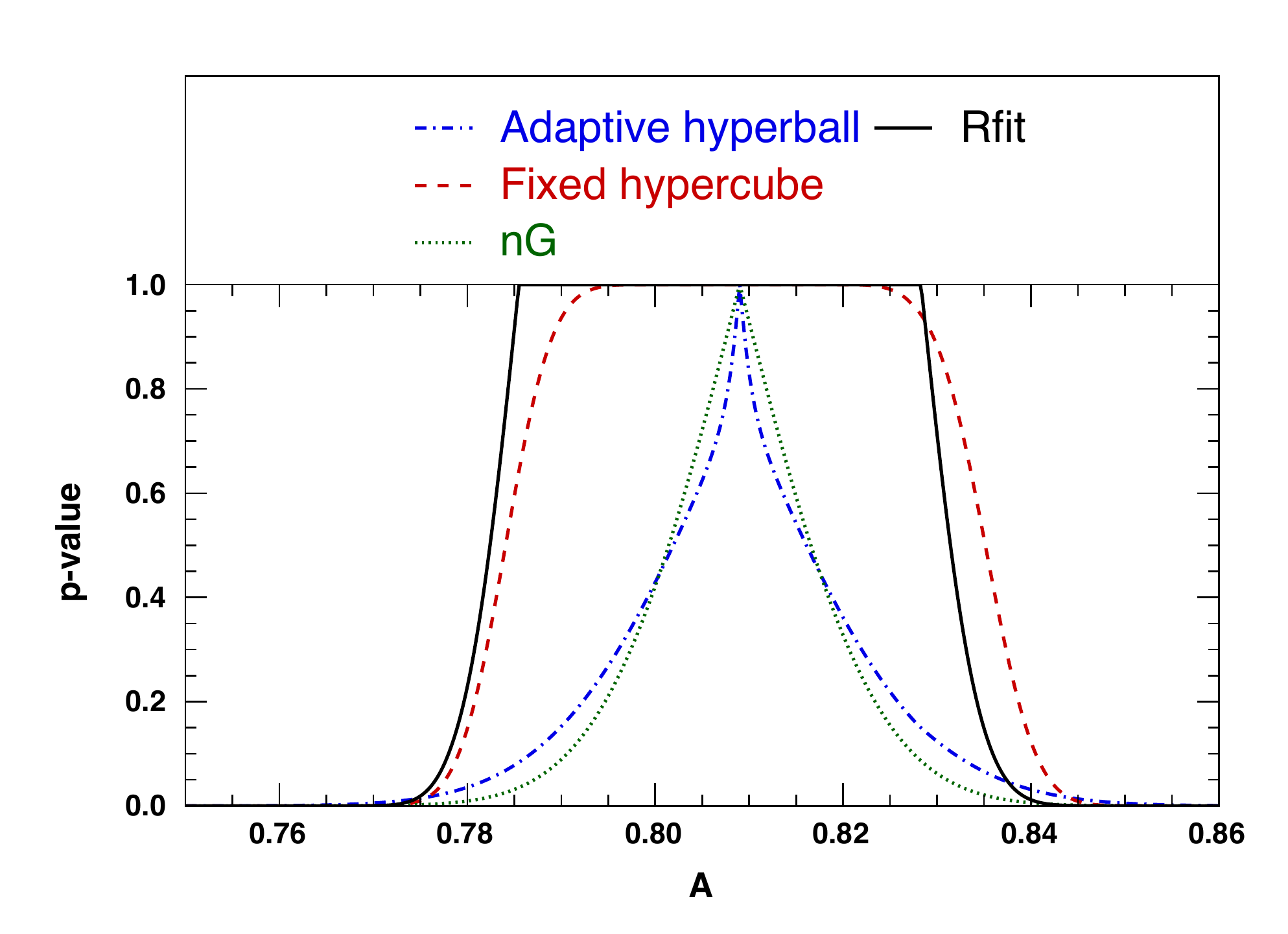}{scenarioA_A}
\pgfdeclareimage[height=5cm]{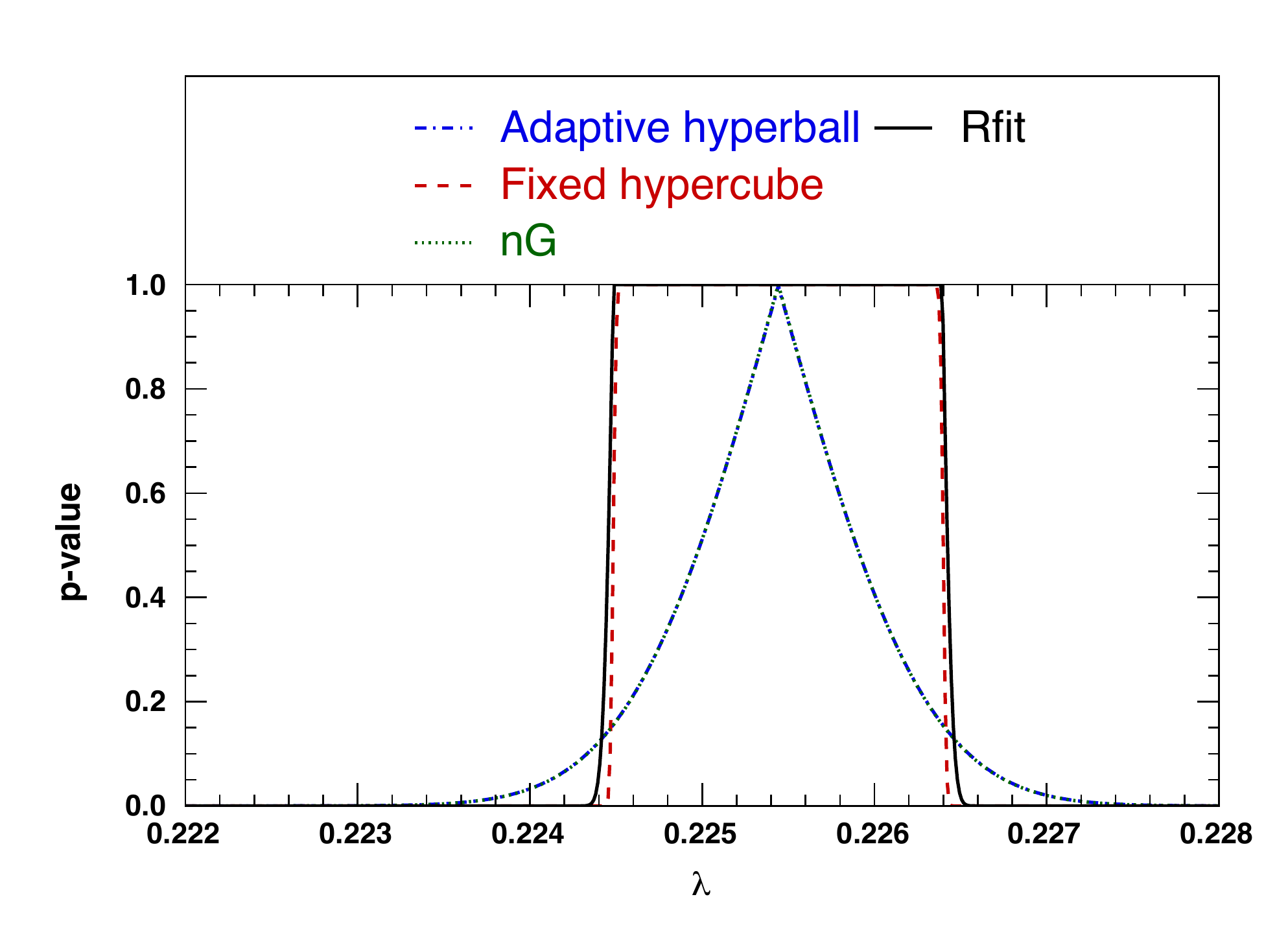}{scenarioA_lambda}
\pgfdeclareimage[height=5cm]{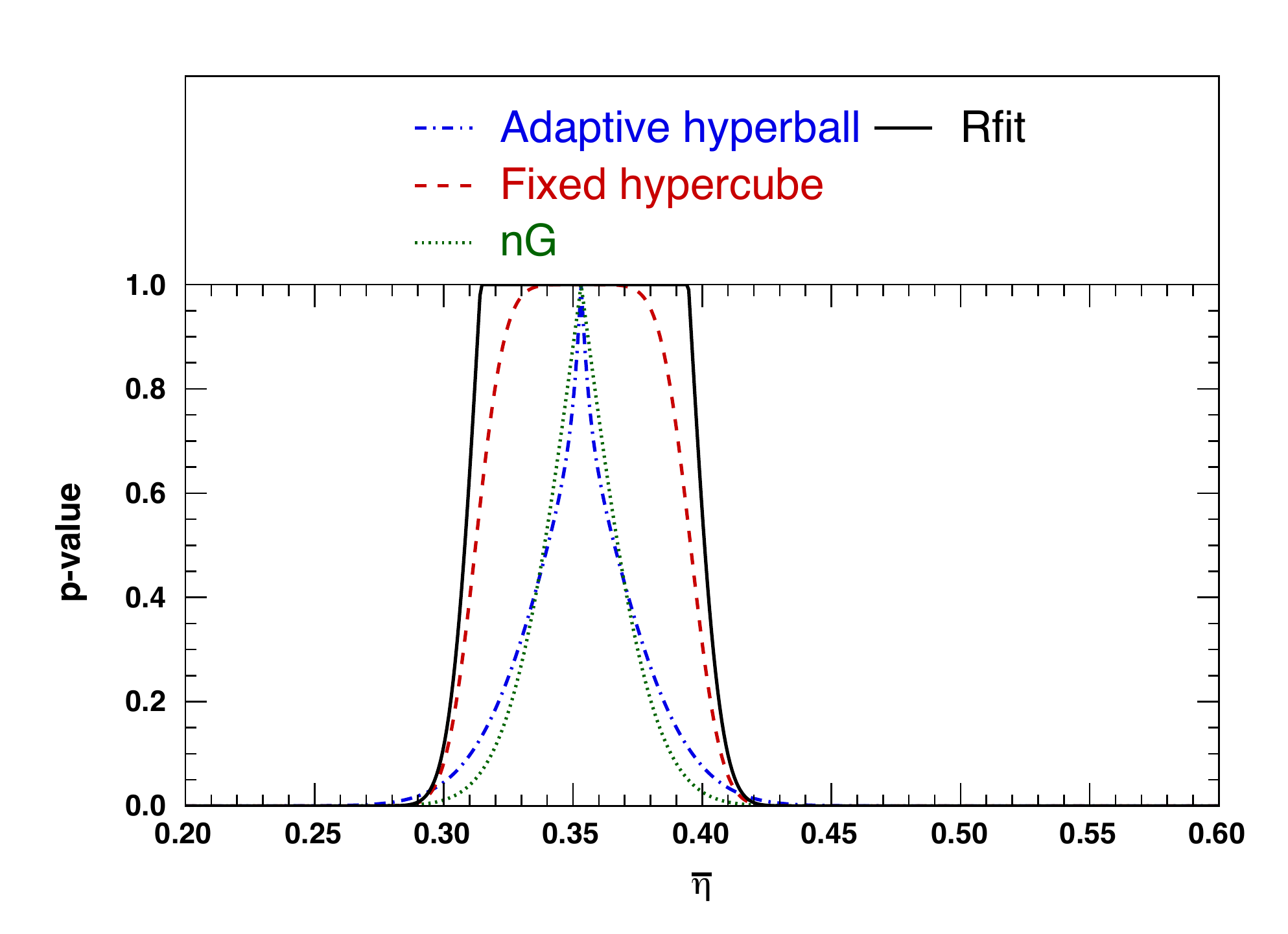}{scenarioA_etabar}
\pgfdeclareimage[height=5cm]{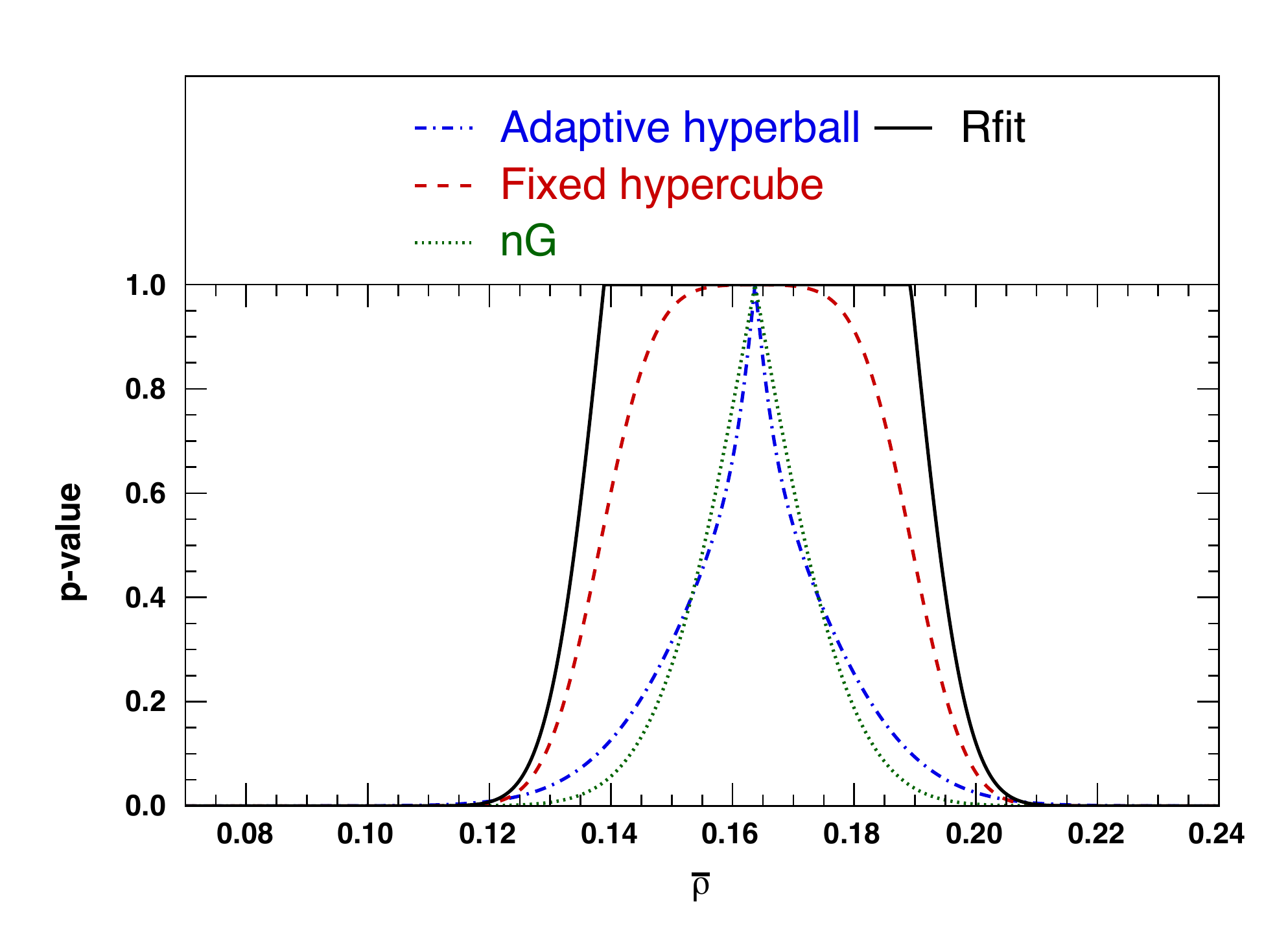}{scenarioA_rhobar}

\pgfdeclareimage[height=5cm]{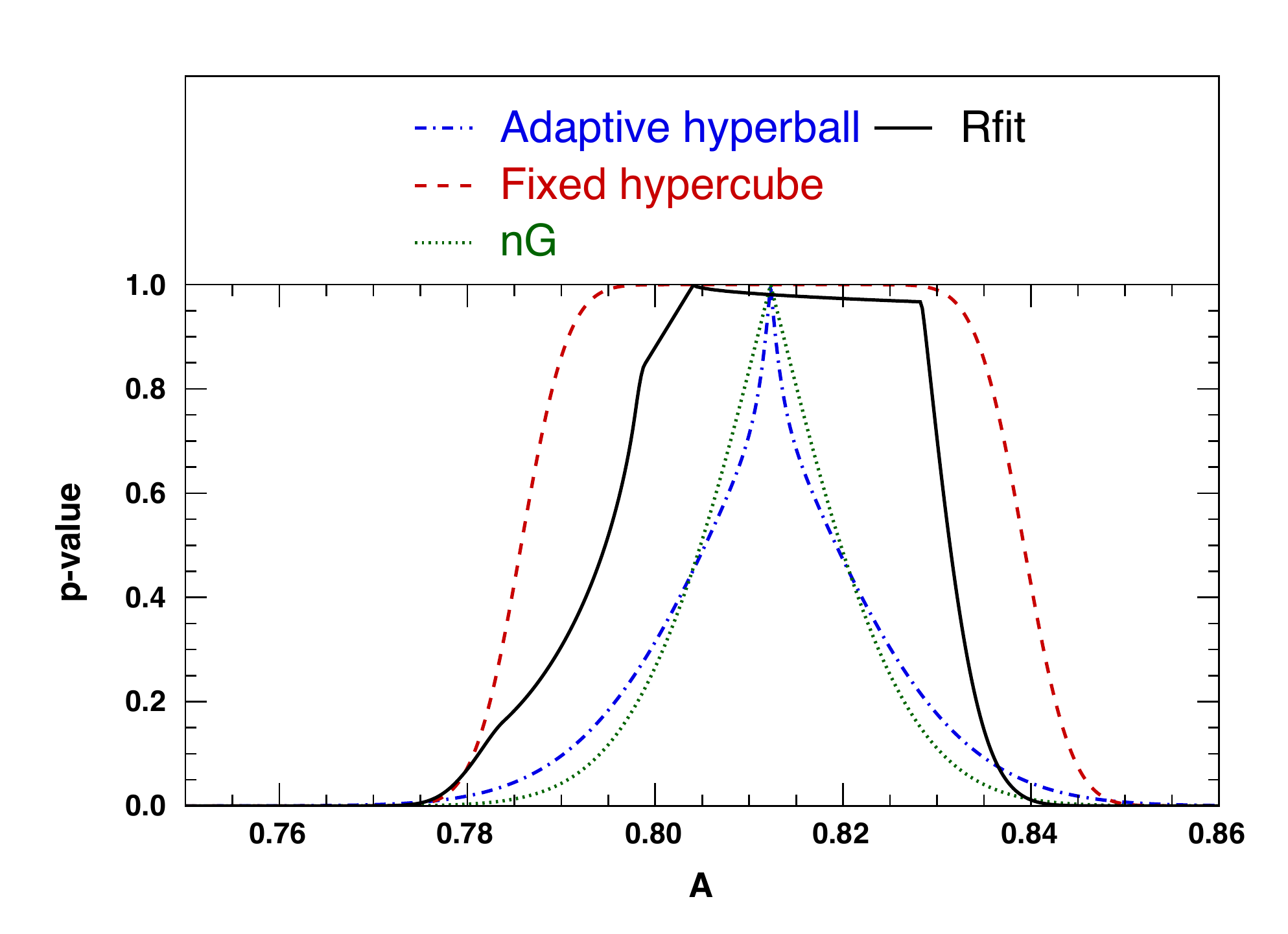}{scenarioB_A}
\pgfdeclareimage[height=5cm]{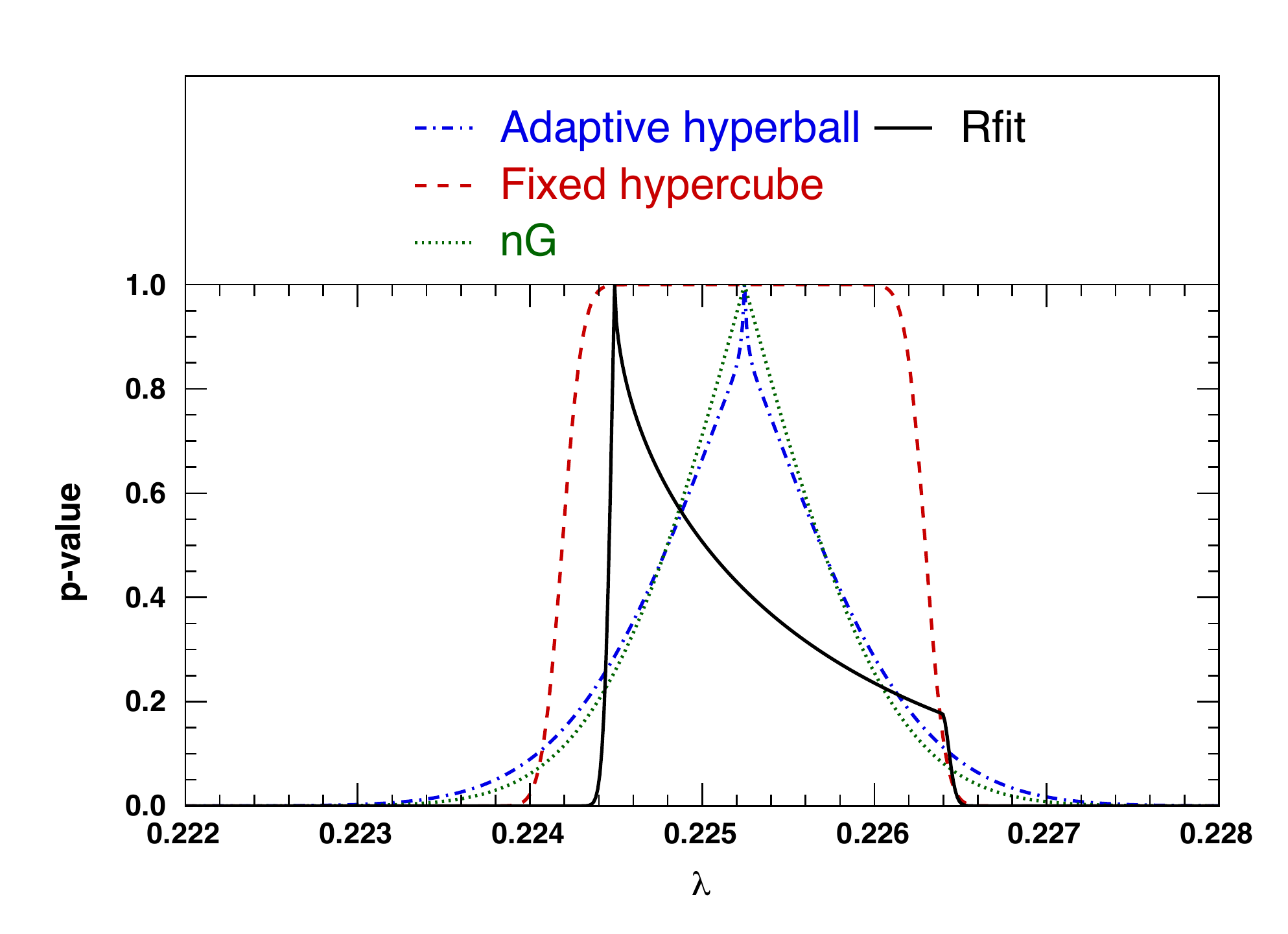}{scenarioB_lambda}
\pgfdeclareimage[height=5cm]{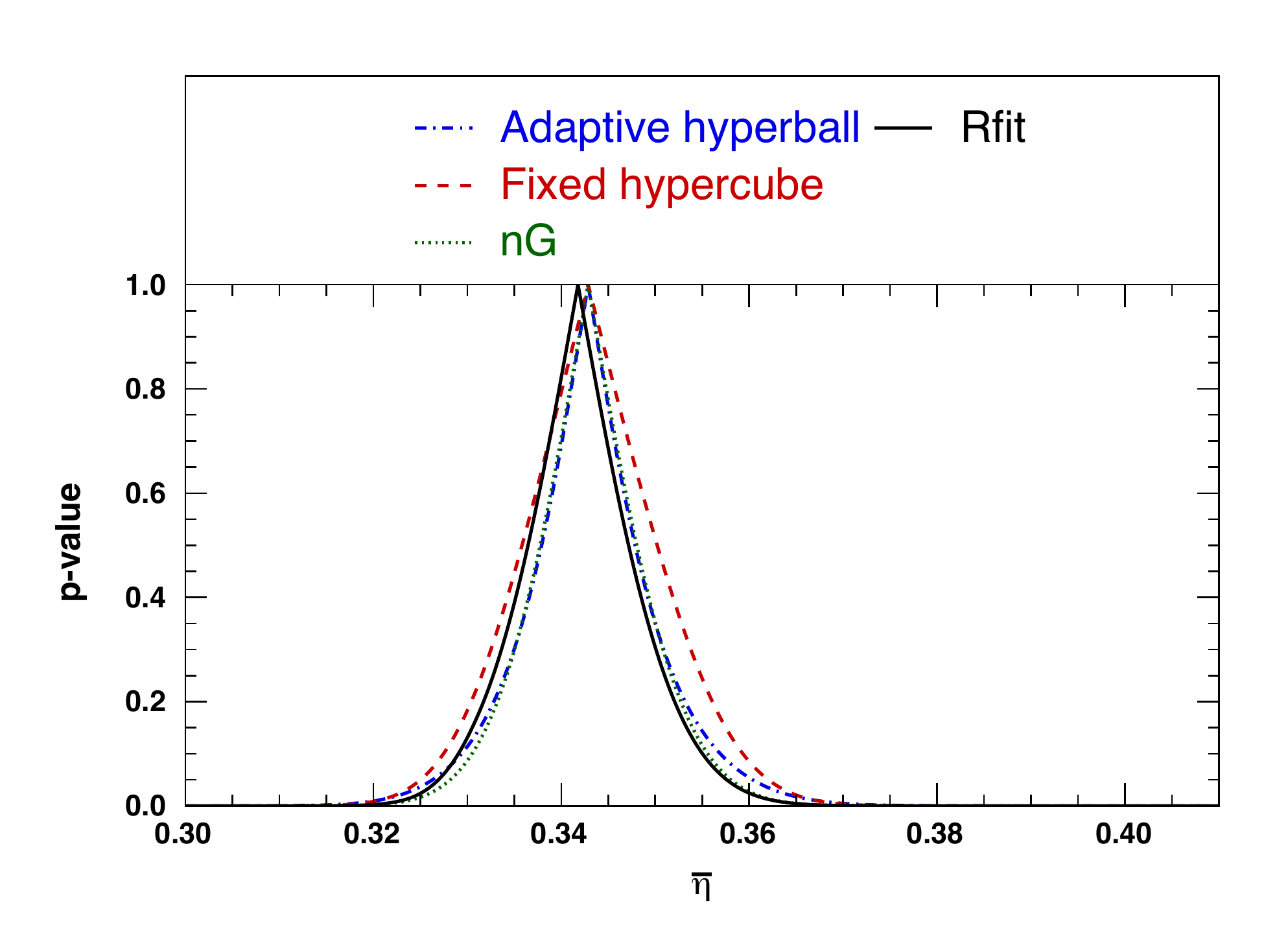}{scenarioB_etabar}
\pgfdeclareimage[height=5cm]{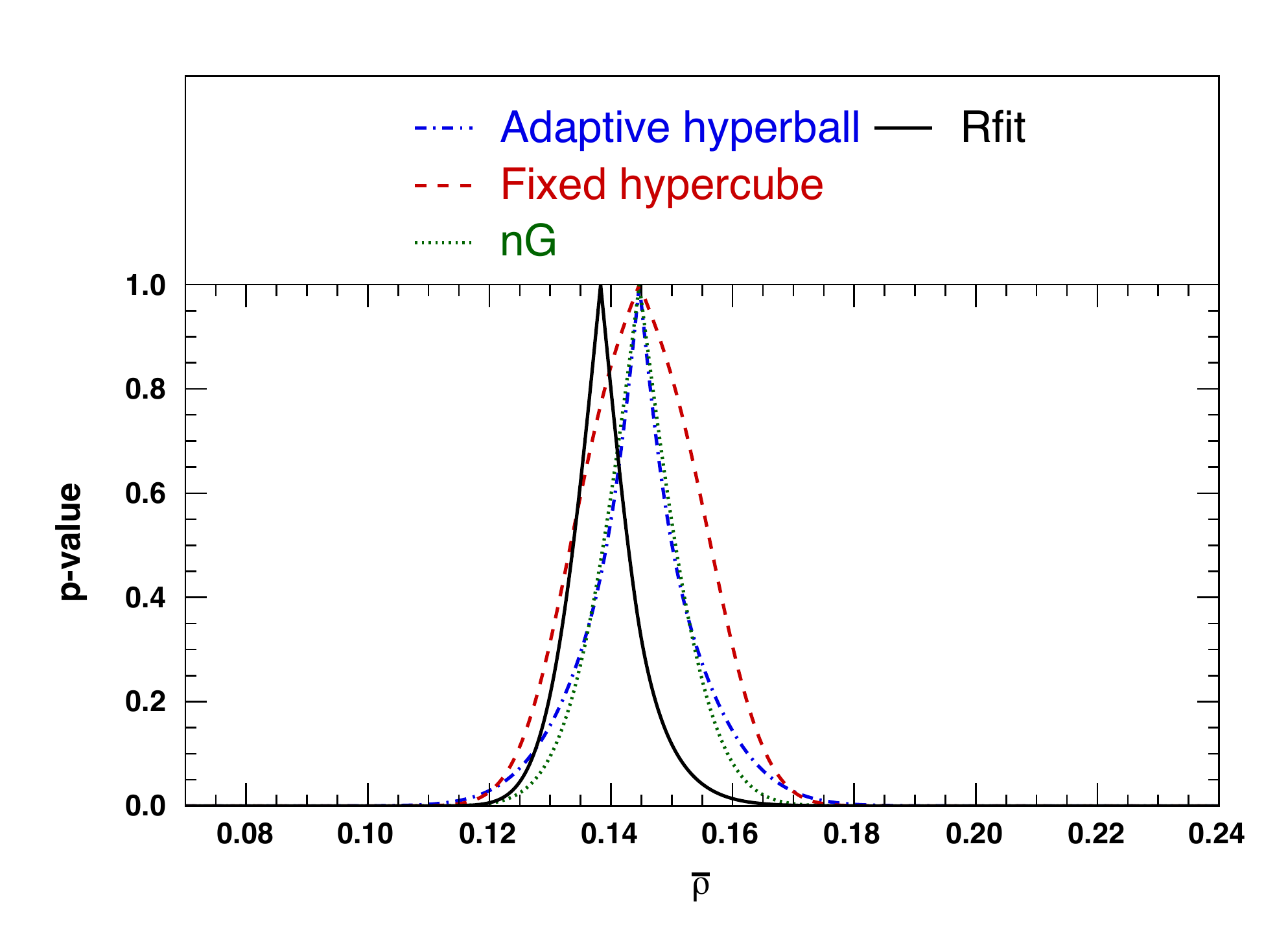}{scenarioB_rhobar}

\usepackage{color}

\usepackage{cite}

\begin{document}
\begin{flushright}
LPT-Orsay-16-65
\end{flushright}

\vspace{0.3cm}

\begin{center}
{\bf {\Large 
Modelling theoretical uncertainties\\ in phenomenological analyses for particle physics}}

\vspace{0.5cm}

J\'er\^ome Charles$^a$, S\'ebastien Descotes-Genon$^{b}$, Valentin Niess$^c$, Luiz Vale Silva$^{b,d,e}$

\vspace{0.2cm}

\emph{$^a$ CNRS, Aix-Marseille Univ, Universit\'e de Toulon, CPT UMR 7332,\\ F-13288 Marseille Cedex 9, France}\\
\emph{$^b$ Laboratoire de Physique Th\'eorique  (UMR 8627), CNRS, Univ. Paris-Sud,\\ Universit\'e Paris-Saclay, 91405 Orsay Cedex, France}\\
\emph{$^c$ Laboratoire de Physique Corpusculaire, CNRS/IN2P3, UMR 6533,\\ Campus des C\'ezeaux, 24 Av. des Landais, F-63177 Aubi\`ere Cedex, France}\\
\emph{$^d$ Groupe de Physique 
Th\'eorique, Institut de Physique Nucl\'eaire, Univ. Paris-Sud, CNRS/IN2P3,\\
Universit\'e Paris-Saclay, 91406 Orsay Cedex, France}\\
\emph{$^e$ J. Stefan Institute, Jamova 39, P. O. Box 3000, 1001 Ljubljana, Slovenia}

\vspace{3\baselineskip}

\vspace*{0.5cm}
\textbf{Abstract}\\
\vspace{1\baselineskip}
\parbox{0.9\textwidth}{The determination of the fundamental parameters of the Standard Model (and its extensions) is often limited by the presence of statistical and theoretical uncertainties. We present several models for the latter uncertainties (random, nuisance, external) in the frequentist framework, and we derive the corresponding $p$-values. In the case of the nuisance approach where theoretical uncertainties are modeled as biases, we highlight the important, but arbitrary, issue of the range of variation chosen for the bias parameters. We introduce the concept of adaptive $p$-value, which is  obtained by adjusting the range of variation for the bias according to the significance considered, and which allows us to tackle metrology and exclusion tests with a single and well-defined unified tool, which exhibits interesting frequentist properties. We discuss how the determination of fundamental parameters is impacted by the model chosen for theoretical uncertainties, illustrating several issues with examples from quark flavour physics.}
\end{center}

\thispagestyle{empty}

\clearpage
\setcounter{page}{1}
\tableofcontents 



\vspace{0.4cm}

In particle physics, an important part of the data analysis is devoted to the interpretation of the data with respect to the Standard Model (SM) or some of its extensions, with the aim of  comparing different alternative models or determining
the fundamental parameters of a given underlying theory~\cite{James:2006zz,Kendall-Stuart,Cowan:2013pha}. In this activity, the role played by uncertainties is essential, since they constitute the limit for the accurate determination of these parameters, and they can prevent from reaching a definite conclusion when comparing several alternative models. In some cases, these uncertainties are from a statistical origin: they are related to the intrinsic variability of the phenomena observed, they decrease as the sample size increases and they can be modeled using random variables. A large part of the experimental uncertainties belong to this first category. However, another kind of uncertainties occurs when one wants to describe inherent limitations of the analysis process, for instance, uncertainties in the calibration or limits of the models used in the analysis. These uncertainties are very often encountered in theoretical computations, for instance when assessing the size of higher orders in perturbation theory or the 
validity of extrapolation formulae. Such 
uncertainties are often called ``systematics'', but they
should be distinguished from less dangerous sources of systematic uncertainties, usually of experimental origin, that roughly scale with the size of the statistical sample and may be reasonably modeled by random variables~\cite{Sinervo:2003wm}. In the following we will thus call them ``theoretical'' uncertainties: by construction,
 they lack both an unambiguous definition (leading to various recipes to determine these uncertainties) and a clear interpretation (beyond the fact that they are not from a statistical origin). It is thus a complicated issue to incorporate their effect properly, even in simple situations often encountered in particle physics~\cite{Schmelling:2000td,Rolke:2004mj,Cousins:2008zz}~\footnote{ 
The issue of theoretical uncertainties is naturally not the only question that arises in the context of statistical analyses. The statistical framework used to perform these analyses is also a matter of choice, with two main approaches, frequentist and Bayesian, adopted in different settings and for various problems in and beyond high-energy physics~\cite{James:2006zz,Kendall-Stuart,Cowan:2013pha,Bolstad,DAgostini:2003bpu,Bevan}.
In this paper, we choose to focus on the frequentist approach to discuss how to model theoretical uncertainties.}.

The relative importance of statistical and theoretical uncertainties might be different depending on the problem considered, and the progress made both by experimentalists and theorists. For instance, statistical uncertainties are the main issue in the analysis of electroweak precision observables~\cite{Schael:2013ita,Baak:2014ora}. On the other hand, in the field of quark flavour physics, theoretical uncertainties play a very important role. Thanks to the $B$-factories and LHCb, many hadronic processes have been very accurately measured~\cite{Bediaga:2012py,Bevan:2014iga}, which can provide stringent constraints on the Cabibbo-Kobayashi-Maskawa matrix (in the Standard Model)~\cite{Hocker:2001xe,Charles:2004jd,Charles:2011va}, and on the scale and structure of New Physics (in SM extensions)~\cite{Deschamps:2009rh,Lenz:2010gu,Lenz:2012az,Charles:2013aka}. However, the translation between hadronic processes and quark-level transitions requires information on hadronisation from strong interaction, encoded in decay constants, form factors, bag parameters\ldots The latter are determined through lattice QCD 
simulations. The remarkable progress in computing power and in 
algorithms over the last 20 years has led to a decrease of statistical uncertainties and a dominance of purely theoretical uncertainties (chiral and heavy-quark extrapolations, scale chosen to set the lattice spacing, finite-volume effects, continuum limit\ldots). As an illustration, the determination of the Wolfenstein parameters of the CKM matrix involves many constraints which are now limited  by theoretical uncertainties (neutral-meson mixing, leptonic and semileptonic decays\ldots)~\cite{Charles:2015gya}.

The purpose of this note is to discuss theoretical uncertainties in more detail in the context of particle physics phenomenology, comparing different models not only from a statistical point of view, but also in relation with the
problems encountered in phenomenological analyses where they play a significant role.
In Sec.~\ref{sec:intro}, we summarise fundamental notions of statistics used in particle physics, in particular $p$-values and test statistics. In Sec.~\ref{sec:threeapproaches}, we list properties that we seek in a good approach for theoretical uncertainties. In Sec.~\ref{sec:1Dillustation}, we propose several approaches and in Sec.~\ref{sec:1Dcompar}, we compare their properties in the most simple one-dimensional case. In Sec.~\ref{sec:nD}, we consider multi-dimensional cases (propagation of theoretical uncertainties, average of several measurements, fits and pulls), which we illustrate using flavour physics examples related to the determination of the CKM matrix in Sec.~\ref{sec:ckm}, before concluding.
An appendix is devoted to several issues connected with the treatment of correlations.

\input{introduction.tex}

\input{threeapproaches.tex}

\input{1Dillustration.tex}

\input{1Dcomparison.tex}

\input{nDcases.tex}

\input{ckm.tex}

\section{Conclusion}

A problem often encountered in particle physics consists in analysing data within the Standard Model (or some of its extensions) in order to extract information on the fundamental parameters of the model. An essential role is played here by uncertainties, which can be classified in two categories, statistical and theoretical. If the former can be treated in a rigorous manner within a given statistical framework, the latter must be described through models. The problem is particularly acute in flavour physics, as theoretical uncertainties often play a central role in the determination of underlying parameters, such as the four parameters describing the CKM matrix in the Standard Model.

This article aims at describing and comparing several approaches that can be implemented in a frequentist framework. After recalling some elements of frequentist analysis, we have discussed three different approaches for theoretical uncertainties: the random-$\delta$ approach treats theoretical uncertainties as random variables, the external-$\delta$ approach considers them as external parameters leading to an infinity of $p$-values to be combined through model averaging, the nuisance-$\delta$ describes them through fixed biases which have to be varied over a reasonable region. These approaches have to be combined with particular choices for the test statistic used to compute the $p$-value. We have illustrated these approaches in the one-dimensional case, recovering the Rfit model used by CKMfitter as a particular case of the external-$\delta$ approach, and discussing the interesting alternative of a quadratic test statistic. 

In the case of the nuisance-$\delta$ approach, one has to decide over which range the bias parameter should be varied. It is possible to compute the $p$-value by taking the supremum of the bias over a fixed range fixed by the size of the theoretical uncertainty to be modeled (fixed nuisance approach). An alluring alternative consists in adjusting the size of the range to the confidence level chosen: the range for a low confidence level can be obtained by varying the bias parameter in a small range, whereas a range for a high confidence level could require a more conservative (and thus larger) range for the bias parameter. We have designed such a scheme, called adaptive nuisance approach. It provides a unified statistical approach to deal with the metrology of the parameters (for low CL ranges) and the exclusion of models (for high CL ranges). 

We have determined the $p$-values associated with each approach for a measurement involving both statistical and theoretical uncertainties. We have also studied the size of error bars, the significance of deviations and the coverage properties. In general, the most conservative approaches correspond to a naive Gaussian treatment (belonging to the random-$\delta$ approach) and the adaptive nuisance approach. The latter is better defined and more conservative than the former in the case where statistical and theoretical approaches are of similar size. Other approaches (fixed nuisance, external) turn out less conservative at large confidence level.

We have then considered extensions to multi-di\-mensional cases, focusing on the linear case where the quantity of interest is a linear combination of observables. Due to the presence of several bias parameters, one has to make another choice concerning the shape of the space over which the bias parameters are varied. Two simple examples are the hypercube and the hyperball, leading to a linear or quadratic combination of theoretical uncertainties respectively. The hypercube is more conservative, as it allows for sets of values of the bias parameters that cannot be reached within the hyperball. On the other hand, the hyperball has the great virtue of associativity, so that one can average different measurements of the same quantity or put all of them in a global fit, without changing its outcome.
It also allows us to include theoretical correlations easily, both in the range of variation of biases to determine errors and in the definition of theoretical correlations for the outcome of a fit.
We have discussed the average of several measurements using the various approaches, including correlations. We considered in detail the case of 100\% correlations leading to a non-invertible covariance matrix.
We also discussed global fits and pulls in a linearised context.
We have then provided several comparisons between the different approaches using examples from flavour physics: averaging theory-dominated measurements, averaging incompatible measurements linear fits to a subset of flavour inputs.

It is now time to determine which choice seems preferable in our case. Random-$\delta$ has no strong statistical basis: its only advantage consists in its simplicity. External-$\delta$
is closer in spirit to the determination of systematics as performed by experimentalists, but it starts
with an inappropriate 
 null hypothesis and tries to combine an infinite set of $p$-values in a single $p$-value. On the contrary, the nuisance-$\delta$ approach starts from the beginning with the correct null hypothesis and deals with a single $p$-value.

This choice is independent from another choice, i.e., the range of variation for the parameter $\delta$. Indeed, when several bias parameters are involved, one may imagine different multidimensional spaces for their variations, in particular the hyperball and the hypercube. As said earlier, the hyperball has the interesting property of associativity when performing averages and avoids fine-tuned solutions where all parameters are pushed in a corner of phase space. The  hypercube is closer in spirit to the Rfit model (even though the latter is not a bias model), but it cannot avoid fine-tuned situations and it does not seem well suited to deal with theoretical correlations, since it is designed from the start to avoid such correlations.

A third choice consists in determining whether one wants to keep the volume of variation fixed (fixed approach), or to modify it depending on the desired confidence level (adaptive approach). 
Adaptive hypercube is in principle the most conservative choice but in practice, it gives too large errors, whereas fixed hyperball would give very small errors. Fixed hypercube is more conservative at low confidence levels (large $p$-values), whereas adaptive hyperball is more conservative at large confidence levels (small $p$-values). 

This overall discussion leads us to consider the nuisance approach with adaptive hyperball as a promising approach to deal with flavour physics problems, which we will investigate in more phenomenological analyses in forthcoming publications~\cite{wip}.

\section*{Acknowledgments}

We would like to thank S.~T'Jampens for collaboration at an early stage of this work, as well as all our collaborators from the CKMfitter group for many useful discussions on the statistical issues covered in this article. We would also like to express a special thanks to the Mainz Institute for Theoretical Physics (MITP) for its hospitality and support during the workshop ``Fundamental parameters from lattice QCD'' where part of this work was presented and discussed. LVS acknowledges financial support from the Labex P2IO (Physique des 2 Infinis et Origines). SDG acknowledges partial support from Contract FPA2014-61478-EXP.
This project has received funding from the European UnionÕs Horizon 2020 research and innovation programme under grant agreements No 690575, No 674896 and No. 692194.

\appendix

\input{singular.tex}

\input{correlation.tex}

\input{reduction.tex}

\input{asymmetric.tex}

\input{ckmappendix.tex}

\end{document}

%% file: introduction.tex
\section{Statistics concepts for particle physics}\label{sec:intro}

We start by briefly recalling frequentist concepts used in particle physics, highlighting the role played by $p$-values in hypothesis testing and how they can be used to define confidence intervals.

\subsection{$p$-values}

\subsubsection{Data fitting and data reduction}

First, we would like to illustrate the concepts of data fitting and data reduction in particle physics, starting with a specific example, namely the observation of the time-dependent CP asymmetry in the decay channel $B^0(t)$ $\to J/\psi K_S$ by the BaBar, Belle and LHCb experiments~\cite{Aubert:2009aw,Adachi:2012et,Aaij:2015vza}. Each experiment collects a sample of observed decay times ${t_i}$ corresponding to the $B$-meson events, where this sample is theoretically known to follow a \textrm{PDF} $f$. The \textrm{PDF} is parameterized in terms of a few physics parameters, among which we assume the ones of interest are the direct and mixing-induced $C$ and $S$ CP asymmetries. The functional form of this \textrm{PDF} is dictated on very general grounds by the CPT invariance and the formalism of two-state mixing (see, \textit{e.g}.,~\cite{Bigi:1987in}), and is independent of the particular underlying phenomenological model (\textit{e.g.} the Standard Model of particle physics). In practice however, detector effects require to be modelled by additional parameters that modify the shape of the \textrm{PDF}. We denote by $\theta$ the set of parameters $\theta=(C,S,\ldots)$ that are needed to specify the \textrm{PDF} completely.
The likelihood for the sample $\{t_i\}$ is defined by
\begin{equation}
\mathcal L_{\{t_i\}}(\theta) = \prod_{i=1}^n f(t_i;\theta)
\end{equation}
and can be used as a test statistic to infer constraints on the parameters
$\theta$, and/or construct estimators for them, as will be discussed in more detail below. The combination of different samples/experiments can be done simply by multiplication of the corresponding likelihoods. On the other hand one can choose to work directly in the framework of a specific phenomenological model, by replacing in $\theta$ the quantities that are predicted by the model in terms of more fundamental parameters: for example in the Standard Model, and neglecting the ``penguin'' contributions, one has the famous relations $C=0$, $S=\sin2\beta$ where $\beta$ is one of the angles of the Unitarity Triangle and can be further expressed in terms of the Cabibbo-Kobayashi-Maskawa couplings.

The latter choice of expressing the experimental likelihood in terms of model-dependent parameters such as $\beta$ has however one technical drawback: the full statistical analysis has to be performed for each model one wants to investigate, e.g., the Standard Model, the Minimal Supersymmetric Standard Model, GUT models\ldots  In addition, building a statistical analysis directly on the initial likelihood requires one to deal with a very large parameter space, depending on the parameters in $\theta$ that are needed to describe the detector response. One common solution to these technical difficulties is a two-step approach. In the first step, the data are \textit{reduced} to a set of model- and detector-independent~\footnote{
It may happen that the detector and/or background effects have a sizeable impact on the fitted quantities $\hat{C}$ and $\hat{S}$; this can be viewed as uncertainties in the modelling of the event \textrm{PDF} $f$. These effects are reported as \textit{systematic uncertainties} and in particle physics, it is customary to treat them on the same footing as the pure statistical uncertainties. Although we will not try to follow this avenue  in the examples discussed here, 
it would be possible to consider these systematic uncertainties as theoretical uncertainties, to be modelled according to the methods that we describe in the following sections.}
random variables that contains the same information as the original likelihood (to a good approximation): in our example the likelihood-based estimators  $\hat C$ and $\hat S$ of the parameters $C$ and $S$ can play the role of such variables  (estimators are functions of the data and thus are random variables). In a second step, one can work
in a particular model, e.g., in the Standard Model, to use  $\hat C$ and $\hat S$ as inputs to a statistical analysis of the parameter $\beta$.
This two-step procedure gives the same result as if the analysis were done in a single step through the expression of the original likelihood in terms of $\beta$. This technique is usually chosen if the \textrm{PDF} $g$ of the estimators $\hat C$ and $\hat S$ can be parameterized in a simple way: for example, if the sample size is sufficiently large, then the \textrm{PDF} can often be modelled by a multivariate normal distribution, where the covariance matrix is approximately independent of the mean vector.

Let us now extend the above discussion to a more general case. A sample of random events is $\{E_i,i=1\ldots n\}$, where each event corresponds to a set of directly measurable quantities (particle energies and momenta, interaction vertices, decay times\ldots). The distribution of these events is described by a \textrm{PDF}, the functional form $f$ of which is supposed to be known. In addition to the event value $E$, the \textrm{PDF} value depends on some fixed parameters ${\theta}$, hence the notation $f(E;\theta)$. The likelihood for the sample $\{E_i\}$ is defined by
\(
\mathcal L_{\{E_i\}}(\theta) = \prod_{i=1}^n f(E_i;\theta).
\)
We want to interpret the event observation in a given phenomenological scenario that predicts 
at least some of the parameters $\theta$ describing the {\rm PDF} 
in terms of a set of more fundamental parameters $\chi$. 

To this aim we first reduce the event observation to a set of model- and detector-indepen\-dent random variables $X$ together with a \textrm{PDF} $g(X;\chi)$, in such a way that the information that one can get on $\chi$ from $g$ is equivalent to the information one can get from $f$, once $\theta$ is expressed in terms of $\chi$ consistently with the phenomenological model of interest. Technically, it amounts to identifying a minimal set of variables $x$ depending on $\theta$ that are independent of both the experimental context and the phenomenological model. One performs an analysis on the sample of events ${E_i}$ to derive estimators $\hat x$ for $x$. The distribution of these estimators
 can be described in terms of a \textrm{PDF} that is written in the $\chi$ parametrization as $g(X;\chi)$, where we have replaced $\hat x$ by the notation $X$, to stress that in the following $X$ will be considered as a new random variable, setting aside how it has been constructed from the original data $\{E_i\}$. Obviously, in our previous example for $B^0(t)$ $\to J/\psi K_S$, $\{t_i\}$ correspond to $\{E_i\}$, $C$ and $S$ to $x$, and $\beta$ to $\chi$.

\subsubsection{Model fitting}\label{sec:pvalues}
From now on we work with one or more observable(s) $x$, with associated random variable $X$, and an associated \textrm{PDF} $g(X;\chi)$ depending on purely theoretical parameters $\chi$. With a slight abuse of notation we include in the symbol $g$ not only the functional form, but also all the needed parameters that are kept fixed and independent of $\chi$. In particular for a one-dimensional Gaussian \textrm{PDF} we have 
\begin{equation}
g(X;\chi)\sim \exp\left[-\frac{1}{2}\left(\frac{X-x(\chi)}{\sigma}\right)^2\right]
\end{equation}
where $X$ is a potential value of the observable $x$ and $x(\chi)$ corresponds to the theoretical prediction of $x$ given $\chi$.
This \textrm{PDF} is obtained from the outcome of an experimental analysis yielding both a central value $X_0$ and an uncertainty $\sigma$, where $\sigma$ is assumed to be independent of the realisation $X_0$ of the observable $x$ and is thus included in the definition of $g$.

Our aim is to derive constraints on the parameters
$\chi$, from the measurement $X_0\pm \sigma$ of the observable $x$. One very general way to
perform
this task is \textit{hypothesis testing}, where one wants to quantify how
much
the data are compatible with the null hypothesis that the true value of $\chi$, $\chi_t$, is equal to some fixed value $\chi$:
\begin{equation}\label{hyp_q}
{\mathcal H}_{\chi}: \chi_t = \chi
\end{equation}
In order to interpret the observed data $X_0$ measured in a given experiment
in light of the distribution of the observables $X$ under the null
hypothesis ${\mathcal H}_\chi$, one
defines a \textit{test statistic} $T(X;\chi)$, that is a scalar function of
the data $X$ that
measures whether the data are in favour or not of the null hypothesis. 
We indicated the dependence of $T$ on $\chi$ explicitly, i.e., the dependence on the null
hypothesis ${\mathcal H}_\chi$. 
The test statistic is generally a definite positive function chosen in a way that
large
values  indicate that the data present evidence against the null hypothesis. By comparing
the actual data value $t=T(X_0;\chi)$ with
the sampling distribution of $T=T(X;\chi)$ under the null hypothesis, one is
able to quantify the degree of agreement of
the data
with the null hypothesis.

\begin{figure}
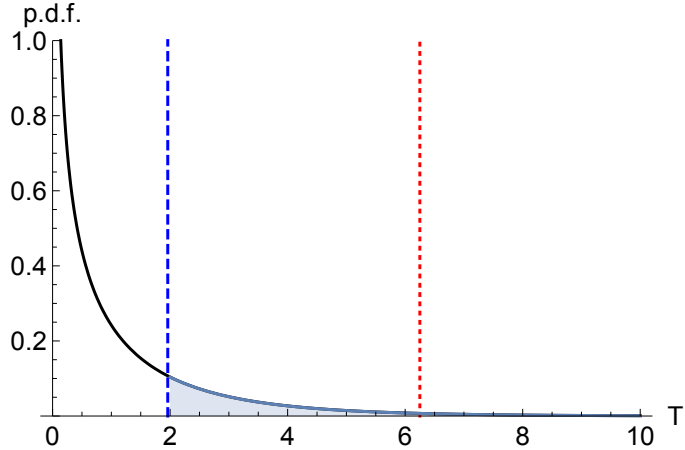

\begin{center}
\pgfuseimage{Pdfchi2}
\end{center}
\caption{Illustration in the simple case where $X$ is predicted as $x(\mu)=\mu$.
Under the hypothesis $\mu_t=\mu$, and having measured $X=0\pm 1$, one can determine the associated $p$-value $p(0;\mu)$ by examining the distribution of the quadratic test statistic $T(X;\mu)=(X-\mu)^2$ assuming $X$ is distributed as a Gaussian random variable with central value $0$ and width 1. The blue dashed line corresponds to the value of $T$ associated with the hypothesis $\mu=-1.4$, with a $p$-value obtained by considering the gray area. The red dotted line corresponds to the hypothesis $\mu=2.5$.}
\label{fig:pvalueexample}
\end{figure}

\begin{figure*}
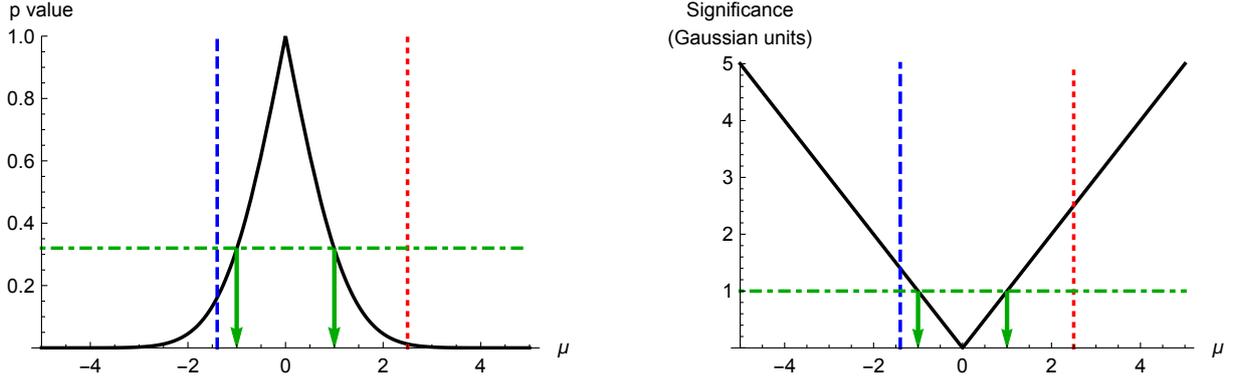

\begin{center}
\pgfuseimage{CLs}
\end{center}
\caption{On the left: for a given observation $X=0\pm1$, $p$-value $p(0;\mu)$ as a function of the value of $\mu$ being tested. Blue dashed and red dotted lines correspond to $\mu=-1.4$ and $\mu=2.5$. A confidence interval for $\mu$ at 68\% CL is obtained by considering the region of $\mu$ with a $p$-value larger than 0.32, as indicated by the green dotted dashed line and arrows. On the right: the same information is expressed in Gaussian units of $\sigma$, where the 68\% CL  interval corresponds to the region below the horizontal line of significance 1.}
\label{fig:CLexample}
\end{figure*}

Mathematically it amounts to defining a \textit{$p$-value}. One calculates the probability
to obtain a value for the test statistic at least as large as the one that was
actually observed, assuming that the null hypothesis is true. This tail probability is used to define the $p$-value of the test for this particular observation
\begin{equation}\label{pvalue}
1-p(X_0;\chi) = \int^{T(X_0;\chi)}_{0} dT\, h(T|{\mathcal H}_\chi)={\mathcal P}[T<T(X_0;\chi)]
\end{equation}
where the \textrm{PDF} $h$ of the test statistic is obtained from the ${\rm PDF}$ $g$ of
the data
as
\begin{equation}\label{eq:pdft}
h(T|{\mathcal H}_\chi) = \int dX\, \delta\left[T-T(X;\chi)\right] g(X;\chi)
\end{equation}
which can be obtained easily from comparing the convolution of $\frac{dT}{dX}h(T)=g(X)$ with a test function of $T$ with the convolution of the r.h.s. of~(\ref{eq:pdft}) with the same test function. A small value of the $p$-value means that
$T(X_0;\chi)$ belongs to the ``large'' region, and thus provides evidence against the null hypothesis.
This is illustrated for a simple example in Figs.~\ref{fig:pvalueexample} and \ref{fig:CLexample}.

From its definition, one sees that $1-p(X_0;\chi)$
is nothing else but the cumulative distribution function of the \textrm{PDF} $h$
\begin{equation}
{\rm CDF}[h](T(X_0;\chi)|{\mathcal H}_\chi) = \int dX\, \theta\left[T(X_0;\chi)-T(X;\chi)\right] g(X;\chi)
\end{equation}
where $\theta$ is the Heaviside function. This expression corresponds to the probability for the test statistic to be smaller than a given value $T(X_0;\chi)$.
The $p$-value in Eq.~(\ref{pvalue}) is defined as a function of $X_0$ and as such, is a random
variable.

Through the simple change of variable $\frac{dp}{dT}\frac{d\mathcal P}{dp}=\frac{d\mathcal P}{dT}$, one obtains that \textit{the null distribution}
(that is, the distribution when the null hypothesis is true) \textit{of a $p$-value is uniform}, i.e., the distribution of values of the p-value is flat between 0 and 1.  
This uniformity is a fundamental property of $p$-values that is at the core of their various interpretations (hypothesis comparison, determination of confidence intervals\ldots)~\cite{James:2006zz,Cowan:2013pha}. 

In the frequentist approach, one wants to design a procedure to decide whether to accept or reject the null hypothesis ${\mathcal H}_\chi$, by avoiding as much as possible either incorrectly rejecting the null hypothesis (Type-I error) or incorrectly accepting it (Type-II error). The standard frequentist procedure consists in selecting a Type-I error $\alpha$ and determining a region of sample space that has the probability $\alpha$ of containing the data under the null hypothesis. If the data fall in this critical region, the hypothesis is rejected. This must be performed before data are known (in contrast to other interpretations, e.g, Fischer's approach of significance testing~\cite{James:2006zz}). In the simplest case, the critical region is defined by a condition of the form $T\geq t_\alpha$, where $t_\alpha$ is a function of $\alpha$ only, which can be rephrased in terms of $p$-value as $p\leq \alpha$. The interest of the frequentist approach depends therefore on the ability to design $p$-values assessing the rate of Type-I error correctly (its understatement is clearly not desirable, but its overstatement yields often a reduction in the ability to determine the truth of an alternative hypothesis), as well as avoiding too large a Type-II error rate.

A major difficulty arises when the hypothesis to be tested is \emph{composite}.
In the case of numerical hypotheses like~(\ref{hyp_q}), one gets
compositeness when one is only interested in a subset $\mu$
of the parameters $\chi$. The remaining parameters are called
\textit{nuisance parameters}~\footnote{
``Nuisance'' does not mean that these parameters are necessarily
unphysical, ``pollution'' parameters. They can be fundamental constants
of Nature, and interesting as such.
} and will be denoted by $\nu$, thus
$\chi=(\mu,\nu)$. In this case the hypothesis
${\mathcal H}_\mu: \mu_t=\mu$ is composite, because determining the distribution of
 the observables requires the knowledge of the true value $\nu_t$
in addition to $\mu$. In this situation, one has to devise a procedure to infer a ``$p$-value'' for ${\mathcal H}_\mu$ out of
$p$-values built for the simple hypotheses where both $\mu$ and $\nu$ are fixed. 
Therefore, in contrast to a simple hypothesis, a composite
hypothesis does not allow one to compute the distribution of the data~\footnote{
We have defined compositeness for numerical hypotheses, since this is our case of interest in the following.
More generally, compositeness also occurs in the case of non-numerical hypotheses such as ``The Standard Model is true'', for which it is not possible to compute the distribution of data either. Indeed assuming that the Standard Model is true does not imply anything on the value of its fundamental parameters, and thus one
cannot compute the distribution of a given observable under this
hypothesis.}.

At this stage, it is not necessarily guaranteed that the
distribution of the $p$-value for ${\mathcal H}_\mu$  is uniform, and one may get different situations:
\begin{eqnarray} \label{eq:pvalue1}
p{\rm\ exact} &:& P(p\leq \alpha | {\mathcal H}_\mu)=\alpha\\
p{\rm\ conservative}&:& P(p\leq \alpha | {\mathcal H}_\mu)<\alpha\\
p{\rm\ liberal}&:& P(p\leq \alpha | {\mathcal H}_\mu)>\alpha \label{eq:pvalue3}
\end{eqnarray}
which may depend on the value of $\alpha$ considered. Naturally, one would like to design as much as possible an exact $p$-value (exact
coverage), or if this is not possible, a (reasonably) conservative one (overcoverage). 
Such $p$-values will be called ``valid'' $p$-values.
In the case of composite hypotheses, the
conservative or liberal nature of a $p$-value may depend not only on $\alpha$, but also on the structure of the problem and of the procedure used to construct the $p$-value, and it has to be checked explicitly~\cite{James:2006zz,Cowan:2013pha}.

\begin{figure}
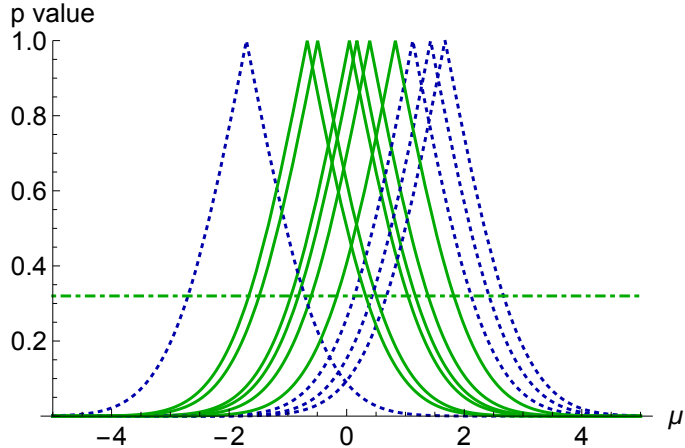

\begin{center}
\pgfuseimage{coverage}
\end{center}
\caption{A $\alpha$-CL interval built from a $p$-value with exact coverage has a probability of $\alpha$ of containing the true value. This is illustrated in the simple case of a quantity $X$ which has a true value $\mu_t=0$ but is measured with an uncertainty $\sigma=1$.
Each time a measurement is performed, it will yield a different value for $X_0$ and thus a different $p$-value curve as a function of the hypothesis tested $\mu_t=\mu$. From each measurement, a 68\% CL interval can be determined by considering the part of the curve above the line $p=0.32$, but this interval may or may not contain the true value $\mu_t=0$. The curves corresponding to the first case (second case) are indicated with 6 green solid lines (4 blue dotted lines). Asymptotically, if the $p$-value has exact coverage, 68\% of these confidence intervals will contain the true value.}
\label{fig:coverageexample}
\end{figure}

Once $p$-values are defined, one can build confidence intervals out of them by using the correspondence between  acceptance regions of tests and confidence sets. Indeed, if we have an exact $p$-value, and 
the critical region $C_\alpha(X)$ is defined
as the region where $p(X;\mu)<\alpha$, the complement of this region turns out to be a confidence set of level $1-\alpha$, i.e., $P[\mu\notin C_\alpha(X)]= 1-\alpha$. This justifies the general use of plotting the $p$-value as a function of $\mu$, and reading  the  68\% or 95\% CL intervals by looking at the ranges where the $p$-value curve is above 0.32 or 0.05. This is illustrated for a simple example in Figs.~\ref{fig:CLexample} and \ref{fig:coverageexample}.
Once again, this discussion is affected by issues of compositeness and nuisance parameters, as well as the requirement of
checking the coverage of the $p$-value used to define these confidence intervals: an overcovering $p$-value will yield too large confidence intervals, which will prove indeed conservative.

A few words about the notation and the vocabulary are in order at this stage. A $p$-value necessarily refers to a null hypothesis, and when the null hypothesis is purely numerical such as~(\ref{hyp_q}) we can consider the $p$-value as a mathematical function of the fundamental parameter $\mu$. This of course does not imply that $\mu$ is a random variable (in frequentist statistics, it is always a fixed, but unknown, number). When the $p$-value as a function of $\mu$ can be described in a simple way by a few parameters, we will often use the notation $\mu=\mu_0\pm \sigma_\mu$. In this case, one can easily build the $p$-value and derive any desired confidence interval. Even though this notation is similar to the measurement of an observable, we stress that this does not mean that the fundamental parameter $\mu$ is a random variable, and it should not be seen as  the definition of a PDF. 
In line with this discussion, we will call {\it uncertainties} the parameters like $\sigma$ that can be given a frequentist meaning, e.g., they can be used to define the PDF of a random variable. On the other hand, we will call {\it errors} the intermediate quantities such as $\sigma_\mu$ that can be used to describe the $p$-value of  a fundamental parameter, but cannot be given a statistical meaning for this parameter.
 
\subsection{Likelihood-ratio test statistic}

Here
we consider test statistics that are constructed from
the logarithm of the likelihood~\footnote{
Strictly speaking, the likelihood is only defined for the actually
measured data $X_0$: ${\mathcal L}_0(\chi)\equiv g(X_0;\chi)$ and
thus is only a function of the parameters $\chi$. Nevertheless it is common
practice to use the word ``likelihood'' for the object $g(X;\chi)$,
considered as a function of both the observables $X$ and the parameters $\chi$.}
\begin{equation}
T \sim -2\ln {\mathcal L}_X(\chi) \qquad {\mathcal L}_X(\chi) \equiv g(X;\chi)
\end{equation}

More precisely, one uses tests based on the likelihood ratio in many instances. Its use is justified by the Neyman-Pearson lemma~\cite{James:2006zz,Cowan:2013pha,Neyman-Pearson} showing that this test has appealing features in a binary model with only two alternatives for $\chi_t$, corresponding to the two simple hypotheses ${\mathcal H}_{\chi_1}$ and ${\mathcal H}_{\chi_2}$. Indeed one can introduce the likelihood ratio
${\mathcal L}_X(\chi_1)/{\mathcal L}_X(\chi_2)$, define the critical region where this likelihood ratio is smaller than 
a given $\alpha$, and decide  that one rejects ${\mathcal H}_{\chi_1}$ whenever the observation falls in this critical region. This test is the most powerful test that can be built~\cite{James:2006zz,Cowan:2013pha
}, in the sense that among all the tests with a given Type-I error $\alpha$ (probability of rejecting ${\mathcal H}_{\chi_1}$ when ${\mathcal H}_{\chi_1}$ is true), the likelihood ratio test
has the smallest Type-II error (probability of accepting ${\mathcal H}_{\chi_1}$ when ${\mathcal H}_{\chi_2}$ is true). 
These two conditions are the two main criteria to determine the performance of a test. 

In the case of a composite hypothesis, there is no such clear-cut approach to choose the most powerful test. 
The Maximum Likelihood Ratio (MLR) is inspired by the Neynman-Pearson lemma, comparing the most 
plausible configuration under ${\mathcal H}_\mu$ with the most plausible one in general: 
\begin{eqnarray}\label{gammatest}
T(X;\mu) &=& -2\ln \frac {{\rm Max}_{\nu_t}
{\mathcal L}_X(\mu,\nu_t)}{{\rm Max}_{\mu_t,\nu_t} {\mathcal L}_X(\mu_t,\nu_t)}\nonumber\\
&=& {\rm Min}_{\nu_t} [-2\ln {\mathcal L}_X(\mu,\nu_t)]-{\rm Min}_{\mu_t,\nu_t} [-2\ln {\mathcal L}_X(\mu_t,\nu_t)] 
\end{eqnarray}
Let us emphasise that even though $T$ is constructed not to depend on  the nuisance parameters $\nu$ explicitly,
its distribution Eq.~(\ref{eq:pdft}) \textit{a priori} depends on them (through the PDF $g$). Even though the Neyman-Pear\-son lemma does not apply here, there is empirical evidence that this test is powerful, and in some cases it exhibits good asymptotic properties (easy computation and distribution independent of nuisance parameters)~\cite{James:2006zz,Cowan:2013pha}.

For the problems considered here, the MLR choice features alluring properties, and in the following we will use test statistics that are derived from this choice. First, if
$g(X;\chi_t)$ is a multidimensional Gaussian function, then the quantity 
$-2\ln{\mathcal L}_X(\chi_t)$ is the sum of the squares of standard normal random variables,
\textit{i.e.,} is distributed as a $\chi^2$ with a number of degrees of
freedom ($N_{dof}$) that is given by ${\rm dim}(X)$. 
Secondly, for linear models, in which the observables $X$
depend linearly on the parameters $\chi_t$, the MLR Eq.~(\ref{gammatest}) is
again a sum of standard normal random variables, and is distributed as a
$\chi^2$ with $N_{dof}={\rm dimension}(\mu )$. Wilks' theorem~\cite{Wilks} states
that this property can be extended to non-Gaussian cases in the asymptotic limit:
under regularity conditions and when
the sample size tends to infinity, the distribution of
 Eq.~(\ref{gammatest}) will converge to the same $\chi^2$ distribution
depending only on the number of parameters tested.

The great virtue of the $\chi^2$-distribution is that it only depends on the
number of degrees of freedom, which means in particular that the
null-distribution of Eq.~(\ref{gammatest}) is independent of the nuisance
parameters $\nu$, whenever the conditions of the Wilks'  theorem apply.
Furthermore the integral~(\ref{pvalue}) can be computed straightforwardly in
terms of complete and incomplete $\Gamma$ functions:
\begin{eqnarray}\label{prob}
p(X_0;\mu)&=&\mathrm{Prob}\left(T(X_0;\mu),N_{dof}\right)\nonumber\\
&\equiv&
\frac{\Gamma(N_{dof}/2,T(X_0;\mu)/2)}{\Gamma(N_{dof}/2)}
\end{eqnarray}

In practice the models we want to analyse, such as the Standard Model,
predict non linear relations between the observables and the parameters.
In this case one has to check whether Wilks' theorem applies, by considering whether the theoretical equations can be approximately linearized~\footnote{
More precisely, the asymptotic limit is reached when the model can be
linearized for all values of the data that contribute significantly to
the integral~(\ref{pvalue}). It corresponds to the situation where the
errors on the parameters derived from computing $p$-values are small with
respect to the typical parameter scales of the problem.
}.

%% file: threeapproaches.tex
\section{Comparing approaches to theoretical uncertainties}\label{sec:threeapproaches}

We have argued before that an appealing test statistic is provided by the likelihood ratio 
Eq.~(\ref{gammatest}) due to its properties in limit cases (linearised theory, asymptotic limit). These properties rely on the fact that the likelihood ratio can be built as a function of random variables 
described by measurements involving only statistical uncertainties.
However, in flavour physics (as in many other fields in particle physics), there are not only statistical but also theoretical uncertainties. 
Indeed, as already indicated in the introduction, these phenomenological analyses combine experimental information and theoretical estimates. In the case of flavour physics, the latter come mainly from QCD-based calculations, which are dominated by theoretical uncertainties.

Unfortunately, the very notion of theoretical uncertainty is ill-defined as ``anything that is not due to the intrinsic variability of data''.
Theoretical uncertainties (model uncertainty) are thus of a different nature with respect to statistical uncertainties (stochastic
uncertainty, i.e. variability in the data), but they can only be \emph{modelled} (except in the somewhat academic case where a bound on the
difference between the exact value and the approximately computed one can be proven). The choice of a model for theoretical uncertainties involves not only the study of its
mathematical properties and its physical implications in specific cases, but also some personal taste.
One can indeed imagine several ways of modelling/treating theoretical uncertainties:
\begin{itemize}
\item one can (contrarily to what has just been said) treat the theoretical uncertainty on the same footing as a statistical
uncertainty; in this case, in order to follow a meaningful frequentist procedure, one has to assume that one lives in a world where
the repeated calculation of a given quantity leads to a distribution of values around the exact one, with some variability that can
be modelled as a PDF (``random-$\delta$ approach''),
\item one can consider that theoretical uncertainties can be modelled as external parameters, and perform a purely
statistical analysis for each point in the theoretical uncertainty parameter space; this leads to an infinite collection of
$p$-values that will have to be combined in some arbitrary way, following a model averaging procedure (``external-$\delta$ approach''),
\item one can take the theoretical uncertainties as fixed asymptotic biases~\footnote{
A bias is defined as the difference between the average of the estimator among a large number of  experiments with finite sample size and the true value. An estimator is said to be consistent if it converges to the true value when the size of the sample tends to infinity (e.g., maximum likelihood estimators). Consistency implies that the bias vanishes asymptotically, while inconsistency may stem from theoretical uncertainties.}, treating them as nuisance parameters that have to be varied in a reasonable region (``nuisance-$\delta$ approach'').
\end{itemize}

There are some desirable properties for a convincing treatment of theoretical uncertainties:
\begin{itemize} 
\item as general as possible, i.e., apply to as many ``kinds'' of theoretical uncertainties as possible (lattice uncertainties,
scale uncertainties) and as many types of physical models as possible,
\item leading to meaningful confidence intervals, in reasonable limit cases: obviously, in the absence of theoretical uncertainties, one
must recover the standard result; one may also consider the type of constraint obtained in the absence of statistical
uncertainties,
\item exhibiting good coverage properties, as it benchmarks the quality of the statistical approach: the comparison of different models
provides interesting information but does not shed light on their respective coverage,
\item associated with a statistically meaningful goodness-of-fit,
\item featuring reasonable asymptotic properties (large samples),
\item yielding the errors as a function of the estimates easily (error propagation), in particular by disentangling the impact of theoretical and statistical contributions,
\item leading to a reasonable procedure to average independent estimates -- if possible, it should be equivalent for any analysis to
include the independent estimates separately or the average alone (associativity). In addition, one may wonder whether the averaging procedure should be conservative
or aggressive (i.e., the average of similar theoretical uncertainties should have a smaller uncertainty or not), and if the procedure should be stationary (the uncertainty of an average should be independent of the central values or not),
\item leading to reasonable results in the case of averages of inconsistent measurements.
\end{itemize}
Finally a technical requirement is the computing power needed to calculate the best fit point and confidence intervals for a large parameter space with a large number of constraints. Even
though it should not be the sole argument in favour of a model, it should be kept in mind (a very complicated model for theoretical uncertainties would not be particularly
interesting if it yields very close results to a much simpler one).

We summarize some of the points mentioned above in Tab.~\ref{tab:comparisonMethods}. As it will be seen, it will however prove challenging to fulfill all these criteria at the same time, and we will have to make compromises along the way.

%% file: 1Dillustration.tex
\section{Illustration of the approaches in the one-dimensional case}\label{sec:1Dillustation}

\subsection{Situation of the problem}

We will now discuss the three different approaches and some of their properties in the simplest case, i.e. with a single measurement (for an experimental quantity) or a single theoretical determination (for a theoretical quantity). Following a fairly conventional abuse of language, we will always refer to this piece of information as a ``measurement'' even though some modelling may be involved in its extraction through data reduction, as discussed in Sec.~\ref{sec:intro}. The main, yet not alone, aim is to model/interpret/exploit a measurement like~\footnote{We discuss how the method can be adapted for asymmetric uncertainties in App.~\ref{app:asym}.} 
\begin{equation}
X=X_0 \pm \sigma\ ({\rm exp}) \pm \Delta ({\rm th})
\end{equation}
to extract information on the value of the associated fundamental parameter $\mu$.
Without theoretical uncertainty ($\Delta =0$), one would use this measurement to build a PDF
\begin{equation}
{\rm PDF}_{\rm no\ th}(X;\mu)={\mathcal N}_{(\mu,\sigma)}(X)
\end{equation}
yielding the MLR test statistic
\begin{equation}
T_{\rm no\ th}=\frac{(X-\mu)^2}{\sigma^2}
\end{equation}
and one can build a $p$-value easily from Eq.~(\ref{pvalue})
\begin{equation}
p_{\rm no\ th}(\mu)=1-{\rm Erf}\left[\frac{|\mu-X_0|}{\sqrt{2}\sigma}\right]
\end{equation}

In the presence of a theoretical uncertainty $\Delta$, the situation is more complicated, as there is no clear definition of what $\Delta$ corresponds to.
A possible first step is to introduce a theoretical uncertainty parameter $\delta$ that describes the shift of the approximate theoretical computation from the exact value, and
that is taken to vary in a region that is defined by the value of $\Delta$. This leads to the PDF
\begin{equation}
{\rm PDF}(X;\mu)={\mathcal N}_{(\mu+\delta,\sigma)}(X)
\end{equation}
in such a way that in the limit of an infinite sample size ($\sigma\to 0$), the measured value of $X$ reduces to $\mu+\delta$. The challenge is to extract some information on $\mu$,
given the fact that the value of $\delta$ remains unknown.

The steps (to be spelt out below) to achieve this goal are:
\begin{itemize}
\item Take a model corresponding to the interpretation of $\delta$: \emph{random variable, external parameter, fixed bias as a nuisance parameter\ldots}

\item Choose a test statistic $T(X;\mu)$ that is consistent with the model and that discriminates the null hypothesis: \emph{Rfit, quadratic, other \ldots}
\item Compute, consistently with the model, the $p$-value that is in general a function of $\mu$ and $\delta$
\item Eliminate the dependence with respect to $\delta$ by some well-defined procedure

\item Exploit the resulting $p$-value (coverage, confidence intervals, goodness-of-fit)
\end{itemize}

\begin{table*}
\begin{center}
{\small
\begin{tabular}{c|ccccc}
     Approach                     & Random-$\delta$ &  Nuisance-$\delta$ & External-$\delta$ \\
\hline
Hypothesis & Random var. &  Composite hyp. & Family of simple hyp.\\
                  & ${\rm PDF}_\Delta(\delta)$ 
                  & ${\mathcal H}_\mu:\mu_t=\mu$& ${\mathcal H}^{(\delta)}_\mu:\mu_t=\mu+\delta$\\
Test & Likelihood ratio &  Quadratic & Quadratic\\
Constraint on $\delta$ & $-$  & $\Omega$ & $\Omega$ \\
\hline
Associativity   & Yes if normal ${\rm PDF}$         &  Yes if $\Omega$ hyperball &  Yes if $\Omega$ hyperball\\
Splitting of errors & Yes if normal  ${\rm PDF}$  &  Yes for all $\Omega$ & Yes for all $\Omega$ \\
Stationarity & Yes &   Yes & Yes\\                          
Simple asympt. lim. &   Yes if normal ${\rm PDF}$          &  Yes & Yes\\     
Simple $\sigma\to 0$ limit & Depends on ${\rm PDF}$ &  $\Omega$ & $\Omega$\\
\hline
Particular cases & naive Gaussian &  Fixed/adaptive nuis. & Scan\\
 if we take & normal {\rm PDF} &   Fixed/adaptive $\Omega$    & Sup over fixed $\Omega$
\end{tabular}
}
\caption{Summary table of various approaches to theoretical uncertainties considered in the text.}\label{tab:comparisonMethods}
\end{center}
\end{table*}

Since we focus on Gaussian experimental uncertainties (the generalization to other shapes is formally straightforward but may be technically more complicated), for all approaches
that we discuss in this note we take the following PDF
\begin{equation}
{\rm PDF}(X;\mu)={\mathcal N}_{(\mu+\delta,\sigma)}(X)
\end{equation}
where, in the limit of an infinite sample size ($\sigma\to 0$), $\mu$ can be interpreted as the exact value of the parameter of interest, and $\mu+\delta$ the
approximately theoretically computed one. The interpretation of $\delta$ will differ depending on the approach considered, which we will discuss now.

\subsection{The random-$\delta$ approach}\label{sec:random-delta}
In the random-$\delta$ approach, $\delta$ would be
related to the variability of theoretical computations, that one can model with some PDF for $\delta$, such as ${\mathcal N}_{(0,\Delta)}$ (normal) or ${\mathcal U}_{(-\Delta,+\Delta)}$ (uniform).
The natural candidate for the test statistic  $T(X;\mu)$ is the MLR  built from the PDF. One considers a model where
 $X=s+\delta$ is the sum of two random variables, $s$ being distributed as a Gaussian of mean $\mu$ and width $\sigma$, and $\delta$ as an additional random variable
with a distribution depending on $\Delta$. 

One may often consider for $\delta$ a variable normally distributed with a mean zero and a width $\Delta$ (denoted naive Gaussian or ``nG'' in the following, corresponding to the most common procedure in the literature of particle physics phenomenology). The resulting PDF for $X$ is then the convolution of two Gaussian PDFs, leading to
\begin{equation}\label{eq:PDFnG}
{\rm PDF}_{\rm nG}(X;\mu)={\mathcal N}_{(\mu,\sqrt{\sigma^2+\Delta^2})}(X)
\end{equation}
to which corresponds the usual quadratic test statistic (obtained from MLR)
\begin{equation}\label{quadstatROTW}
T_{\rm nG}=\frac{(X-\mu)^2}{\sigma^2+\Delta^2}
\end{equation}
recovering the $p$ value that would be obtained when the two uncertainties are added in quadrature
\begin{equation}
p_{\rm nG}(\mu)=1-{\rm Erf}\left[\frac{|\mu-X|}{\sqrt{2}\sqrt{\sigma^2+\Delta^2}}\right]
\end{equation}

We should stress that considering $\delta$ as a random variable corresponds to a rather strange frequentist world~\footnote{On the other hand, this is natural in the Bayesian approach, where incomplete information is modelled as a PDF for the unknown parameters associated with a theoretical uncertainty~\cite{Bolstad,DAgostini:2003bpu,Bevan}.}, and there is no strong argument that would help to choose
the associated PDF (for instance, $\delta$ could be
a variable uniformly distributed over $[-\Delta,\Delta]$). However for a general PDF, the $p$-value has no simple analytic formula and it must be computed numerically from Eq.~(\ref{pvalue}). In the following, we will only consider the case of a Gaussian PDF when we discuss the random-$\delta$ approach.

\subsection{The nuisance-$\delta$ approach}\label{sec:nuisance-delta}

In the nuisance approach,  $\delta$ is not interpreted as a random variable but as a fixed parameter so that in the limit of an infinite sample size, the estimator does not converge to the true value $\mu_t$, but to $\mu_t+\delta$. The distinction between statistical and theoretical uncertainties is thus related to their effect as the sample size increases, statistical uncertainties decreasing while theoretical uncertainties remaining of the same size (see Refs.~\cite{Porter:2008uw,Fichet:2015xla,Fichet:2016gvx} for other illustrations in the context of particle physics). 
 One works with the null hypothesis $\mathcal{H}_{\mu}: \mu_t=\mu$, and one has then to determine which test statistic is to be built. 

In the frequentist approach, the choice of the test statistic is arbitrary as long as it models the null hypothesis correctly, i.e., the smaller the value of the test statistic, the better the agreement of the data with the hypothesis. A particularly simple possibility consists in the quadratic statistic already introduced earlier:
\begin{equation}\label{quadstat}
T_{\rm nuisance}={\rm Min}_\delta\left[\left(\frac{X-\mu-\delta}{\sigma}\right)^2+\left(\frac{\delta}{\Delta}\right)^2\right]
 =\frac{(X-\mu)^2}{\sigma^2+\Delta^2}
\end{equation}
where the minimum is not taken over a fixed range, but on the whole space. The great virtue of the quadratic shape is that in linear models it remains quadratic after minimization over any subset of parameters, in contrast with alternative, non-quadratic, test statistics.

The PDF for $X$ is normal, with mean $\mu+\delta$ and variance $\sigma^2$
\begin{equation}\label{eq:PDFnuisance}
{\rm PDF}_{\rm nuisance}(X;\mu)={\mathcal N}_{(\mu+\delta,\sigma)}(X)
\end{equation}
Although we choose test statistics for the random-$\delta$ and nuisance-$\delta$ of the same form,
Eqs.~(\ref{quadstatROTW}) and (\ref{quadstat}), the different PDFs Eqs.~(\ref{eq:PDFnG}) and (\ref{eq:PDFnuisance}) imply very different constructions for the $p$-values and the resulting statistical outcomes. Indeed, with this PDF for the nuisance-$\delta$ approach,
$T$ is distributed as a rescaled, non-central $\chi^2$ distribution
with a non-centrality parameter $(\delta/\sigma)^2$ (this non-centrality parameter illustrates that the test statistic is centered around $\mu$ whereas the distribution of $X$ is centered around $\mu+\delta$). $\delta$ is then a genuine asymptotic bias, implying inconsistency: in the limit of an infinite sample size, the estimator constructed from $T$ is $\mu$, whereas the
true value is $\mu+\delta$. 
Using the previous expressions, one can easily compute the cumulative distribution function of this test statistic
\begin{equation}\label{eq:pvaluedelta}
1-\mathrm{CDF}_\delta(\mu)=1+\frac{1}{2}{\rm Erf}\left(\frac{\delta-|\mu-X|}{\sqrt{2}\sigma}\right) -\frac{1}{2}{\rm Erf}\left(\frac{\delta+|\mu-X|}{\sqrt{2}\sigma}\right)\end{equation}
which depends explicitly on $\delta$ but not on $\Delta$ (as indicated before, even if $T$ is built to be independent of nuisance parameters, its PDF  depends on them  \textit{a priori}).

To infer the $p$-value one can take the supremum value for $\delta$ over some interval $\Omega$
\begin{equation}
p_{\Omega}={\rm Max}_{\delta\in \Omega} [1-\mathrm{CDF}_\delta(\mu)]
\end{equation}
The interpretation is the following: if the (unknown) true value of $\delta$ belongs to $\Omega$, then
$p_{\Omega}$ is a valid $p$-value for $\mu$, from which one can infer confidence intervals for $\mu$.
This space cannot be the whole space (as one would get $p=1$ trivially for all values of $\mu$), but there is no natural candidate (i.e., coming from the derivation of the test statistic). More specifically, 
should the interval $\Omega$ be kept fixed or should it be rescaled when investigating confidence intervals at different levels (\textit{e.g.} 68\% vs 95\%)?
\begin{itemize}
\item
 If one wants to keep it fixed, $\Omega_r=r[-\Delta,\Delta]$:
\begin{equation}\label{pfixed}
p_{\rm fixed\ \Omega_r}={\rm Max}_{\delta\in \Omega_r}[1- \mathrm{CDF}_\delta(\mu)]
\end{equation}
One may wonder what is the best choice for $r$, as the $p$-value gets very large if one works with the reasonable $r=3$, while the choice $r=1$ may appear as non-conservative. We will call this treatment the \textit{fixed} $r$-nuisance approach.
\item
One can then wonder whether one would like to let $\Omega$ depend on the value considered for $p$.
In other words, if we are looking at a $k\,\sigma$ range, we could consider the equivalent range for $\delta$.
This would correspond to
\begin{equation}\label{padapt1}
p_{\rm adapt\ \Omega}={\rm Max}_{\delta\in \Omega_{k_\sigma( p)}}[1- \mathrm{CDF}_\delta(\mu)]
\end{equation}
where $k_\sigma(p )$ is the ``number of sigma'' corresponding to $p$
\begin{equation}\label{padapt2}
k_\sigma(p )^2={\rm Prob}^{-1}(p,N_{\rm dof}=1)
\end{equation} 
where the function Prob has been defined in Eq.~(\ref{prob}).
We will call this treatment the \textit{adaptive} nuisance approach.
The correct interpretation of this $p$-value is: $p$ is a valid $p$-value if the true
(unknown) value of $\delta/\Delta$ belongs to the ``would be'' $1-p$ confidence interval
around 0. This is not a standard coverage criterion: one can use \emph{adaptive coverage}, and \emph{adaptively
valid $p$-value}, to name this new concept. Note that Eqs.~(\ref{padapt1})-(\ref{padapt2}) constitute a non-algebraic implicit equation, that has to be solved by numerical means. 
\end{itemize}

Let us emphasise that the fixed interval is very close 
to the original `Rfit' method of the CKMfitter group~\cite{Hocker:2001xe,Charles:2004jd} in spirit, but not numerically, as will be shown below by an explicit comparison. In contrast the adaptive choice is more aggressive in the region of $\delta$ close
to zero, but allows this parameter to take large values, provided one is interested in computing  small $p$-values accordingly. In this sense, the adaptive approach provides a unified approach to deal with two different issues of importance, namely the metrology of parameters (at 1 or 2 $\sigma$) and exclusion tests (at 3 or 5 $\sigma$).

\subsection{The external-$\delta$ approach} \label{sec:external-delta}

In this approach, the parameter $\delta$ is also considered as a fixed parameter. 
The idea behind this approach is very simple, and it is close to what
experimentalists often do to estimate systematic effects: in a first step one considers that $\delta$ is a fixed constant, and one performs a standard, purely statistical analysis that leads
to a $p$-value that explicitly depends on $\delta$.
If one takes $X\sim\mathcal{N}_{(\mu+\delta,\sigma)}$ and $T$ quadratic [either $(X-\mu-\delta)^2/\sigma^2$ or  $(X-\mu-\delta)^2/(\sigma^2+\Delta^2)$]~\footnote{The choice of the weight in the denominator of the test statistic will be discussed in the multidimensional case in Sec.~\ref{sec:otheraverages}, but it does not impact the result for the $p$-value in one dimension where it plays only the role of an overall normalisation that cancels when computing the $p$-value.}:
\begin{equation}
p_\delta(\mu)=1-{\rm Erf}\left[\frac{|X-\mu-\delta|}{\sqrt{2}\sigma}\right]\,.
\end{equation}
Note that this procedure actually corresponds to the simple null hypothesis $\mathcal{H}^{(\delta)}_{\mu}: \mu_t=\mu+\delta$ instead of
$\mathcal{H}_{\mu}$: $\mu_t=\mu$, hence one gets an infinite collection of $p$-values instead of a single one related to the aimed constraint on $\mu$.

Since $\delta$ is unknown one has to define a procedure to average all the $p_\delta(\mu)$ obtained. The simplest possibility is to take the envelope (i.e., the maximum) of $p_\delta(\mu)$ for $\delta$ in a definite interval (\textit{e.g.} $[-\Delta,+\Delta]$), leading to:
\begin{eqnarray}\label{eq:rfit}
p_{\rm nRfit} &=& 1 \qquad\qquad\qquad\qquad\qquad \ {\rm if}\ |X-\mu|\leq \Delta\\
 &=& 1-{\rm Erf}\left[\frac{|X-\mu\pm\Delta|}{\sqrt{2}\sigma}\right]\ \qquad \ {\rm otherwise}
\end{eqnarray}
By analogy with the previous case, we will call this treatment the \textit{fixed} $r$-external approach for $\delta\in\Omega_r$. This is equivalent to the Rfit ansatz used by CKMfitter~\cite{Hocker:2001xe,Charles:2004jd} in the one-dimensional case (but not in higher dimensions), proposed to treat theoretical uncertainties in a different way from statistical uncertainties, treating all values within $[-\Delta,\Delta]$ on an equal footing. We recall that the Rfit ansatz was obtained starting from a well test statistic, with a flat bottom with a width given by the theoretical error and parabolic walls given by statististical uncertainty.

A related method, called the Scan method, has been developed in the context of flavour physics~\cite{DuboisFelsmann:2003jd,Eigen:2013cv}. It is however slightly different from the case discussed here. First, the test statistic chosen is not the same, since the Scan method uses the likelihood rather than the likelihood ratio, i.e. it relies on the test statistic $T=-2\log{\cal L(\mu,\nu})$ which is interpreted assuming that $T$ follows a $\chi^2$-law with the corresponding number of degrees of freedom $N$, including both parameters of interest and nuisance parameters~\footnote{Such a test statistic tends typically to be less sensitive  to discrepancies in a global fit than the likelihood ratio. In presence of quantities having no or little dependence on the scanned parameters, the impact of discrepancies is diluted in the case of the likelihood statistic.}. Then the $1-\alpha$ confidence region is then determined by varying nuisance parameters in given intervals (typically $\Omega_1$), but 
accepting only points where $T\leq T_c$, where $T_c$ is a critical value so that $P(T\geq T_c;N|H_0)\geq \alpha$ (generally taken as $\alpha=0.05$). This latter condition acts as a test of compatibility between a given choice of nuisance parameters and the data.

%% file: 1Dcomparison.tex
\section{Comparison of the methods in the one-dimensional case}\label{sec:1Dcompar}

In the following, we will discuss properties of the different approaches in the case of one dimension. More specifically, we will consider:
\begin{itemize}
\item the random-$\delta$ approach with a Gaussian random variable, or naive Gaussian (nG), see Sec.~\ref{sec:random-delta},
\item the nuisance-$\delta$ approach with quadratic statistic and fixed range, or fixed nuisance, see Sec.~\ref{sec:nuisance-delta},
\item the nuisance-$\delta$ approach with quadratic statistic and adaptive range, or adaptive nuisance, see Sec.~\ref{sec:nuisance-delta},
\item the external-$\delta$ approach with quadratic statistic and fixed range, equivalent to the  Rfit approach in one dimension, see Sec.~\ref{sec:external-delta}.
\end{itemize}
Note that we will not consider other (non quadratic) statistics. Finally, we consider 
\begin{equation}
X=0\pm \sigma\pm \Delta \qquad\qquad \sigma^2+\Delta^2=1
\end{equation}
varying $\Delta/\sigma$ as an indication of the relative size of the experimental and theoretical uncertainties.

\subsection{$p$-values and confidence intervals}\label{sec:sizeCLintervals}

\begin{figure*}[t]
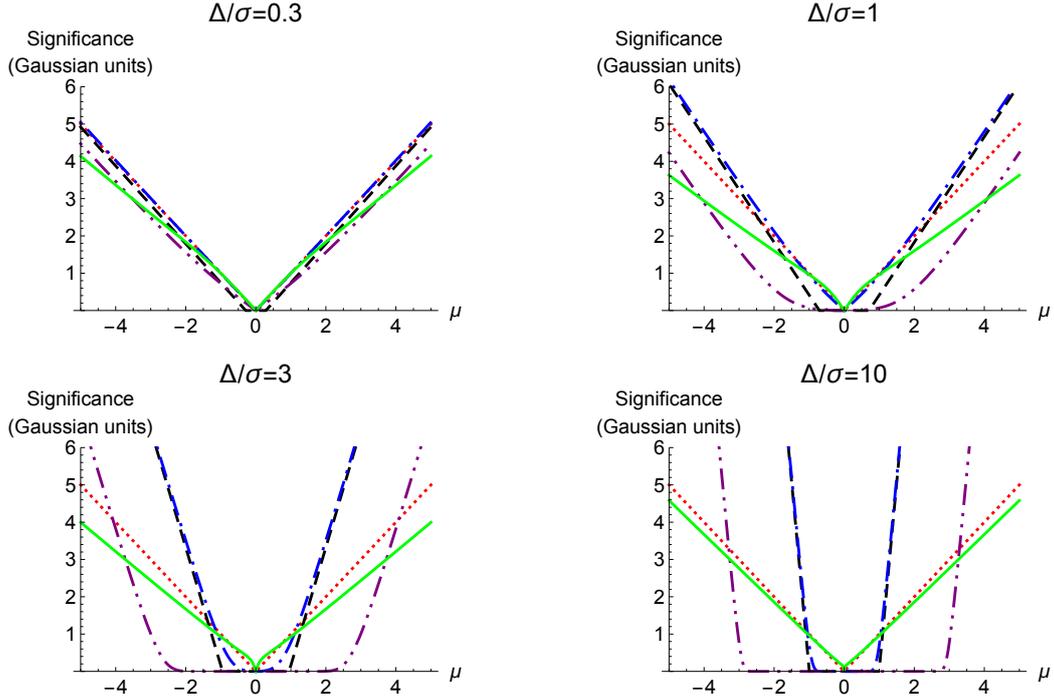

\begin{center}
\pgfuseimage{Signif2}
\caption{Comparison of different treatments of theoretical uncertainties of the measurement $X=0 \pm \sigma\ ({\rm exp}) \pm \Delta ({\rm th})$, with different values of $\Delta/\sigma$ (with the normalisation $\sqrt{\Delta^2+\sigma^2}=1$).  The $p$-values have been converted into a significance in Gaussian units of $\sigma$ following the particle physics conventions.
The various approaches are: nG (dotted, red), Rfit or $1$-external (dashed, black), fixed $1$-nuisance   (dotted-dashed, blue), fixed $3$-nuisance (dotted-dotted-dashed, purple), adaptive nuisance (solid, green).} \label{fig:comp}
\end{center}
\end{figure*}

\begin{figure}
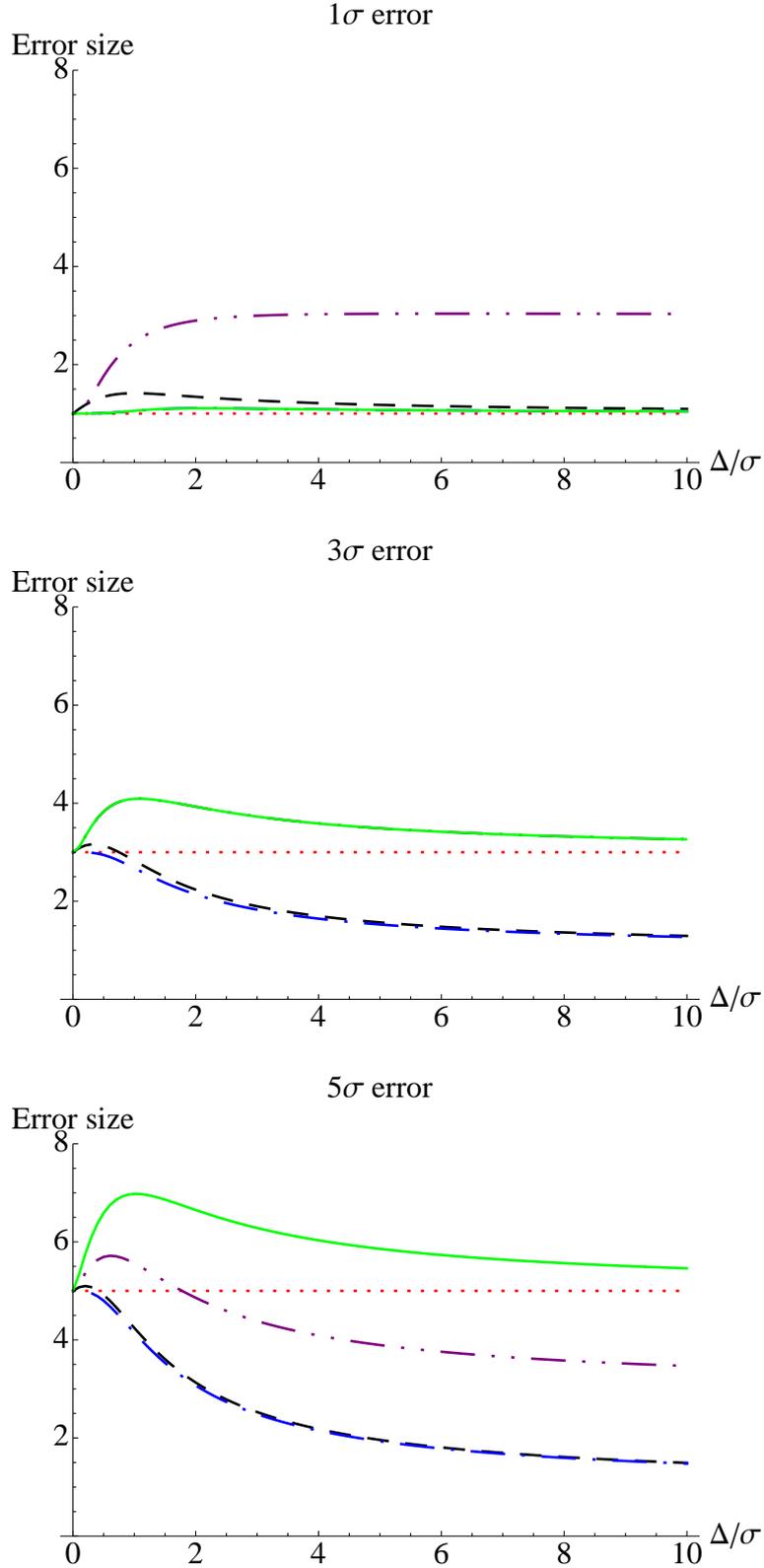

\begin{center}
\pgfuseimage{CI1}

\vspace{0.5cm}

\pgfuseimage{CI2}

\vspace{0.5cm}

\pgfuseimage{CI3}

\caption{Comparison of the size of the $(1, 3, 5)\sigma$ errors (upper, central and lower panels respectively)
as a function of $\Delta/\sigma$. Different approaches are shown: nG (dotted, red), Rfit or $1$-external (dashed, black), fixed $1$-nuisance   (dotted-dashed, blue), fixed $3$-nuisance (dotted-dotted-dashed, purple), adaptive nuisance (solid, green). In the upper panel ($1\sigma$ confidence level), the adaptive and fixed $1$-nuisance approaches yield the same result by construction,
and the two curves cannot be distinguished (only the adaptive one is shown). The same situation occurs in the central panel corresponding to 3 $\sigma$ with the adaptive and fixed $3$-nuisance approaches.
}\label{fig:errors}
\end{center}
\end{figure}

We can follow the discussion of the previous section and plot the results for the $p$-values obtained from the various methods discussed above in Fig.~\ref{fig:comp}, where we compare nG, Rfit, fixed nuisance and adaptive nuisance approaches.
From these $p$-values, we can infer confidence intervals at a given significance level and a 
given value of $\Delta/\sigma$, and determine the length of the (symmetric) confidence interval (see Tab.~\ref{tab:compar1Derrors}).
We notice the following points:
\begin{itemize}
\item by construction, nG always provides the same errors whatever the relative proportion of theoretical and statistical uncertainties, and all the approaches provide the same answer in the limit of no theoretical uncertainty $\Delta=0$.
\item by construction, for a given $n\sigma$ confidence level, the interval provided by the adaptive nuisance approach is identical to the one obtained using the fixed nuisance approach with a $[-n,n]$ interval. This explains why the adaptive nuisance approach yields identical results to the fixed 1-nuisance approach at 1 $\sigma$ (and similarly for the fixed 3-nuisance approach at 3 $\sigma$).
The corresponding curves cannot be distinguished on the upper and central panels of Fig.~\ref{fig:errors}.
\item the adaptive nuisance approach is numerically quite close to the nG method; the maximum difference occurs for $\Delta/\sigma=1$ (up to 40\% larger error size for 5 $\sigma$ intervals).
    \item the $p$-value from the fixed-nuisance approach has a very wide plateau if one works with the `reasonable' range $[-3\Delta,+3\Delta]$, while the choice of $[-\Delta,+\Delta]$ might be considered as non conservative.
\item the 1-external  and fixed 1-nuisance approaches are close to each other and less conservative than the adaptive approach, which is expected, but also than nG, for confidence intervals at 3 or 5 $\sigma$ when theory uncertainties dominate.
\item when dominated by theoretical uncertainties ($\Delta/\sigma$ large), all approaches provide 3 and 5 $\sigma$ errors smaller than the nG approach, apart from the adaptive nuisance approach.
\end{itemize}

\begin{table*}
\begin{center}
\begin{tabular}{c|cccc}
 & \quad\qquad nG \qquad\qquad & $1$-nuisance & adaptive nuisance & $1$-external \\
 \hline
$\Delta/\sigma=0.3$ & & & & \\
$1\sigma$ & 1.0 & 1.0 & 1.0 & 1.2 \\
$3\sigma$ & 3.0 & 3.0 & 3.5 & 3.2 \\
$5\sigma$ & 5.0 & 5.0 & 6.1 & 5.1 \\
\hline
$\Delta/\sigma=1$ & & & & \\
$1\sigma$ & 1.0 & 1.1 & 1.1 & 1.4 \\
$3\sigma$ & 3.0 & 2.7 & 4.1 & 2.8 \\
$5\sigma$ & 5.0 & 4.1 & 7.0 & 4.2 \\
\hline
$\Delta/\sigma=3$ & & & & \\
$1\sigma$ & 1.0 & 1.1 & 1.1 & 1.3 \\
$3\sigma$ & 3.0 & 1.8 & 3.7 & 1.9 \\
$5\sigma$ & 5.0 & 2.5 & 6.3 & 2.5 \\
\hline
$\Delta/\sigma=10$ & & & & \\
$1\sigma$ & 1.0 & 1.0 & 1.0 & 1.1 \\
$3\sigma$ & 3.0 & 1.3 & 3.3 & 1.3 \\
$5\sigma$ & 5.0 & 1.5 & 5.5 & 1.5 \\

\end{tabular}
\caption{Comparison of the size of one-dimensional confidence intervals at $1,3,5\sigma$ for various methods and various values of $\Delta/\sigma$.
\label{tab:compar1Derrors}}
\end{center}
\end{table*}

\subsection{Significance thresholds}\label{sec:signifthresh}

Another way of comparing methods consists in taking the value of $\mu$ for which the $p$-value corresponds to $1,3,5\ \sigma$ (in significance scale) in a given method, and compute the corresponding $p$-values for the other methods. The results are gathered in Tabs.~\ref{sigthresholds1} and \ref{sigthresholds2}. Qualitatively, the comparison of significances can be seen from Fig.~\ref{fig:comp}: if the size of the error is fixed,  the different approaches quote
different significances for this same error.

In agreement with the previous discussion, we see that fixed 1-nuisance and 1-external yield similar results for 3 and 5 $\sigma$, independently of the relative size of statistical and theoretical effects. Moreover, they are prompter to claim a tension than nG, the most conservative method in this respect being the adaptive nuisance approach.

As a physical illustration of this problem, we can consider the current situation for the anomalous magnetic moment of the muon, namely the difference between the experimental measurement and the theoretical computation in the Standard Model~\cite{Agashe:2014kda}:
\begin{equation}
(a_\mu^{exp}-a_\mu^{SM})\times 10^{11}=288\pm 63_{stat}\pm 49_{th}
\end{equation}
This discrepancy has a different significance depending on the model chosen for theoretical uncertainties, which can be computed from the associated $p$-value (under the hypothesis that the true value of $a_\mu^{SM}-a_\mu^{exp}$ is $\mu=0$).~\footnote{In full generality, one should have kept the different sources of theoretical uncertainties separated, as their combination in a single theoretical uncertainty depends on the precise model used for theoretical uncertainties. We consider here the result of ref.~\cite{Agashe:2014kda} where all theoretical uncertainties are already combined.}
The nG method yields 3.6~$\sigma$, the $1$-external approach 3.8~$\sigma$, the 1-nuisance approach 4.0~$\sigma$, and the adaptive nuisance approach 2.7~$\sigma$.
The overall pattern is similar to what can be seen from the above tables, with a significance of the discrepancy which depends on the model used for theoretical uncertainties.

\begin{table*}
\begin{center}
\begin{tabular}{c|cccc}
1$\sigma$ signif. threshold  & \quad\qquad nG \qquad\qquad & $1$-nuisance & adaptive nuisance & $1$-external \\
\hline
nG & 1 & 0.9 & 1.0 & 0.4 \\
$1$-nuisance & 1.1 & 1 & 1.0 & 0.5 \\
adaptive nuisance & 1.1 & 1.0 & 1 & 0.5 \\
$1$-external & 1.4 & 1.4 & 1.2 & 1 
\\
\\3$\sigma$  signif. threshold  & \quad\qquad nG \qquad\qquad & $1$-nuisance & adaptive nuisance & $1$-external \\
\hline
nG & 3 & 3.4 & 2.3 & 3.2 \\
$1$-nuisance & 2.7 & 3 & 2.0 & 2.8 \\
adaptive nuisance & 4.1 & 4.9 & 3 & 4.8 \\
$1$-external & 2.8 & 3.2 & 2.1 & 3 
\\
\\
5$\sigma$  signif. threshold  & \quad\qquad nG \qquad\qquad & $1$-nuisance & adaptive nuisance & $1$-external \\
\hline
nG & 5 & 6.2 & 3.6 & 6.1 \\
$1$-nuisance & 4.1 & 5 & 3.0 & 4.9 \\
adaptive nuisance & 7.0 & $\infty$ & 5 & $\infty$ \\
$1$-external & 4.2 & 5.1 & 3.1 & 5 
\end{tabular}
\end{center}
\caption{Comparison of 1D $1,3,5\sigma$ significance thresholds for $\Delta/\sigma=1$. For instance, the first line should read: if with nG a $p$-value=1 $\sigma$ is found, then the corresponding values for the three other methods are 0.9/1.0/0.4 $\sigma$.  $\infty$ means that the corresponding $p$-value
was numerically zero (corresponding to more than 8 $\sigma$).}\label{sigthresholds1}
\end{table*}

\begin{table*}
\begin{center}
\begin{tabular}{c|cccc}
1$\sigma$ signif. threshold & \quad\qquad nG \qquad\qquad & $1$-nuisance & adaptive nuisance & $1$-external \\
\hline
nG & 1 & 0.8 & 0.9 & 0.2 \\
$1$-nuisance & 1.1 & 1 & 1.0 & 0.5 \\
adaptive nuisance & 1.1 & 1.0 & 1 & 0.5 \\
$1$-external & 1.3 & 1.4 & 1.1 & 1 
\\
\\
3$\sigma$ signif. threshold & \quad\qquad nG \qquad\qquad & $1$-nuisance & adaptive nuisance & $1$-external \\
\hline
nG & 3 & 6.6 & 2.4 & 6.5 \\
$1$-nuisance & 1.8 & 3 & 1.5 & 2.8 \\
adaptive nuisance & 3.7 & $\infty$ & 3 & $\infty$ \\
$1$-external & 1.9 & 3.2 & 1.6 & 3 
\\
\\
5$\sigma$ signif. threshold & \quad\qquad nG \qquad\qquad & $1$-nuisance & adaptive nuisance & $1$-external \\
\hline
nG & 5 & $\infty$ & 4.0 & $\infty$ \\
$1$-nuisance & 2.5 & 5 & 2.0 & 4.9 \\
adaptive nuisance & 6.3 & $\infty$ & 5 & $\infty$ \\
$1$-external  & 2.5 & 5.1 & 2.1 & 5 
\end{tabular}
\end{center}
\caption{Comparison of 1D $1,3,5\sigma$ significance thresholds for $\Delta/\sigma=3$. Same comments as in the previous table.}\label{sigthresholds2}
\end{table*}

\subsection{Coverage properties}

As indicated in Sec.~\ref{sec:pvalues}, $p$-values are interesting objects if they cover exactly or  slightly overcover in the domain where they should be used corresponding to a given significance, see Eqs.~(\ref{eq:pvalue1})-(\ref{eq:pvalue3}). If coverage can be ensured for a simple hypothesis~\cite{James:2006zz,Cowan:2013pha}, this property is far from trivial and should be checked explicitly in the case of composite hypotheses, where compositeness comes from nuisance parameters that can be related to theoretical uncertainties, or other parameters of the problem.

For all methods we study coverage properties in the standard way: one first fixes the true values of the parameters $\mu$ and $\delta$ (which are not assumed to be random variables), from which one generates a
large sample of toy experiments $X_i$. Then for each toy experiment one computes the $p$-value
at the true value of $\mu$. The shape of the
distribution of $p$-values indicates over, exact or under coverage. More specifically, one can determine $P(p\geq 1-\alpha)$ for a CL of $\alpha$: if it is larger (smaller) than $\alpha$, the method overcovers (undercovers) for this particular CL, i.e. it is conservative (liberal). We emphasise that this property is \textit{a priori} dependent on the chosen CL.

\begin{table*}
\begin{center}
\begin{tabular}{c|ccc|ccc}
& 68.27\% CL & 95.45\% CL & 99.73\% CL & 68.27\% CL & 95.45\% CL & 99.73\% CL\\
\hline
& \multicolumn{3}{c}{$\Delta/\sigma=1$, $\delta/\Delta=1$} & \multicolumn{3}{|c}{$\Delta/\sigma=1$, $\delta/\Delta=0$} \\\hline
nG & 65.2\% & 96.6\% & 99.9\% & 84.1\% & 99.5\% & 100.0\% \\
$1$-nuisance & 68.2\% & 95.4\% & 99.7\% & 86.5\% & 99.3\% & 100.0\% \\
adaptive nuisance & 68.3\% & 99.6\% & 100.0\% &86.4\% & 100.0\% & 100.0\% \\
$1$-external  & 83.9\% & 97.8\% & 99.9\% & 95.4\% & 99.7\% & 100.0\% \\
$1$-ext. (excl. $p\equiv 1$) & 69.2\% & 95.7\% & 99.8 \% & 85.5\% & 99.1\% & 100.0 \% \\
\hline
& \multicolumn{3}{c}{$\Delta/\sigma=1$, $\delta/\Delta=3$} & \multicolumn{3}{|c}{$\Delta/\sigma=3$, $\delta/\Delta=0$} \\
\hline
nG & 5.76\% & 43.2\% & 89.1\% & 99.8\% & 100.0\% & 100.0\% \\
$1$-nuisance & 6.60\% & 38.0\% & 78.4\% & 100.0\% & 100.0\% & 100.0\% \\
adaptive nuisance & 6.53\% & 75.4\% & 99.8\% &99.9\% & 100.0\% & 100.0\% \\
$1$-external & 16.0\% & 50.3\% & 84.2\% & 100.0\% & 100.0\% & 100.0\% \\
$1$-ext. (excl. $p\equiv 1$) & 14.0\% & 49.1\% & 83.8 \%  & 98.5\% & 100.0\% & 100.0 \% \\
\hline
& \multicolumn{3}{c}{$\Delta/\sigma=3$, $\delta/\Delta=3$} & \multicolumn{3}{|c}{$\Delta/\sigma=3$, $\delta/\Delta=1$}\\ 
\hline
nG & 0.00\% & 0.35\% & 68.7\% & 56.3\% & 100.0\% & 100.0\% \\
$1$-nuisance & 0.00\% & 0.00\% & 0.07\% & 68.1\% & 95.5\% & 99.7\% \\
adaptive nuisance & 0.00\% & 9.60\% & 99.8\% & 68.2\% & 100.0\% & 100.0\% \\
$1$-external & 0.00\% & 0.00\% & 0.13\% & 84.1\% & 97.7\% & 99.9 \% \\
$1$-ext. (excl. $p\equiv 1$) & 0.00\% & 0.00\% & 0.13\% & 68.2\% & 95.4\% & 99.7\%  
\end{tabular}
\end{center}
\caption{Coverage properties of the various methods at 68.27, 95.45 and 99.73\% CL,
for different true values of $\delta/\Delta$ contained in, at the border of, or outside the fixed volume $\Omega$, and for various relative sizes of statistical and theoretical uncertainties $\Delta/\sigma$.}\label{tab:coverage}
\end{table*}

\begin{figure*}
\begin{center}
\begin{tabular}{cc}
$\Delta/\sigma=1$, $\delta/\Delta=1$ & $\Delta/\sigma=1$, $\delta/\Delta=0$\\
\pgfuseimage{coverage11} & \pgfuseimage{coverage10}\\
\\
$\Delta/\sigma=1$, $\delta/\Delta=3$ & $\Delta/\sigma=3$, $\delta/\Delta=0$\\
\pgfuseimage{coverage13} & \pgfuseimage{coverage30}\\
\\
$\Delta/\sigma=3$, $\delta/\Delta=3$ & $\Delta/\sigma=3$, $\delta/\Delta=1$ \\
\pgfuseimage{coverage33} & \pgfuseimage{coverage31}
\end{tabular}
\end{center}
\caption{Distribution of $p$-value (for a fixed total number of events) for different true values $\delta/\Delta$ and various relative sizes of statistical and theoretical uncertainties $\Delta/\sigma$. The following approaches are shown: nG (dotted, red), Rfit or $1$-external (dashed, black), fixed $1$-nuisance   (dotted-dashed, blue), adaptive nuisance (solid, green).  Since the  1-external approach produces clusters of $p=1$ $p$-values, the coverage values excluding these clusters are also shown, as well as the distribution of $p$-values (dotted-dotted-dashed, grey). Note that the behaviour of the 1-external $p$-value  around $p=1$ is smoothened by the graphical representation.}\label{fig:coverage}
\end{figure*}

In order to compare the different situations, we take $\sigma^2+\Delta^2=1$ for all methods, and compute for each method the coverage fraction (the number of times the confidence level interval includes the true value of the parameter being extracted) for various confidence levels and for various values of $\Delta/\sigma$. Note that the coverage depends also on the true value of $\delta/\Delta$ (the normalized bias). The results are gathered in Tab.~\ref{tab:coverage} and Fig.~\ref{fig:coverage}. We also indicate the distribution of $p$ values obtained for the different methods. 

One notices in particular that the 1-external approach has a cluster of values for $p=1$, which is expected due to the presence of a plateau in the $p$-value. This behaviour makes the interpretation of the coverage more difficult, and as a comparison, we also include the results when we consider the same distribution with the $p=1$ values removed. Indeed one could imagine a situation where reasonable coverage values could only be due to the $p=1$ clustering, while other values of $p$ would 
systematically undercover: such a behaviour would either yield no constraints or too liberal constraints on the parameters depending on the data.

The results are the following:
\begin{itemize}
\item if $\Omega$ is fixed and does not contain the true value of
$\delta/\Delta$ (``unfortunate'' case), both external-$\delta$ and nuisance-$\delta$ approaches 
 lead to undercoverage; the size of the effect depends on
the distance of $\delta/\Delta$ with respect to $\Omega$. This is also the case for nG.
\item if $\Omega$ is fixed and contains the true value of $\delta/\Delta$ (``fortunate'' case), 
both the external-$\delta$ and nuisance-$\delta$ approaches overcover. 
This is also the case for nG.
\item if $\Omega$ is adaptive, for a fixed true value of $\delta$, a $p$-value becomes
valid if it is sufficiently small so that the corresponding interval contains $\delta$. Therefore,  
for the adaptive nuisance-$\delta$ approach,  there is always a maximum
value of CL above which all $p$-values are conservative; this maximum value is
given by $1-\mathrm{Erf}[\delta/(\sqrt{2}\Delta)]$.
\end{itemize}

To interpret the pattern of coverage seen above in the external and nuisance approaches, note that one starts with a $p$-value that has exact coverage under the individual simple hypotheses when $\delta$ is fixed. Therefore, as long as the true value $\delta$ lies within the range over which one takes the supremum, this procedure yields a conservative envelope. This explains the
 overcoverage/undercoverage properties for the external-$\delta$ and nuisance-$\delta$ approaches given above.

\subsection{Conclusions of the uni-dimensional case}

It should be stressed that, by construction, all methods are conservative if the true value of the $\delta$ parameter satisfy the assumption that has been made for the computation of the $p$-value. Therefore coverage properties are not the only criterion to investigate  in this situation in order to assess the methods: in particular one has to study the robustness of the $p$-value when the assumption set on the true value of $\delta$  is not true.  The adaptive approach provides a means to deal with \textit{a priori} unexpected true values of $\delta$, provided one is interested in a small enough $p$-value, that is, a large enough significance effect. Other considerations (size of confidence intervals, significance thresholds) suggest that the adaptive approach provides an interesting and fairly conservative framework to deal with theoretical uncertainties. We are going to consider the different approaches in the more general multi-dimensional case, putting emphasis on the adaptive nuisance-$\delta$ approach and the quadratic test statistic.

%% file: nDcases.tex
\section{Generalization to multi-dimensional cases}\label{sec:nD}

Up to here we only have discussed the simplest example of a single measurement $X$ linearly related to a single model parameter $\mu$. Obviously the general case is multi-dimensional, where we deal with several observables, depending on several underlying parameters, possibly in a non-linear way, with several measurements involving different sources of theoretical uncertainty. Typical situations correspond to averaging different measurements of the same quantity, and performing fits to extract confidence regions for fundamental parameters from the measurement of observables. In this section we will discuss the case of an arbitrary number of observables in a linear model with an arbitrary number of parameters, where we are particularly interested in a one-dimensional or two-dimensional subset of these parameters.

\subsection{General formulae}

We start by defining the following quadratic test statistic
\begin{equation} \label{eq:masterTstat}
T(X;\chi,\delta)= (X-x(\chi)-\Delta\tilde\delta)^T\cdot W_s \cdot (X-x(\chi)-\Delta\tilde\delta)+\tilde\delta^T \cdot\widetilde{W}_t \cdot \tilde\delta
\end{equation}
where 
$X=(X_i,\ i=1,\ldots,n)$ is the $n$-vector of measurements,
$x=(x_i,\ i=1,\ldots,n)$ is the $n$-vector of model predictions for the $X_i$ that depends on
$\chi=(\chi_j,\ j=1,\ldots, n_\chi)$, the $n_\chi$-vector of model parameters,
$\tilde\delta$ is the $m$-vector of (dimensionless) theoretical biases,
$W_s$ is the (possibly non diagonal) $n\times n$ inverse of the statistical covariance matrix $C_s$,  
$\widetilde{W}_t$ is the inverse of the (possibly non diagonal) $m\times m$ theoretical correlation matrix $\widetilde{C}_t$,  
$\Delta$ is the $n\times m$-matrix of theoretical uncertainties $\Delta_{i\alpha}$, so that the reduced biases $\tilde\delta_\alpha$ have a range of variation within $[-1,1]$ (this explains the notation with tildes for the reduced quantities rescaled to be dimensionless). 

After minimization over the $\tilde\delta_\alpha$, $T$ can be recast into the canonical form
\begin{equation}\label{Tmu}
T(X;\chi) = (X-x(\chi))^T \cdot\bar W \cdot (X-x(\chi))
\end{equation}
where
\begin{equation}\label{eq:averageWbar}
\bar{W}=W_s - B^T \cdot A^{-1} \cdot B
\end{equation} 
with
\begin{equation}
B = (W_s \Delta)^T \qquad \qquad A = \widetilde{W}_t + B \Delta
\end{equation}
The definition of $\bar{W}$ involves the inverse of matrices that can be singular. This may occur in particular in cases where the statistical uncertainties are negligible and some of the theoretical uncertainties are assumed to be 100\% correlated. This requires us to define a generalised inverse, including singular cases, which is described in detail in App.~\ref{app:singcov} and corresponds to a variation of the approach presented in Ref.~\cite{Schmelling:2000td}. Ambiguities and simplifications that can occur in the definition of $T$ are further discussed in App.~\ref{app:reduction}. In particular, one can reduce  the test statistic to the case $m=n$ with a diagonal $\Delta$ matrix without losing information.
In the case where both correlation/covariance matrices are regular, Eq.~(\ref{eq:averageWbar}) boils down to $\bar{W}=[C_s+C_t]^{-1}$ with $C_t =\Delta \widetilde{C}_t \Delta^T$. This structure is reminiscent of the discussion of theoretical uncertainties as biases and the corresponding weights given in Ref.~\cite{Porter:2008uw}, but it extends it to the case where correlations yield singular matrices.

We will focus here on the case where the model is \textit{linear}, i.e., the predictions $x_i$ depend linearly on the parameters $\chi_j$:
\begin{equation}\label{eq:linearmod}
x_i(\chi) = \sum_{k=1}^{n_\chi} a_{ik} \chi_k +b_i
\end{equation}
where $a_{ik}$ and $b_i$ are constants. We leave the phenomenologically important non-linear case and its approximate linearisation for a dedicated discussion in a separate paper~\cite{wip}.

Following the one-dimensional examples in the previous sections, we always assume that the measurements $X_i$ have Gaussian distributions for the statistical part. We will consider two main cases of interest in our field: averaging measurements and determining confidence intervals for several parameters.

\subsection{Averaging measurements}

We start by considering the averages of several measurements of a single quantity, each with both statistical and theoretical uncertainties, with possible correlations. We will focus mainly on the nuisance-$\delta$ approach, starting with two measurements before moving to other possibilities.

\subsubsection{Averaging two measurements and the choice of a hypervolume}\label{sec:average2}

A first usual issue consists in the case of two uncorrelated measurements $X_1\pm \sigma_1\pm \Delta_1$ and $X_2\pm \sigma_2\pm \Delta_2$ that we want to combine.
The procedure is well defined in the case of purely statistical uncertainties, but it depends obviously on the way theoretical uncertainties are treated. 
As discussed in Sec.~\ref{sec:threeapproaches}, associativity is a particulary appealing property for such a problem as it allows one to replace 
a series of measurements by its average without loss of information. 

Averaging two measurements amounts to combining them in the test statistic. 
The nuisance-$\delta$ approach, together with the quadratic statistic Eq.~(\ref{eq:masterTstat}), in the absence of correlations yields:
\begin{equation}
T=\frac{(X_1-\mu)^2}{\sigma_1^2+\Delta_1^2}+\frac{(X_2-\mu)^2}{\sigma_2^2+\Delta_2^2}
=(\mu-\hat\mu)^2(w_1+w_2)+T_{\rm min}
\end{equation}
with 
\begin{equation}
\hat\mu=\frac{w_1X_1+w_2X_2}{w_1+w_2}\qquad w_i=\frac{1}{\sigma_i^2+\Delta_i^2}\qquad
T_{\rm min}=\frac{(X_1-X_2)^2}{\sigma_1^2+\Delta_1^2+\sigma_2^2+\Delta_2^2}
\end{equation}
$\hat\mu$ is a linear combination of Gaussian random variables, and is thus distributed according to 
 a Gaussian p.d.f, with  mean $\mu +\delta_\mu$ and  variance $\sigma_\mu^2$
\begin{equation}\label{eq:averagenuis}
\delta_\mu=\frac{w_1\delta_1+w_2\delta_2}{w_1+w_2}\qquad
\sigma_\mu^2=\frac{w_1^2\sigma_1^2+w_2^2\sigma_2^2}{(w_1+w_2)^2}
\end{equation}
Therefore, $T-T_{\rm min}$ is distributed as a rescaled uni-dimensional non-central $\chi^2$ distribution with non-cen\-trality parameter $(\delta_\mu/\sigma_\mu)^2$.

$\sigma_\mu$ corresponds to the statistical part of the error on $\mu$. $\delta_1$ and $\delta_2$ remain unknown by construction, and the combined theory error can only be
obtained once a region of variation is chosen for the $\delta$'s (as a generalisation of the [-1,1] interval in the one-dimension case).
If one maximises the $p$-value over a rectangle ${\mathcal C}$ (called ``hypercube case'' in the following, in reference to its multidimensional generalisation),  $\delta_\mu$ varies in $\Delta_\mu$, with
\begin{equation}\label{Deltamu}
\Delta_\mu=\frac{w_1\Delta_1+w_2\Delta_2}{w_1+w_2}
\end{equation}
recovering the proposal in Ref.~\cite{Porter:2008uw} for the treatment of systematic uncertainties. In this case, $\delta_1$ and $\delta_2$ are allowed to be varied separately, without introducing any relation in their values, and can assume both extremal values.
On the other hand, if one performs the maximisation over a disk (referred to as the ``hyperball case'' for the same reasons as above) one has the range
\begin{equation}
\Delta_\mu=\frac{\sqrt{w_1^2\Delta_1^2+w_2^2\Delta_2^2}}{w_1+w_2}
\end{equation}
In this case, the values of $\delta_1$ and $\delta_2$ are somehow related, since they cannot both reach extremal values simultaneously.

Each choice of volume provides an average with different properties. As discussed earlier, associativity is a very desirable property: one can average different observations of the same quantity prior to the full fit, since it gives the same
result as keeping all individual inputs. The hyperball choice indeed fulfills associativity. On the other hand, the hypercube case does not: the combination of the inputs 1 and 2 yields the following test statistic:
$(w_1+w_2)(\mu-\hat{\mu})^2$, 
whereas the resulting combination $\hat\mu \pm \sigma_\mu\pm \Delta_\mu$ has the statistic
$(\mu-\hat{\mu})^2/(\sigma_\mu^2+\Delta_\mu^2)$.
The two statistics are proportional and hence lead to the same $p$-value, but they are not equivalent 
when added to other terms in a larger combination. 

A comment is also in order concerning the size of the uncertainties for the average.
 In the case of the hypercube, the resulting linear addition scheme  is the only
one where the average of different determinations of the same quantity cannot lead
to a weighted theoretical uncertainty that is smaller than the smallest uncertainty
among all determinations~\footnote{This is true at least for approaches where theoretical errors are modelled by fixed bias parameters: the combined error on the quantity of interest is a weighted sum as in Eq.~(\ref{Deltamu}), and the maximal value of this quantity can only be made always larger than each individual contribution if the corners of the hypercube are included in the maximisation region.}. In the case of the hyperball, it may occur that the average of different determinations of the same quantity yields a weighted theoretical uncertainty smaller than the smallest uncertainty among all determinations.

Whatever the choice of the volume, a very important and alluring property of our approach is the clean separation between the statistical and theoretical contribution to the uncertainty on the parameter of interest. This is actually a general property that directly follows from the choice of a quadratic statistic, and in the linear case it allows one to perform global fits while keeping a clear distinction between various sources of uncertainty.

\subsubsection{Averaging $n$ measurements with biases in a hyperball}\label{sec:averageN}

We will now consider here the problem of averaging $n$, possibly correlated, determinations of the same quantity, each individual determination coming with both a Gaussian statistical uncertainty, and a number of different sources of theoretical uncertainty. We focus first on the nuisance$-\delta$ approach, as it is possible to provide closed analytic expressions in this case. We will first discuss the variation of the biases over a hyperball, before  discussing other approaches, which will be illustrated and compared with examples from flavour physics in Sec.~\ref{sec:ckm}. 

We use the  test statistic Eq.~(\ref{eq:masterTstat}) for $\mu$, with $x(\chi)$ simply replaced by $\mu U$,
where $U$ is the $n$-vector $(1,\ldots,1)$.
After minimization over the $\tilde\delta_\alpha$, $T$ can be recast into the canonical form
\begin{equation}\label{Tmu-average}
T(\mu) = (X-\mu U)^T \cdot\bar W \cdot (X-\mu U)
\end{equation}
The minimization of Eq.~(\ref{Tmu-average}) over $\mu$ leads to an estimator $\hat\mu$ of the average in terms of the measurements $X_i$
\begin{equation}
\hat\mu = \sum_i w_i X_i \qquad
w_i = \sum_j \bar W_{ij} \times \left[\sum_{i,j}\bar W_{ij}\right]^{-1} 
\end{equation}
that allows one to compute the statistical uncertainty $\sigma_\mu$ in the following way
\begin{equation}\label{eq:nDstat}
\sigma_\mu^2 = \sum_{i,j} (C_s)_{ij}w_iw_j 
\end{equation}
The theoretical bias is given by $\delta_\mu=\sum_{i,\alpha} w_i \Delta_{i\alpha} \tilde\delta_\alpha$.
We would like to vary $\tilde\delta_\alpha$ in ranges required to infer the theoretical uncerainty, identifying 
the combination of biases that is uncorrelated. This is a well known problem of statistics, and 
it can be easily achieved in a linear manner by noticing that the relevant combination is $ \Delta^T \tilde{C}_t \Delta $, cf. Eq.~\eqref{eq:averageWbar}, and by introducing the Cholesky decomposition 
for the theoretical correlation matrix $\widetilde{C}_t=P\cdot P^T$, with $P$ a lower triangular matrix with diagonal positive entries. This yields the expression for the bias
\begin{equation}
\delta_\mu=\sum_{i,\alpha} w_i \Delta_{i\alpha} \tilde\delta_\alpha=\sum_{i,\alpha,\beta} w_i \Delta_{i\alpha} P_{\alpha\beta} (P^{-1}\tilde\delta)_\beta
\end{equation}
where $(P^{-1}\tilde\delta)_\beta$ are uncorrelated biases. If the latter biases are varied over a hyperball, the 
biases $\tilde\delta$ are varied over a hyperellipsoid elongated along the directions corresponding to strong correlations (see App.~\ref{app:theocorrelation} for illustrations) and one gets
\begin{equation}\label{eq:nDtheo}
\Delta_\mu = \sqrt{\sum_\beta \left(\sum_{i,\alpha} w_i \Delta_{i\alpha} P_{\alpha\beta} \right)^2}=\sqrt{w^T C_t w}\qquad\qquad ({\rm hyperball}) 
\end{equation}

Known (linear) statistical correlations between two measurements are straightforward to implement, by using the full covariance matrix in the test statistic Eq.~(\ref{eq:nDstat}).
On the other hand, in the physical problems considered here (involving hadronic inputs from lattice QCD simulations),
 it often happens that two \textit{a priori} independent calculations of the same quantity are statistically correlated, because they use the same (completely or partially) ensemble of gauge configurations. The correlation is not perfect of course, since usually different non linear actions are used to perform the computation. However the accurate calculation of the full covariance matrix is difficult, and in many cases it is not available in the literature. For definiteness, we will assume that if two lattice calculations are statistically correlated, then the (linear) correlation coefficient is one. In such a case the covariance matrix is singular, and its inverse $W_s$ is ill-defined, as well as all quantities that are defined above in terms of $W_s$. A similar question arises for fully correlated theoretical uncertainties (coming from the same method), leading to ambiguities in the definition of $\widetilde{W}_t$. Details on these issues are given in Apps.~ \ref{app:singcov} and \ref{app:theocorrelation}.

Statistical uncertainties are assumed here to be strictly Gaussian and hence symmetric (see App.~\ref{app:asym} for more detail on the asymmetric case). In contrast, in the nuisance approach, a theoretical uncertainty that is modelled by a bias parameter $\delta$ may be asymmetric: that is, the region in which $\delta$ is varied may depend on the sign of $\delta$, \textit{e.g.}, $\delta\in [-\Delta_-,+\Delta_+]$ in one dimension with the fixed hypercube approach ($\Delta_\pm\ge 0$). In order to keep the stationarity property that follows from the quadratic statistic, we take the conservative choice $\Delta={\rm Max}(\Delta_+,\Delta_-)$ in the definition Eq.~(\ref{eq:masterTstat}). Let us emphasise that this symmetrisation of the test statistic is independent of the range in which $\delta$ is varied: if theoretical uncertainties are asymmetric, one computes Eqs.~(\ref{eq:nDstat})-(\ref{eq:nDtheo}) to express the asymmetric combined uncertainties $\Delta_{\mu,\pm}$ in terms of the $\Delta_{i\alpha,\pm}$.

\subsubsection{Averages with other approaches}\label{sec:otheraverages}

In Sec.~\ref{sec:average2}, we indicated that other domains can be chosen in principle in order to perform the averages of measurements, for instance a hypercube rather than a hyperball. If we do not try to take into account theoretical correlations in the range of variation, it is quite easy to determine the result for $\Delta$
\begin{equation}\label{eq:nDtheocubenocorr}
\Delta_\mu = \sum_\alpha \left|\sum_i w_i \Delta_{i\alpha} \right|\qquad\qquad ({\rm hypercube,\ no\ theoretical\ correlation}) 
\end{equation}
reminiscent of the formulae derived in Ref.~\cite{Porter:2008uw}. However, we encountered severe difficulties when trying to include theoretical correlations in the discussion. Similarly to the hyperball case, it would be interesting to consider a linear transformation 
$P$ of the biases (for instance, the Cholesky decomposition of $C_t$, but the discussion is more general), so that $(P^{-1}\tilde\delta)_\beta$ are uncorrelated biases varied within a hypercube. This would lead to $\tilde{\delta}$ varied within a deformed hypercube, which corresponds to cutting the hypercube by
a set of $(\tilde{\delta}_i,\tilde{\delta}_j)$ hyperplanes. It can take a rather complicated convex polygonal shape that is not symmetric along the diagonal in the $(\tilde{\delta}_i,\tilde{\delta}_j)$ plane, leading to the unpleasant feature that the order in which the measurements are considered in the average matters to define the range of variation of the biases (an illustration is given in App.~ \ref{app:theocorrelation})~\footnote{This problem does not occur in the hyperball case, where the section of the hyperellipsoid by a hyperplane always yields an ellipse symmetric along the diagonal, with an elongation according to the theoretical correlation between the biases.}.
 
 As indicated before, this discussion occurs for any linear transformation $P$ and is not limited to the Cholesky decomposition. We have not been able to find other procedures that would avoid these difficulties while paralleling the hypercube case. In the following, we will thus use Eq.~(\ref{eq:nDtheocubenocorr}) even in the presence of theoretical correlations: therefore, the latter will be taken into account in the definition of $T$ through $\bar{W}$, but not in the definition of the range of variations to compute the error $\Delta$. We also notice that the problems that we encounter are somehow due to contradicting expectations concerning the hypercube approach. In Sec.~\ref{sec:average2}, the hypercube corresponds to values of $\delta_1$ and $\delta_2$ left free to vary without relation among them (contrary to the hyperball case). It seems therefore difficult to introduce correlations in this case which was designed to avoid them initially. Our failure to introduce correlations in this case might be related to 
the fact that the hypercube is somehow designed to avoid 
such correlations from the start and cannot accomodate them easily.

In the case of the external-$\delta$ approach, the scan method leads to the same discussion as for the nuisance case, provided that one uses the following statistic: $T = (X-\mu-\delta)^2/(\sigma^2+\Delta^2)$.
This choice is different from Ref.~\cite{DuboisFelsmann:2003jd} by the normalisation ($\sigma^2+\Delta^2$ rather than $\sigma^2$) in order to take into account of the importance of both uncertainties when
combining measurements (damping measurements which are unprecise in one way or the other).
As indicated in Sec.~\ref{sec:external-delta}, the difference of normalisation of the test statistic does not affect the determination of the $p$-value in the uni-dimensional case, but it has an impact once several determinations are combined. The choice above corresponds to the usual one when $\Delta$ is of statistical nature. It gives a reasonable balance when two or more inputs are combined, that all come with both statistical and theoretical uncertainties. 

A similar discussion holds for the random-$\delta$ approach. However, if the combined errors $\sigma_\mu$ and $\Delta_\mu$ are the same between the nuisance-$\delta$ (with hyperball), the random-$\delta$ and the external-$\delta$ (with hyperball) approaches, we emphasise that the $p$-value for $\mu$ built from these errors is different and yields different uncertainties for a given confidence level for each approach, as discussed in Sec.~\ref{sec:1Dillustation}.

 \subsubsection{Other approaches in the literature}\label{sec:otheraverageslit}

There are other approaches available in the literature, often starting from the random-$\delta$ approach (i.e., modeling all uncertainties as random variables). 

The Heavy Flavour Averaging Group~\cite{Amhis:2014hma} choose to perform the average including correlations. In the absence of knowledge on the correlation coefficient between uncertainties of two measurements (typically coming from the same method), they tune the correlation coefficient so that the resulting uncertainty is maximal (which is not $\rho=1$ in the case where the correlated uncertainties have a different size and are combined assuming a statistical origin, see App.~\ref{app:geninv}). This choice is certainly the most conservative one when there is no knowledge concerning correlations.

The Flavour Lattice Averaging Group~\cite{Aoki:2016frl} follows the proposal in Ref.~\cite{Schmelling:1994pz}: they build a covariance matrix where correlated sources of uncertainties are included with 100\% correlation, and they perform the average by choosing weights $w_i$ that are not optimal but are well defined even in the presence of $\rho=\pm 1$ correlation coefficients. As discussed in App.~ \ref{app:geninv}, our approach to singular covariance matrices is similar but more general and guarantees that we recover the weights advocated in Ref.~\cite{Schmelling:1994pz} for averages of fully correlated measurements.

Finally, the PDG approach~\cite{Agashe:2014kda} combines all uncertainties in a single covariance matrix. In the case of inconsistent measurements, one may then obtain an average with an uncertainty that may be interpreted as  `too small' (notice however that the weighted uncertainty does not increase with the incompatibility of the measurements).
This problem occurs quite often in particle physics and cannot be solved by purely statistical considerations (even in the absence of theoretical uncertainties).  If the model is assumed to be correct, one may invoke an underestimation of the uncertainties. A (commonly used) recipe in the pure statistical case has been adopted by the Particle Data Group, which consists in computing a factor $S=\sqrt{\chi^2/(N_{dof}-1)}$ and rescaling all uncertainties by this factor. A drawback of this approach is the lack of associativity: the inconsistency is either  removed or kept as it is, depending on whether the average is performed before any further analysis, or inside a global fit. Furthermore since the ultimate goal of statistical analyses is indeed to exclude the null hypothesis (\textit{e.g.} the Standard Model), it looks counter-intuitive to first wash out possible discrepancies by an \textit{ad hoc} procedure. Therefore we refrain to define a $S$ factor in presence of theoretical uncertainties, and leave the discussion of discrepancies between independent determinations of the same quantity to a case-by-case basis, based on physical (and not statistical) grounds. 

In the case of the Rfit approach adopted by the CKMfitter group~\cite{Charles:2004jd,Hocker:2001xe}, a specific recipe was chosen to avoid underestimating combined uncertainties in the case of marginally compatible values.
The idea is first combine the statistical uncertainties by combining the likelihoods
restricted to their statistical part, then assign to this combination the smallest
of the individual theoretical uncertainties. This is justified by the following two points:
the present state of the art is assumed not to allow one to reach a better theoretical
   accuracy than the best of all estimates, and this best estimate should not be penalized by less precise methods. In contrast with the plain (or naive) Rfit approach for averages (consisting in just combining Rfit likelihoods without further treatment), this method of combining uncertainties was called educated Rfit and is used by the CKMfitter group for averages~\cite{Charles:2011va,Lenz:2010gu,Charles:2015gya}.
Let us note finally  that the calculation of pull values, discussed in Sec.~\ref{sec:globalfit}, is a crucial step for assessing the size of discrepancies.

\subsection{Global fit}\label{sec:globalfit}

\subsubsection{Estimators and errors}

Another prominent example of multi-dimensional problem is the extraction of a constraint on a particular parameter of the model from the measured observables. If the model is linear, Eq.~(\ref{eq:linearmod}), the discussion follows closely that of Sec.~\ref{sec:averageN}. In the case where there is a single parameter of interest $\mu$, we do not write explicitly the calculations and refer to Sec.~\ref{sec:ckm} for numerical examples.

We start from the test statistic Eq.~(\ref{eq:masterTstat}) in the linear case defined in Eq.~(\ref{eq:linearmod}), reducing the number of theoretical biases to the case $m=n$ as indicated in App.~\ref{app:reduction}. Following the same discussion as in Sec.~\ref{sec:averageN}, we can minimise with respect to  $\tilde\delta_\alpha$, leading to the canonical form
\begin{equation}\label{eq:minchi1}
T(X;\chi)=(X-x(\chi))^T\cdot \bar{W} \cdot (X-x(\chi))
\end{equation}
The  minimum of this function is found at the point $\hat\chi_k$ where
\begin{equation}\label{eq:minchi2}
\left.\frac{\partial T}{\partial \chi_q}\right|_{\chi=\hat\chi} = 0, \qquad \hat\chi=(a^T\bar{W} a)^{-1}\cdot (a^T\bar{W}(X-b))
\end{equation}
so that we have
\begin{equation}\label{eq:minchi3}
\hat\chi_q=\sum w^{(q)}_i (X_i-b_i),\ w^{(q)}_i=[(a^T\bar{W} a)^{-1}a^T\bar{W}]_{qi}
\end{equation}
The minimum $\hat\chi_q$ is thus linearly related to the measured observables $X_i$ and their statistical properties are closely related. The test statistic for a particular parameter $\mu=\chi_q$ will 
lead to $T(X;\mu)=(\mu-\hat\chi_q)^2 \times (a^T\bar{W}a)_{qq}$, so that the discussion of the $p$-value for $\mu$ follows exactly the discussion for uni-dimensional measurements~\footnote{As discussed in Sec.~\ref{sec:external-delta}, the overall normalisation of $T(X;\mu)$ is irrelevant to derive uni-dimensional $p$-values.}.

For instance, if the observables $X_i$ have central values $X_{i0}$ and variances $\sigma^2_{X_i}$, the central value and the variance for $\hat\chi_q$ (corresponding also to the central value and statistical uncertainty for the $p$-value for $\mu=\chi_q$), can readily be obtained from
\begin{eqnarray} \label{eq:minlinstat}
\mu_0=\hat\chi_{q0}&=&\sum_{i,j,l=1}^n (a^T\bar{W}a)^{-1}_{qj}\times
\left[a_{ij}\bar{W}_{il}\right]\times (X_{l0}-b_l)=w^{(q)T}  (X_{0}-b)\\
\sigma_{\mu}^2=\hat\sigma_{\chi_q0}^2 &=&\sum_{i,j,l=1}^n\left[(a^T\bar{W}a)^{-1}_{qj}\right]^2\times
\left[a_{ij}\bar{W}_{il}\right]^2\times \left(\sigma_{X_l}\right)^2=w^{(q)T}C_s w^{(q)}
\end{eqnarray}
Similarly to what was presented in the previous section, the theoretical uncertainty on  $\mu=\chi_q$ is obtained in the hyperball case as
\begin{equation} \label{eq:minlintheo}
\Delta_{\mu} = \sqrt{\sum_{i,j,l=1}^n\left[(a^T\bar{W}a)^{-1}_{qj}\right]^2\times
\left[a_{ij}\bar{W}_{il}\right]^2\times \left(\Delta_{X_i}\right)^2}
  =\sqrt{w^{(q)T}C_t w^{(q)}} \qquad ({\rm hyperball})
\end{equation}
It remains to determine how to define the theoretical correlation in this framework, denoted $\kappa_{qr}$ corresponding to the actual parameters of interest. This can be seen as trying to infer a scalar product on the vectors
$[w^{(q)}\Delta P]_i$ from the knowledge of a norm, here $L^2$. We will thus define the theoretical correlation in the following way
\begin{eqnarray}\label{eq:nDcorrtheo}
\kappa_{qr}&=&\frac{w^{(q)T} C_t w^{(r)}}{\sqrt{w^{(q)T} C_t w^{(q)}}\sqrt{w^{(r)T} C_t w^{(r)}}} \qquad ({\rm hyperball})
\end{eqnarray}

In Sec.~\ref{sec:averageN} we encountered difficulties in extending the discussion to the hypercube case. We can define
errors varying the biases without correlations in the definition of the hypercube
\begin{equation} \label{eq:minlintheo-hypercube}
\Delta_{\mu} = \sum_{i=1}^n\left|\sum_{j,l=1}^n (a^T\bar{W}a)^{-1}_{qj}\times
\left[a_{ij}\bar{W}_{il}\right]\times \Delta_{X_i}\right| \qquad  ({\rm hypercube\ no\ correlation})
\end{equation}
but we could not determine a way of defining this hypercube taking into account theoretical correlations. Moreover, there is no obvious way to extend the definition of theoretical correlation for the hypercube in a similar way to Eq.~(\ref{eq:nDcorrtheo}), as there is no scalar product associated to the $L^1$-norm. We will thus not quote theoretical correlations for the hypercube case.

\subsection{Goodness-of-fit}

We would like also to compute the distribution of $T_{\rm min}$ in presence of biases and extract a goodness-of-fit value. Coming back to the initial problem, we see that $T_{\min}$ can be written as
\begin{equation}
T_{\rm min}=(X-b)^T (\bar{W}-\bar{W}a(a^T\bar{W}a)^{-1} a^T \bar{W}) (X-b)=(X-b)^T M (X-b)
\end{equation}
where $X$ are distributed following a multivariate normal distribution, with central value $a \chi+b+\Delta\tilde\delta$ and correlation matrix $C_s$. 
The CDF $H_{\tilde\delta}(t)$ for $T_{\rm min}$ at fixed $\tilde\delta$ can thus be rephrased in the following way: considering a vector $Y$ distributed according to a multivariate normal distribution of covariance $C_s$ centred around 0, $H_{\tilde\delta}(t)$ is the probability $P[(Y-a \chi-\Delta\tilde\delta)^T M (Y-a \chi-\Delta\tilde\delta)\leq t]$.

We are able to reexpress this problem as a linear combination of non-central $\chi^2$ distributions. Indeed, we can define
\begin{equation}
C_s=LL^T,\qquad L^TML=K\alpha K^T,\qquad \beta= K^T L^{-1}(a \chi-\Delta\tilde\delta)
\end{equation}
with $L$ lower triangular (using the Cholesky decomposition), $\alpha$ is diagonal and $K$ orthogonal (so that $\alpha$
are the (positive) eigenvalues of $L^TML$ and thus of $MC_s$). Let us note that $\alpha$ does depend only on $C_s$ and $ C_t$, whereas the dependence on the true value of $\chi$ and $\tilde\delta$ is only present in $\beta$. The problem is then equivalent to
considering a vector $Z$ distributed according to a multivariate normal distribution of covariance identity centred around 0, and computing $P[(Z-\beta)^T \alpha (Z-\beta)\leq t]$. This is the CDF of a linear combination of the form $\sum_i \alpha_i X_i^2$ corresponding to non-central $\chi^2$ distributions.

In the case where $\alpha$ is proportional to identity, the CDF can be expressed in terms of the generalised Marcum Q-function
\begin{equation}
H_{\tilde\delta}(t)=1-Q_{n/2}\left(\sqrt{\lambda},\sqrt{t/\alpha}\right)
\end{equation}
with the non-centrality parameter $\lambda=\sum_i \beta_i^2$. In the general case, the answer can be found in various articles, for instance in Ref.~\cite{Ruben}, as a linear combination of infinitely many (central or non-central) $\chi^2$ distribution functions, and in Ref.~\cite{Castano}, where an expansion in terms of Laguerre polynomials is provided for a fast numerical evaluation. We can thus infer the corresponding $p$ value as
\begin{equation}
p_\Omega=\max_{\tilde\delta\in\Omega} [1-H_{\tilde\delta}(t)]
\end{equation}
where $\tilde\delta$ has to be varied in a hyperball or a hypercube depending on the volume chosen, and $\chi_q$ are replaced by their estimated values $\mu_{\chi_q}$.

\subsection{Pull parameters}

In addition to the general indication given by goodness-of-fit indicators, it is useful to determine the agreement between individual measurements and the model. One way of quantifying this agreement consists in determining the pull of each quantity. Indeed, the agreement between the indirect fit prediction and the
direct determination of some observable $X$ is measured by its pull, which can be determined by considering the difference of minimum values of the test statistic including or not the observables~\cite{Charles:2015gya}. In the absence of non-Gaussian effects or correlations, the pulls are random variables of vanishing mean and unit variance.

The pull of an observable $X_m$ can be conveniently computed
by introducing an additional \textit{pull parameter} $p_{X_m}$ in the test statistic
$T(X_0;\chi,p_{X_m})$
\begin{equation}
T=(X_{0}-x(\chi) - P_m)^T \bar W (X_{0}-x(\chi) - P_m)\qquad
(P_m)_i=\delta_{mi}\ p_{X_m}/\sqrt{\bar{W}_{mm}}
\end{equation}

The pull parameter $p_{X_m}$ is a dimensionless fit parameter for which one can compute confidence intervals, or errors and uncertainties. Its best-fit value is  a random variable that measures the distance of the indirect prediction (determined by the global fit) from the direct measurement, in units of $\sigma$. The $p$-value for the null hypothesis $p_{X_m} = 0$ is by definition the pull for $X_i$. It can be understood as a comparison of the best-fit value of the test statistic reached letting $p_{X_a}$ free (corresponding to a global fit without the measurement $X_m$) with the case setting $p_{X_m}=0$ (corresponding to a global fit including the measurement $X_m$).

As far as the test statistic is concerned, the pull parameter can be treated on the same footing as the parameters $\chi$, and it can be determined in the same way as in the previous section, first solving the minimisation condition $\partial T/\partial p_{X_m}=0$, and plugging the result for $p_{X_m}$ into $T$, leading to the same expression for $T$ as in Eq.~(\ref{eq:minchi1}), but with $\bar{W}$ replaced by the matrix
\begin{equation}
\bar{W}^{(m)}_{ij}=\bar{W}_{ij}-\frac{\bar{W}_{im}\bar{W}_{jm}}{\bar{W}_{mm}}
\end{equation}
which can be solved as before for $\hat{\chi}$, leading to the expression for $\hat{p}_{X_m}$
\begin{equation}
\hat{p}_{X_m}=\sum_i y_i^{(m)} X_i \qquad
  y_{i}^{(m)}=\frac{1}{\sqrt{\bar{W}_{mm}}}[\bar{W}-\bar{W}a(a^T\bar{W}^{(k)}a)^{-1}a^T\bar{W}^{(k)}]_{mi}
\end{equation}

If the statistical method allows one to separate the statistical  and theoretical  contributions to the
error on $p_{X_i}$, one can report the values for the errors $\Delta_{p_{X_i}}$ and $\sigma_{p_{X_i}}$ in addition to the pull itself: this gives an indication of how independent from theoretical uncertainties the underlying tested hypothesis is. One can also extend this notion for $N$ parameters, introducing $N$ distinct pull parameters and determining the $p$-value for the null hypothesis where all pull parameters vanish simultaneously.

As an illustration in a simple case, one can  compute the pulls associated with the average of $n$ measurements, introducing the modified test statistic compared to Eq.~(\ref{Tmu-average}):
\begin{equation}
T(\mu,p_{X_m}) =(X-\mu U - P_m)^T \bar{W} (X-\mu U - P_m)
\end{equation}
corresponding to the case with only one parameter $\chi=\mu$, $a=U$, $b=0$.
The minimisation with respect to both parameters yields an estimator of of the pull parameter in this particular case
\begin{equation}
\hat{p}_{X_m}=\sum_i y^{(m)}_i X_i\qquad\qquad
  y^{(m)}_i=\sqrt{\bar{W}_{mm}}\frac{\left(\sum_j \bar{W}_{mj}\right)\left(\sum_j \bar{W}_{ji}\right)
     -\bar{W}_{mi}\sum_{jl}\bar{W}_{jl}}{\left(\sum_j \bar{W}_{mj}\right)^2-\bar{W}_{mm} \sum_{jl}\bar{W}_{jl}}
\end{equation}
allowing a propagation of errors in a similar way to the average of several measurements discussed in secs. \ref{sec:average2} and \ref{sec:averageN}. Numerical examples are presented in Sec.~\ref{sec:ckm}.

\subsection{Conclusions of the multi-dimensional case}

We have discussed several situations where a multi-dimensional approach is needed in phenomenology analysis. 
In addition to the issues already encountered in one dimension, a further arbitrary choice 
must be performed  in the multi-dimensional case for nuisance and external approaches concerning the shape of the volume in which the biases are varied: two simple cases are given by the hypercube and the hyperball, corresponding respectively to the well-known linear and quadratic combination of uncertainties. We have then discussed how to average two (or several) measurements, emphasising the case of the nuisance approach.  We have finally illustrated how a fit could be performed in order to determine confidence regions. Beyond the metrology of the model, we can also determine the agreement between model and experiments thanks to the pull parameters associated with each observable.

 The uni-dimensional case (stationarity of the qua\-dratic test statistic under minimisation, coverage properties) has led us to prefer the adaptive nuisance approach, even though the fixed nuisance approach could also be considered. In the multidimensional case, the hyperball in conjunction with the quadratic test statistic allows us to keep associativity when performing averages, so that it is rigourously equivalent from the statistical point of view to keep several measurements of a given observable or to average them in a single value. 
 We have also been able to discuss theoretical correlations using the hyperball case at two different stages: including the correlations among observables in the domain of variations of the biases when computing the errors $\Delta$, and providing a meaningful definition for the theoretical correlation among parameters of the fit.
 We have not found a way to keep these properties in the case of the hypercube. Moreover, choosing the hypercube may favour best-fit configurations where all the biases are at the border of their allowed regions, whereas the hyperball prevents such `fine-tuned' solutions from occurring.
 
 For comparison, in the following we will focus on two nuisance approaches: fixed 1-hypercube and adaptive hyperball with a preference for the latter. The other combinations would yield far too conservative (adaptive hypercube) or too liberal (fixed 1-hyperball) ranges of variations for the biases.

%% file: ckm.tex
\begin{table}[t]
\begin{center}
{\small
\begin{tabular}{c|ccccc}
Reference     &  & $N_f$ & Mean &  Stat &  Theo\\
\hline
ETMC10 & \cite{Constantinou:2010qv} & 2 & 0.532 & $\pm$ 0.019 & $\pm0.003\pm 0.007\pm 0.003\pm 0.008\pm 0.005$ \\
LVdW11& \cite{Laiho:2011np} & 2+1 & 0.5572 & $\pm$ 0.0028 & $\pm0.0045\pm 0.0033\pm 0.0039\pm 0.0006\pm 0.0134$ \\
BMW11 & \cite{Durr:2011ap} & 2+1 & 0.5644 & $\pm$ 0.0059 &  $\pm0.0022\pm 0.0008\pm 0.0006\pm 0.0006\pm 0.0002\pm 0.0056$\\
RBC-UKQCD12   & \cite{Arthur:2012opa} & 2+1 &  0.554 &  $\pm$ 0.008 &  $\pm0.007 \pm0.003\pm 0.012$\\
SWME14 & \cite{Bae:2014sja} & 2+1 & 0.5388 & $\pm$ 0.0034 & $\pm0.0237\pm 0.0048\pm 0.0005\pm 0.0108\pm 0.0022\pm 0.0016\pm 0.0005$
\end{tabular}

\vspace{0.4cm}

\begin{tabular}{c|ccccc}
Method & Average & 1 $\sigma$ CI & 2 $\sigma$ CI & 3 $\sigma$ CI &5 $\sigma$ CI \\
\hline
nG & $0.5577 \pm 0.0063 \pm 0$ & $0.5577 \pm 0.0063$ & $0.5577 \pm 0.0126$ & $0.5577 \pm 0.0189$ & $0.5577 \pm 0.0315$\\
naive Rfit & $0.5562\pm0.0120 \pm 0.0018 $ & $0.5562 \pm 0.0138$ & $0.5562 \pm 0.0258$ &
$0.5562 \pm 0.0379$ & $0.5562 \pm 0.0619$\\
educ Rfit & $0.5562 \pm 0.0020 \pm 0.0100$ & $0.5562 \pm 0.0120$ & $0.5562 \pm0.0139$& $0.5562 \pm0.0159$& $0.5562 \pm0.0198$\\
1-hypercube & $0.5577\pm0.0038  \pm0.0176$ & $0.5577 \pm0.0193$ & $0.5577\pm0.0240$ & $0.5577 \pm0.0281$ & $0.5577 \pm0.0360$\\
adapt hyperball & $0.5577\pm0.0038  \pm0.0050$ & $0.5577 \pm0.0068$ & $0.5577 \pm0.0165$ & $0.5577 \pm0.0257$ & $0.5577 \pm0.0436$
\end{tabular}

\vspace{0.4cm}

\begin{tabular}{c|cccccc}
Pull & nG & (e)Rfit & 1-hypercube & adaptive hyperball\\
\hline 
ETMC10 & $-1.22\pm1.04\pm 0\ (1.2\sigma)$ & $(0.0 \sigma)$& $-1.22\pm0.85\pm1.88\ (0.3\sigma)$& $-1.22\pm0.85\pm 0.60\ (1.1\sigma)$\\
LVdW11& $-0.04\pm1.10\pm 0\ (0.0\sigma)$ & $(0.0 \sigma)$& $-0.04\pm0.35\pm2.71\ (0.0\sigma)$& $-0.04\pm 0.35\pm 1.04\ (0.1\sigma)$\\
BMW11 & $\ 1.74\pm1.49\pm 0\ (1.2\sigma)$ & $(0.0 \sigma)$& $\ 1.74\pm 0.86\pm 4.32\ (0.0\sigma)$& $\ 1.74\pm0.86 \pm1.21 \ (1.0 \sigma)$\\
RBC-UKQCD12  & $-0.27\pm1.08\pm 0\ (0.2\sigma)$ & $(0.0 \sigma)$& $-0.27\pm0.55\pm2.38\ (0.0\sigma)$& $-0.27\pm0.56\pm0.93\ (0.4\sigma)$\\
SWME14  & $-0.75\pm1.03\pm 0\ (0.7\sigma)$ & $(0.0 \sigma)$& $-0.75\pm 0.19\pm 2.24\ (0.0\sigma)$& $-0.75\pm 0.19\pm 1.01\ (0.7\sigma)$
\end{tabular}}
\end{center}

\caption{Top: Lattice determinations of the kaon bag parameter $B_K^{\bar{\rm MS}}(2{\rm GeV})$.
Middle: Averages according to the various methods, and corresponding confidence intervals for various significances. Bottom: Pulls associated to each measurement for each method.
For Rfit methods, we quote only the significance of the pull, whereas other methods yield the pull parameter as well as the pull itself under the form $p\pm \sigma\pm \Delta$ (significance of the pull).}\label{tab:inputsBK}
\end{table}

\begin{table}[t]
\begin{center}
{\small
\begin{tabular}{c|ccccc}
Reference     &  & $N_f$ & Mean &  Stat &  Theo\\
\hline
ETMC09 & \cite{Blossier:2009bx} & 2    & 244 & $\pm$ 3 & $\pm 2 \pm 7$\\ 
HPQCD10 & \cite{Davies:2010ip} & 2+1 & 248.0 & $\pm$ 1.4 &  $\pm 0.4 \pm 1.4 \pm 1.0 \pm 0.8 \pm 0.3 \pm 0.3 \pm 0.3$\\ 
FNAL-MILC11 & \cite{Bazavov:2011aa} & 2+1 & 260.1 & $\pm$ 8.9 & $\pm 2.2 \pm 1.6\pm  1.0\pm  1.4 \pm 2.8 \pm 2.0  \pm 3.4 \pm 1.8$\\ 
FNAL-MILC14 & \cite{Bazavov:2014wgs} & 2+1+1 & 248.8 &$\pm$  0.3 & $\pm 1.2 \pm 0.2\pm  0.1 \pm 0.4$\\
ETMC14 & \cite{Carrasco:2014poa} & 2+1+1 & 247.2 & $\pm$ 3.9 & $\pm  0.7 \pm 1.2 \pm 0.3$  
\end{tabular}

\vspace{0.4cm}

\begin{tabular}{c|ccccc}
Method & Average & 1 $\sigma$ CI & 2 $\sigma$ CI & 3 $\sigma$ CI &5 $\sigma$ CI \\
\hline
nG & $248.5\pm1.1 \pm0 $ & $248.5 \pm1.1$ & $248.5 \pm2.2$ & $248.5 \pm3.3$ & $248.5 \pm5.5$\\
naive Rfit & $248.1\pm0.9  \pm1.3 $ & $248.1\pm2.2$ & $248.1 \pm3.1$ & $248.1 \pm4.1$ & $248.1\pm5.9$ \\
educ Rfit & $248.1\pm0.3 \pm 1.9$ & $248.1 \pm2.2$ & $248.1 \pm2.5$ & $248.1 \pm2.8$ & $248.1 \pm3.4$\\
1-hypercube & $248.5 \pm 0.5 \pm 2.7 $ & $248.5 \pm 3.0$ & $248.5 \pm 3.5$ & $248.5 \pm 4.0$ & $248.5 \pm 5.0$\\
adapt hyperball & $248.5\pm0.5 \pm 1.0 $ & $248.5 \pm 1.2$ & $248.5 \pm 2.8$ & $248.5 \pm 4.3$ & $248.5 \pm 7.2$
\end{tabular}

\vspace{0.4cm}

\begin{tabular}{c|cccccc}
Pull & nG & (e)Rfit & 1-hypercube & adaptive hyperball\\
\hline
ETMC09 & $-0.59\pm 1.01\pm0\ (0.6\sigma)$ & $(0.0 \sigma)$&  $-0.59\pm0.39\pm1.47\ (0.0\sigma)$& $-0.59\pm0.39\pm0.93\ (0.6\sigma)$\\
HPQCD10 & $-0.28\pm1.12\pm0\ (0.3\sigma)$ & $(0.0 \sigma)$& $-0.28\pm0.60\pm2.77 \ (0.0\sigma)$& $-0.28\pm0.60\pm0.95\  (0.4\sigma)$\\
FNAL-MILC11 & $1.08\pm1.00\pm0\ (1.1\sigma)$ & $(0.0 \sigma)$& $1.08\pm 0.82\pm1.74 \ (0.3\sigma)$& 
$1.08\pm 0.83\pm0.57\ (1.0\sigma)$\\
FNAL-MILC14 &  $0.63\pm1.82\pm0\ (0.3\sigma)$ & $(0.0 \sigma)$& $0.63\pm1.05\pm4.97 \ (0.0\sigma)$& $0.63\pm1.05\pm1.48 \ (0.5\sigma)$\\
ETMC14 &  $-0.35\pm1.04\pm0\ (0.3\sigma)$ & $(0.0 \sigma)$& $-0.35\pm0.94\pm 1.20\ (0.2\sigma)$& $-0.35\pm0.94\pm0.43 \ (0.4\sigma)$\\
\end{tabular}}
\end{center}
\caption{Top: Lattice determinations of the  $D_s$-meson decay constant $f_{D_s}$ (in MeV). Middle:
Averages according to the various methods, and corresponding confidence intervals for various significances. Bottom: Pull associated to each measurement for each method.
For Rfit methods, we quote only the significance of the pull, whereas other methods yield the pull parameter as well as the pull itself under the form $p\pm \sigma\pm \Delta$ (significance of the pull).}\label{tab:inputsfDs}
\end{table}

\begin{figure}[t]
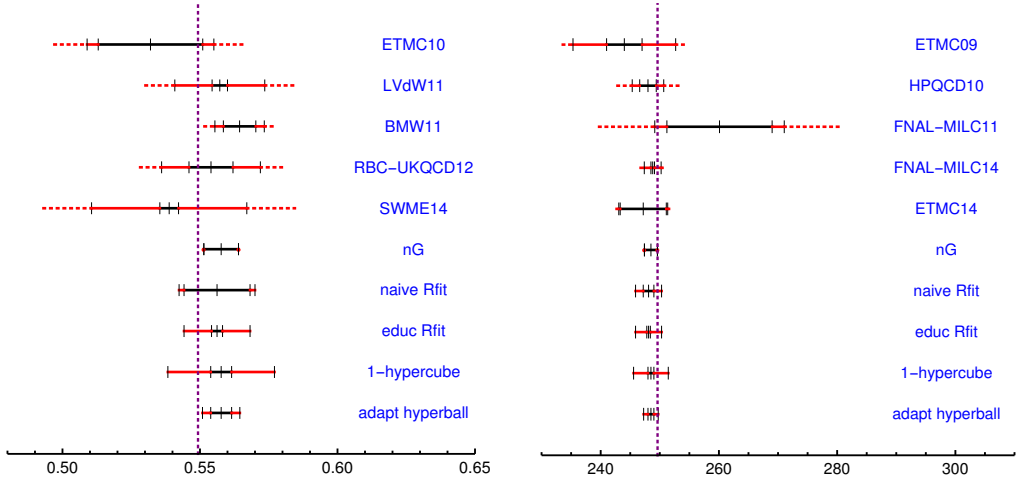

\begin{center}
\pgfuseimage{Image_BK} \hspace{0.4cm}
\pgfuseimage{Image_Ds}
\end{center}
\caption{Left: Inputs for $B_K^{\bar{\rm MS}}(2{\rm GeV})$ and the averages resulting from the different models considered here. Right: Same for the lattice determinations of the $D_s$-meson decay constant (in MeV). The black range gives the statistical error. For each individual input, the solid red range indicates the $1\sigma$ interval according to the adaptive hyperball approach (combining theoretical errors in quadratically) and the dashed red range according to the 1-fixed hypercube approach (combining theoretical errors linearly). For average according to the different approaches, the black range corresponds again to the statistical error, whereas the red range corresponds to the $1\sigma$ interval following the corresponding approach. The comparison between black and red ranges illustrates the relative importance of statistical and theoretical errors.
 Finally, for illustrative purposes, the vertical purple line gives the arithmetic average of the inputs (same weight for all central values).}
\label{fig:comparisonBKandDs}
\end{figure}

\begin{table}[t]
\begin{center}
{\small \begin{tabular}{c|cccc}
Reference & & Mean & Stat ($ \times 10^{-3} $) & Theo ($ \times 10^{-3} $) \\
\hline
ALEPH-j \& s & \cite{Dissertori:2009ik} & 0.1224 & $ \pm $ 0.9 $ \pm $ 0.9 $ \pm $ 1.2 & $ \pm $ 3.5 \\
OPAL-j \& s & \cite{OPAL:2011aa} & 0.1189 & $ \pm $ 0.8  $ \pm $ 1.6 $ \pm $ 1.0 & $ \pm $ 3.6  \\
JADE-j \& s & \cite{Bethke:2008hf} & 0.1172 & $ \pm $ 0.6 $ \pm $ 2.0 $ \pm $ 3.5 & $ \pm $ 3.0 \\
Dissertori-3j & \cite{Dissertori:2009qa} & 0.1175 & $ \pm $ 2.0 & $ \pm $ 1.5  \\
JADE-3j & \cite{Schieck:2012mp} & 0.1199 & $ \pm $ 1.0 $ \pm $ 2.1 $ \pm $ 5.4 & $ \pm $ 0.7  \\
BS-T & \cite{Becher:2008cf} & 0.1172 & $ \pm $ 1.0 $ \pm $ 0.8 & $ \pm $ 1.2 $ \pm $ 1.2  \\
DW-T & \cite{Davison:2008vx} & 0.1165 & $ \pm $ 2.2 & $ \pm $ 1.7 \\
AFHMS-T & \cite{Abbate:2010xh} & 0.1135 & $ \pm $ 0.2 & $ \pm $ 0.5 $ \pm $ 0.9  \\
GLM-T & \cite{Gehrmann:2012sc} & 0.1134 &  $ \pm$ 2.5  & $\pm$ 0.6  \\
HKMS-C & \cite{Hoang:2015hka} & 0.1123 & $ \pm $ 0.2 & $ \pm $ 0.7 $ \pm $ 1.4 \\
\end{tabular}

\vspace{0.4cm}

\begin{tabular}{c|ccccc}
Method & Average & 1 $\sigma$ CI & 2 $\sigma$ CI & 3 $\sigma$ CI &5 $\sigma$ CI \\
\hline
nG& $0.1143 \pm 0.0010 \pm 0$ &  $0.1143 \pm 0.0010$ & $0.1143 \pm 0.0020$ & $0.1143 \pm 0.0030$ & $0.1143 \pm 0.0050$\\
naive Rfit& $0.1145\pm0.0002 \pm 0$ &  $0.1145 \pm0.0002$ & $0.1145 \pm0.0004$ & $0.1145 \pm0.0006$ & $0.1145 \pm0.0011$\\
educ Rfit & 
$0.1145\pm0.0001 \pm0.0006$ & $0.1145 \pm0.0007$ & $0.1145 \pm0.0009$ & $0.1145 \pm0.0010$ &
$0.1145 \pm0.0013$\\
1-hypercube & 
$0.1143\pm0.0005 \pm0.0018$ & $0.1143 \pm0.0020$ & $0.1143 \pm0.0026$ & $0.1143 \pm0.0031$ & $0.1143 \pm0.0041$\\
adapt hyperball & 
$0.1143\pm0.0005 \pm0.0009$ & $0.1143 \pm0.0011$ & $0.1143 \pm0.0026$ & $0.1143 \pm0.0039$ & $0.1143 \pm0.0067$
\end{tabular}

\vspace{0.4cm}

\begin{tabular}{c|cccccc}
Pull & nG & (e)Rfit & 1-hypercube & adaptive hyperball\\
\hline
ALEPH-j \& s & $1.30 \pm 0.69 \pm 0\  (1.9\sigma)$ & (2.5$\sigma$)& $1.30 \pm 0.26 \pm 0.91\ (1.8\sigma)$ & $1.30 \pm 0.26 \pm 0.63\ (1.6\sigma)$ \\
OPAL-j \& s & $0.93 \pm 0.69 \pm 0\ (1.3\sigma)$
 & (0.4$\sigma$)& $0.93 \pm 0.29 \pm 0.89\ (0.7\sigma)$ & $0.93 \pm 0.29 \pm 0.63\ (1.2\sigma)$ \\
JADE-j \& s & $0.76 \pm 0.79 \pm 0\ (0.9\sigma)$ & (0.0$\sigma$)& $0.76 \pm 0.55 \pm 0.84\ (0.6\sigma)$ & $0.76 \pm 0.55 \pm 0.57\ (0.9\sigma)$ \\
Dissertori-3j & $1.13 \pm 0.77 \pm 0\ (1.4\sigma)$ & (0.9$\sigma$)& $1.13 \pm 0.58 \pm 0.95\ (0.9\sigma)$ & $1.13 \pm 0.58 \pm 0.51\ (1.3\sigma)$ \\
JADE-3j &$1.10 \pm 1.00 \pm 0\ (1.1\sigma)$ & (0.8$\sigma$)& $1.10 \pm 0.98 \pm 0.46\ (1.0\sigma)$ & $1.10 \pm 0.98 \pm 0.22\ (1.1\sigma)$ \\
BS-T &$0.36 \pm 0.92 \pm 0\ (0.4\sigma)$ & (0.2$\sigma$)& $0.36 \pm 0.88 \pm 0.41\ (0.4\sigma)$ & $0.36 \pm 0.88 \pm 0.26\ (0.4\sigma)$ \\
DW-T & $0.15 \pm 0.97 \pm 0\ (0.2\sigma)$ & (0.1$\sigma$)& $0.15 \pm 0.96 \pm 0.18\ (0.2\sigma)$ & $0.15 \pm 0.96 \pm 0.10\ (0.2\sigma)$ \\
AFHMS-T & $-0.24 \pm 0.78 \pm 0\ (0.3\sigma)$ & (0.0$\sigma$)& $-0.24 \pm 0.57 \pm 1.00\ (0.1\sigma)$ & $-0.24 \pm 0.57 \pm 0.53\ (0.4\sigma)$ \\
GLM-T & $-0.29 \pm 0.95 \pm 0\ (0.3\sigma)$ & (0.2$\sigma$)& $-0.28 \pm 0.88 \pm 0.73\ (0.2\sigma)$ & $-0.28 \pm 0.88 \pm 0.36\ (0.3\sigma)$ \\
HKMS-C &$-2.27 \pm 1.35 \pm 0\ (1.7\sigma)$ & (1.4$\sigma$)& $-2.27 \pm 0.72 \pm 2.27\ (0.7\sigma)$ & $-2.27 \pm 0.72 \pm 1.14\ (1.4\sigma)$
\end{tabular}}
\end{center}
\caption{Top: Determinations of $\alpha_S(M_Z)$ using  $ e^+ e^- $ annihilation, taken from Ref.~\cite{Agashe:2014kda}. Middle: Averages for  $\alpha_S(M_Z)$ from $e^+e^-$ annihilation according to the various methods, and corresponding confidence intervals for various significances. Bottom: Pull associated to each measurement  for each method.
For Rfit methods, we quote only the significance of the pull, whereas other methods yield the pull parameter as well as the pull itself under the form $p\pm \sigma\pm \Delta$.}\label{tab:inputsalphaSmZee}\label{tab:averagesasMZ}
\end{table}

\begin{figure}[t]
\begin{center}
\pgfuseimage{Image_alphaSMZ}
\end{center}
\caption{Determinations of the strong coupling constant at $M_Z$ through $e^+e^-$ annihilation, and the averages resulting from the different models considered here. The intervals are given at 1 $\sigma$. See Fig.~\ref{fig:comparisonBKandDs} for the legend.}
\label{fig:comparisonalphaS}
\end{figure}

\section{CKM-related examples}\label{sec:ckm}

We will now consider the differences between the various approaches considered using several examples from quark flavour physics.
These examples will be only for illustrative purposes, and we refer the reader to other works~\cite{Hocker:2001xe,Charles:2004jd,Charles:2015gya,wip} for a more thorough discussion of the physics and the inputs involved.
From the previous discussion, we could consider a large set of approaches for theoretical uncertainties. 

We will restrict to a few cases compared to the previous sections. First, we will consider 
educated Rfit (Rfit with specific treatment of uncertainties for averages),  as used by the CKMfitter analyses and described in Sec.~\ref{sec:otheraverageslit}, while the naive Rfit approach will only be shown for the sake of comparison and is not understood as an appropriate model.
We will also consider two nuisance approaches, namely  the adaptive hyperball and
the 1-hypercube cases.  Our examples will be chosen in the context of CKM fits, and correspond approximately to the situation for Summer 2014 conferences. However, for pedagogical purposes, we have simplified 
intentionally some of the inputs compared to actual phenomenological analyses performed in flavour physics~\cite{wip}.

\subsection{Averaging theory-dominated measurements}

We start by illustrating the case of measurements dominated by theoretical uncertainties, which is the case for the lattice determinations. We take the case of $B_K$, which is needed to discuss $K\bar{K}$ mixing, and has been the subject of important debates concerning its agreement (or not) with the rest of the global fit. We have selected a particular list of lattice determinations given in Tab.~\ref{tab:inputsBK} (top). For each measurement, we have kept the various theoretical uncertainties separate, since their combination (linear or quadratic) depends on the method used. For purposes of illustration, we perform an average over measurements performed with different lattice gauge actions, we symmetrise the results having asymmetric uncertainties~\footnote{In this section, we will not deal with asymmetric uncertainties, and for illustrative purpose, we symmetrise all uncertainties, statistical and theoretical, following  Eq.~(\ref{eq:asymunc}).}  and we neglect all correlations. We stress that this is 
done only for purposes of 
illustration, and that an extended list of lattice QCD results with asymmetric uncertainties and correlations will be taken into account in forthcoming phenomenological applications~\cite{wip}.

The results for each method are given in Tab.~\ref{tab:inputsBK} (middle). The first column corresponds to the outcome of the averaging procedure. In all the approaches considered, we can split statistical and theoretical uncertainties. In the case of naive Rfit, one combines the measurements by adding the well statistic corresponding to each measurement: the resulting test statistic $T$ is a well with a bottom, the width of which can be interpreted as a theoretical uncertainty, whereas the width at $T_{\min}+1$ determines the statistical uncertainty~\footnote{In general, for naive Rfit, the tails of the resulting test statistic $T$ are neither Gaussian nor symmetric. However, our approximation is valid to a good accuracy for our illustrative purposes and the examples discussed in this section.}. The case of educated Rfit was described
in Sec.~\ref{sec:otheraverageslit}. The confidence intervals are obtained from the $p$-value determined from the ``average'' column.

We compute the pulls in the same way in both cases, interpreting the difference of $T_{min}$ with and without the observables as a random variable distributed according to a $\chi^2$ law with $N_{dof}=1$. The propagation of uncertainties for the quadratic statistic was detailed in Secs.~\ref{sec:average2} and \ref{sec:averageN} where the separate extraction of statistical and theoretical uncertainties was described.
The tables are obtained by plugging the average into the 1-dimensional $p$-value associated with the method, and reading from the $p$-value the corresponding confidence interval at the chosen significance. The associated pulls are given in Tab.~\ref{tab:inputsBK} (bottom).

We present the same analysis in the case of the $D_s$-meson decay constant $f_{D_s}$ in Tab.~\ref{tab:inputsfDs} (with the same caveat concerning the selected inputs, asymmetries and correlations), while graphical comparisons of the different averages in both cases can be seen at $1 \sigma$ in Fig. \ref{fig:comparisonBKandDs} (a similar plot at $3\sigma$ is given in Fig.~\ref{fig:comparisonBKandDs3sigma} in App.~\ref{app:ckm3sigma}). 

For both quantities $B_K$ and $f_{D_s}$ at large confidence level (3 $\sigma$ and above), the most conservative method is the adaptive hyperball nuisance approach, whereas the one leading to the smallest uncertainties is the educated Rfit approach. Below 3 $\sigma$, the 1-hypercube approach is more conservative than the adaptive hyperball nuisance approach, and it becomes less conservative above that threshold. The most important differences are observed at large CL/sig\-nificance. The statistical uncertainty obtained in the nG approach is by construction identical to the combination in quadrature of the statistical and theoretical uncertainties obtained in the adaptive hyperball approach. However, one can notice that the confidence intervals for high significances in the two approaches are different, with nG being less conservative. The overall very good agreement of lattice determinations means vanishing pulls for Rfit methods (since all the wells have a common bottom with a vanishing $T_{\rm min}$). For the other methods, the pull parameter has statistical and theoretical errors of similar size in the adaptive hyperball case, whereas theoretical errors tend to dominate in the 1-hypercube method. This yields smaller pulls in the latter approach.

A last illustration, which does not come solely from lattice simulations, is provided by the determination the strong coupling constant $\alpha_S(M_Z)$.
The subject is covered extensively by recent reviews~\cite{Agashe:2014kda,d'Enterria:2015toz}, and we stress that we do not claim to provide an accurate alternative average to these reviews which requires a careful assessment of the various determinations and their correlations. As a purely illustrative example, we will focus on the average of determinations from $e^+e^-$ annihilation under a set of simplistic hypotheses for the separation between statistical and theoretical uncertainties. In order to allow for a closer comparison with Refs.~\cite{Agashe:2014kda,Bethke:2015etp}, we try to assess correlations this time. We assume that theoretical uncertainties for the same set of observables ($j\&s$, $3j$, $T$), but from different experiments, are 100\% correlated, and the statistical uncertainties for determinations from similar experimental data are 100\% correlated (BS-T, DW-T, AFHMS-T)~\footnote{In addition, we have made further choices concerning the separation of statistical and theoretical uncertainties based on the following considerations. Ref.~\cite{Gehrmann:2012sc} discusses the sources of uncertainties (scales, function parameters, b-quark mass) within a fit leading to uncertainties assumed to be of statistical nature, with a
further systematic uncertainty coming from the difference between the two different schemes. The systematic uncertainties in Ref.~\cite{Becher:2008cf} are assumed to be of statistical nature in the absence of any opposite statement.
For the first two classes (j \& s and 3j) hadronisation is taken into account by Monte Carlo methods, while for the last two classes (T and C) analytic analyses are made: in the former (latter) case, the hadronic uncertainties are treated as statistical (theoretical).}

We perform the average in the different cases considered, see Tab.~\ref{tab:averagesasMZ} (middle), which are represented graphically in Fig. \ref{fig:comparisonalphaS} (a similar plot at $3\sigma$ is given in Fig.~\ref{fig:comparisonalphaS3sigma} in App.~\ref{app:ckm3sigma}). We notice that the various approaches yield results with similar central values to the nG case. The pulls for individual quantities are mostly around 1 $\sigma$, and they are smaller in the adaptive hyperball approach compared to the nG one, showing better consistency.
Refs.~\cite{Agashe:2014kda,Bethke:2015etp} take a different approach, ``range averaging'', which amounts to considering the spread of the central values for the various determinations, leading to $\alpha_S(M_Z)=0.1174  \pm 0.0051$ for the determination from $e^+e^-$ annihilation data considered here~\cite{Bethke:2015etp}. This approach is motivated in Ref.~\cite{Agashe:2014kda} by the complicated pattern of correlations and the limited compatibility between some of the inputs and, more importantly, it does not take into account that the different determinations have different accuracies according to the uncertainties quoted. The approach in Refs.~\cite{Agashe:2014kda,Bethke:2015etp} conservatively accounts for the possibility that some uncertainties are underestimated.
On the contrary, our averages given  in Tab.~\ref{tab:averagesasMZ} and Fig. \ref{fig:comparisonalphaS} assume that all the inputs should be taken into account and averaged according to the uncertainties given in the original articles.
The difference in the underlying hypotheses for the averages explain the large difference observed between our results and the ones in Refs.~\cite{Agashe:2014kda,Bethke:2015etp}. Note however that our numerics directly follow from the use of the different averaging methods, and lack the necessary critical assessment of the individual determinations of $\alpha_S(m_Z)$ performed in Refs.~\cite{Agashe:2014kda,Bethke:2015etp}.

\begin{table}[t]
\begin{center}
{\small
\begin{tabular}{c|ccccc}
Reference     &   & Mean &  Stat &  Theo\\
\hline
Exclusive & CKMfitter Summer 14 & 3.28 & $\pm$ 0.15 &$\pm$  0.26\\ 
Inclusive & CKMfitter Summer 14& 4.359 & $\pm$ 0.180 & $\pm 0.013 \pm 0.027 \pm 0.037\pm 0.161 \pm 0.200$\\ 
\end{tabular}

\vspace{0.4cm}

\begin{tabular}{c|ccccc}
Method & Average & 1 $\sigma$ CI & 2 $\sigma$ CI & 3 $\sigma$ CI &5 $\sigma$ CI \\
\hline
nG & $3.79\pm0.22   \pm0 $ & $ 3.79 \pm0.22$ & $ 3.79 \pm0.44$ & $ 3.79 \pm0.65$ & $ 3.79 \pm1.1$\\
naive Rfit & $3.70\pm0.12   \pm0$ & $3.70 \pm0.12$ & $3.70 \pm0.23$ & $3.70 \pm0.35$ & $3.70 \pm0.58$\\
educ Rfit & $3.70\pm0.11   \pm0.26$ & $3.70 \pm0.38$ & $3.70 \pm0.49$ & $3.70 \pm0.61$ & $3.70 \pm0.84$\\
1-hypercube & $ 3.79\pm0.12   \pm0.34$ & $ 3.79 \pm0.40$ & $ 3.79 \pm0.54$ & $ 3.79 \pm0.67$ & $ 3.79 \pm0.91$\\
adapt hyperball & $ 3.79\pm0.12   \pm0.18 $ & $ 3.79 \pm0.24$ & $ 3.79 \pm0.57$ & $ 3.79 \pm0.88$ & $ 3.79 \pm1.49$
\end{tabular}

\vspace{0.4cm}

\begin{tabular}{c|cccccc}
Pull & nG & (e)Rfit & 1-hypercube & adaptive hyperball\\
\hline
Exclusive  & $-3.60\pm 1.46\pm0 \ (2.5\sigma)$ & $(1.6 \sigma)$& $-3.60\pm0.78\pm 2.31 \ (1.9\sigma)$& $-3.60\pm0.78\pm1.23 \ (1.9\sigma)$\\
Inclusive & $3.40\pm 1.38\pm0 \ (2.5\sigma)$ & $(1.6 \sigma)$& $3.40\pm0.74\pm 2.20 \ (1.9\sigma)$& $3.40\pm0.74\pm1.16 \ (1.9\sigma)$
\end{tabular}}
\end{center}
\caption{Top: Determinations of $|V_{ub}|\cdot 10^3$ from semileptonic decays.
Middle: Averages  according to the various methods, and corresponding confidence intervals for various significances. Bottom: Pulls associated to each determination for each method. For Rfit methods, we quote only the significance of the pull, whereas other methods yield the pull parameter as well as the pull itself under the form $p\pm \sigma\pm \Delta$ (significance of the pull).}\label{tab:inputsVub}
\end{table}

\begin{table}[t]
\begin{center}
{\small
\begin{tabular}{c|ccccc}
Reference     &   & Mean &  Stat &  Theo\\
\hline
Exclusive & CKMfitter Summer 14 & 38.99 & $\pm 0.49$ & $\pm 0.04\pm  0.21\pm  0.13\pm  0.39\pm  0.17\pm  0.04\pm 0.19$\\ 
Inclusive & CKMfitter Summer 14 & 42.42& $\pm 0.44$ & $\pm 0.74$\\ 
\end{tabular}

\vspace{0.4cm}

\begin{tabular}{c|ccccc}
Method & Average & 1 $\sigma$ CI & 2 $\sigma$ CI & 3 $\sigma$ CI &5 $\sigma$ CI \\
\hline
nG & $40.41\pm0.55  \pm0$ & $40.41\pm0.55$ & $40.41 \pm1.11$ & $40.41 \pm1.66$ & $40.41 \pm2.77$\\
naive Rfit & $41.00\pm0.33 \pm0 $ & $41.00\pm0.32$ & $41.00 \pm0.65$ & $41.00 \pm0.98$ & $41.00\pm1.64$\\
educ Rfit & $41.00\pm0.33 \pm0.74$ & $41.00 \pm1.07$ & $41.00 \pm1.39$ & $41.00 \pm1.72$ & $41.00 \pm2.38$\\
1-hypercube & $40.41\pm0.34  \pm0.99$ & $40.41 \pm1.15$ & $40.41 \pm1.57$ & $40.41 \pm1.94$ & $40.41\pm2.65$\\
adapt hyperball & $40.41\pm0.34 \pm0.44$ & $40.41 \pm0.60$ & $40.41 \pm1.45$ & $40.41 \pm2.26$ & $40.41 \pm3.84$
\end{tabular}

\vspace{0.4cm}

\begin{tabular}{c|cccccc}
Pull & nG & (e)Rfit & 1-hypercube & adaptive hyperball\\
\hline
Exclusive  & $-4.75\pm1.56\pm0 \ (3.1\sigma)$ & $(2.3\sigma)$& $-4.75\pm0.91\pm2.65 \ (2.6\sigma)$& $-4.75\pm 0.91\pm 1.26 \ (2.3\sigma)$\\
Inclusive & $3.98\pm1.30\pm0 \ (3.1\sigma)$ & $(2.3\sigma)$& $3.98\pm0.77\pm2.22 \ (2.6\sigma)$& $3.98\pm 0.77\pm 0.74 \ (2.3\sigma)$
\end{tabular}}
\end{center}
\caption{Top: Determinations of $|V_{cb}|\cdot 10^3$ from semileptonic decays.
Middle: Averages  according to the various methods, and corresponding confidence intervals for various significances. Bottom: Pulls associated to each determination for each method. For Rfit methods, we quote only the significance of the pull, whereas other methods yield the pull parameter as well as the pull itself under the form $p\pm \sigma\pm \Delta$ (significance of the pull).}\label{tab:inputsVcb}
\end{table}

\begin{figure}[t]
\begin{center}
\pgfuseimage{Image_Vub} \hspace{0.4cm}
 \pgfuseimage{Image_Vcb}
\end{center}
\caption{Left: Inclusive and exclusive inputs for the CKM matrix element $ \vert V_{ub} \vert $ (times $ 10^{3} $) and the averages resulting from the different models considered here. Right: Same for the determinations of $ \vert V_{cb} \vert $ (times $ 10^{3} $) CKM matrix element. The intervals are given at 1 $\sigma$. See Fig.~\ref{fig:comparisonBKandDs} for the legend.}
\label{fig:comparisonVubandVcb}
\end{figure}

\subsection{Averaging incompatible or barely compatible measurements}

Another important issue occurs when one wants to combine barely compatible measurements. This is for instance the case for
$|V_{ub}|$ and $|V_{cb}|$ from semileptonic decays, where inclusive and exclusive determinations are not in very good agreement. The list of determinations used for illustrative purposes and the results for each method are  given in Tabs.~\ref{tab:inputsVub} and \ref{tab:inputsVcb}, together with the corresponding graphical comparisons in Fig. \ref{fig:comparisonVubandVcb} (a similar plot at $3\sigma$ is given in Fig.~\ref{fig:comparisonVubandVcb3sigma} in App.~\ref{app:ckm3sigma}).
Our inputs are slightly different from Ref.~\cite{Amhis:2014hma} for several reasons. 
The inclusive determination of $|V_{ub}|$ corresponds to the BLNP approach~\cite{Lange:2005yw}, and we
consider  the theoretical uncertainties from shape functions (leading and subleading),
weak annihilation, and heavy-quark expansion uncertainties on matching and $m_b$. We use only
branching fractions measured for $B\to \pi\ell\nu$ and average the 
unquenched lattice calculations quoted in Ref.~\cite{Amhis:2014hma}. For $|V_{cb}|$ exclusive we also split the various sources of theoretical uncertainties coming from the determination of the form factors. We assume that there are no correlations among all these uncertainties.

The lack of compatibility between the two types of determination means in particular that the naive Rfit combined likelihood has not flat bottom, and thus no theoretical uncertainty. This behaviour was one of the reasons to propose the educated Rfit approach, where the theoretical uncertainty of the combination cannot be smaller than any of the individual measurements. 

The same pattern of conservative and aggressive approaches can be observed, with a fairly good agreement at 3 $\sigma$ level (apart from the naive Rfit approach, already discussed). At 5 $\sigma$, the adaptive hyperball proves again rather conservative, even though the theoretical error of the averages are smaller than the 1-hypercube nuisance and the educated Rfit approaches. The analysis of the pulls yields similar conclusions, with discrepancies at the 2 $\sigma$ for $|V_{ub}|$ and between 2 and 3 $\sigma$ for $|V_{cb}|$. Once again, theoretical errors for the pull parameters are larger in the 1-hypercube approach than in the adaptive hyperball case. Let us also notice that in both cases, there are only two quantities to combine, so that the two pull parameters are by construction opposite to each other up to an irrelevant scaling factor, leading to the same pull for both quantities.

\begin{table}[t]
\begin{center}
{\small \begin{tabular}{c|ccc}
            & $\sin(2\beta_{\rm eff})$ &      $\Delta S=\sin(2\beta_{\rm eff})-\sin(2\beta)$ & $\sin(2\beta)$\\
\hline
$\pi^0 K_S$   & $0.57\pm 0.17\pm 0$       &      $  0.085 \pm 0\pm 0.065$ &$0.485\pm 0.17\pm 0.065$\\
$\rho^0 K_S$ & $0.525 \pm 0.195\pm 0$   &    $  -0.135 \pm  0\pm0.155$ &$0.66 \pm 0.195\pm0.155$\\
$\eta' K_S$  & $0.63\pm 0.06\pm 0$         &   $   0.015 \pm  0\pm0.015$ &$0.615\pm 0.06\pm0.015$\\
$\phi K_S$   & $0.73 \pm 0.12\pm 0$          &   $     0.03 \pm  0\pm0.02$ &$0.7 \pm 0.12\pm0.02$\\
$\omega K_S$ & $0.71 \pm 0.21\pm 0$      &     $   0.11 \pm 0\pm 0.10$ &$0.6 \pm 0.21\pm 0.10$\\
$(c\bar{c}) K_S$ &  $0.689\pm 0.018$ & 0 &$0.689\pm 0.018\pm 0$
\end{tabular}

\vspace{0.4cm}

\begin{tabular}{c|ccccc}
Method & Average & 1 $\sigma$ CI & 2 $\sigma$ CI & 3 $\sigma$ CI &5 $\sigma$ CI \\
\hline
nG & $0.681\pm0.017\pm0$ & $0.681 \pm 0.017$&$0.681 \pm 0.034$&$0.681 \pm 0.051$&$0.681 \pm 0.085$
 \\
naive Rfit & $0.683 \pm 0.017 \pm 0$ & $0.683 \pm 0.017$ & $0.683 \pm 0.034$ & $0.683 \pm 0.051$ & $0.683 \pm 0.085$
 \\
educ Rfit & $0.683\pm0.017\pm0.$ & $0.683 \pm0.017$ & $0.683 \pm0.034$ & $0.683\pm0.051$ & $0.683 \pm0.084$\\
1-hypercube & $0.681\pm0.017\pm0.003$ & $0.681 \pm0.017$ & $0.681\pm0.034$ & $0.681\pm0.052$ & $0.681 \pm0.086$\\
adapt hyperball & $0.681\pm0.017\pm0.002$ & $0.681 \pm0.017$ & $0.681 \pm0.034$ & $0.681\pm0.052$ & $0.681\pm0.090$
\end{tabular}

\vspace{0.4cm}

\begin{tabular}{c|cccccc}
Pull & nG & (e)Rfit & 1-hypercube & adaptive hyperball\\
\hline
$\pi^0 K_S$   & $-1.09\pm1.00\pm0 \ (1.1\sigma)$ & $(0.8 \sigma)$& $-1.09\pm0.94\pm0.37 \ (1.1\sigma)$& $-1.09\pm0.94\pm 0.36\ (1.1\sigma)$\\
$\rho^0 K_S$ & $-0.09\pm1.00\pm0 \ (0.1\sigma)$ & $(0.0 \sigma)$& $-0.09\pm0.79\pm0.63 \ (0.1\sigma)$& $-0.09\pm0.79\pm 0.62\ (0.1\sigma)$\\
$\eta' K_S$& $-1.16\pm1.04\pm0 \ (1.1\sigma)$ & $(0.9 \sigma)$& $-1.16\pm1.01\pm0.28 \ (1.1\sigma)$& $-1.16\pm1.01\pm 0.24 \ (1.1\sigma)$\\
$\phi K_S$ & $0.16\pm1.01\pm0 \ (0.1\sigma)$ & $(0.0 \sigma)$& $0.16\pm1.00\pm0.19 \ (0.2\sigma)$& $0.16\pm1.00\pm 0.17\ (0.2\sigma)$\\
$\omega K_S$& $-0.35\pm1.00\pm0 \ (0.3\sigma)$ & $(0.0 \sigma)$& $-0.35\pm0.91\pm0.44 \ (0.3\sigma)$& $-0.35\pm0.91\pm 0.43\ (0.4\sigma)$\\
$(c\bar{c}) K_S$ & $3.79\pm2.97\pm0 \ (1.3\sigma)$ & $(1.1 \sigma)$& $3.79\pm2.87\pm1.63 \ (1.1\sigma)$& $3.79\pm2.87\pm 0.78\ (1.2\sigma)$
\end{tabular}}
\end{center}
\caption{Top: Symmetrised determinations of $\sin(2\beta_{\rm eff})$ from various penguin $b\to q\bar{q}s$ modes and from charmonia modes~\cite{Amhis:2014hma}, and estimate within QCD factorisation of the correction from penguin pollution in the Standard Model (symmetrised range quoted in Tab.~1 in Ref.~\cite{Beneke:2005pu}). We neglect any penguin pollution in the case of the charmonium extraction of $\sin(2\beta)$. Middle: Averages according to the various methods, and corresponding confidence intervals for various significances. Bottom: Pulls associated to each determination for each method. For Rfit methods, we quote only the significance of the pull, whereas other methods yield the pull parameter as well as the pull itself under the form $p\pm \sigma\pm \Delta$ (significance of the pull).
}\label{tab:inputssin2beta}
\end{table}

\subsection{Averaging quantities dominated by different types of uncertainties}

In order to illustrate the role played by statistical and theoretical uncertainties, we consider the question of averaging quantities dominated by one or the other. This happens for instance when one wants to compare a theoretically clean determination with other determination potentially affected by large theoretical uncertainties. This situation occurs in flavour physics for instance when one compares the extraction of $\sin(2\beta)$ from time-dependent asymmetries in $b\to c\bar{c}s$ and $b\to q\bar{q}s$ decays (let us recall that for the CKM global fit, only charmonium input is used for $\sin(2\beta)$). The first have a very small penguin pollution, which we will neglect, whereas the latter is significantly affected by such a pollution. The corresponding estimates of $\sin(2\beta)$ have large theoretical uncertainties, and for illustration we use the computation done in Ref.~\cite{Beneke:2005pu}. 

The results are collected in Tab.~\ref{tab:inputssin2beta}, which were computed neglecting all possible correlations between the different extractions.
One can see that the resulting theoretical uncertainty from the combination of the various inputs remains small, so that most of the approaches yield a very similar result for the confidence intervals. The corresponding pulls show a global consistency concerning the observables that deviate by $1\sigma$.

\begin{table}[t]
\begin{center}
{\small \begin{tabular}{c|cccc}
Method & Fit result & 1 $\sigma$ & 2 $\sigma$ & 3 $\sigma$ \\
\hline \multicolumn{5}{c}{$A$\qquad Scenario A}\\
\hline
nG & $0.809\pm 0.011$ & $0.809\pm 0.011$ & $0.809 \pm 0.023$ & $0.809\pm 0.034$ \\
Rfit & $0.807 \pm 0.026$ & $0.807 \pm 0.026$ & $0.807\pm 0.031$ & $0.807 \pm 0.035$\\
1-hypercube & $0.809 \pm 0.004\pm 0.025$ & $0.809\pm 0.028$ & $0.809\pm 0.033$ & $0.809\pm 0.037$\\
adaptive hyperball & $0.809 \pm 0.004\pm 0.010$ & $0.809\pm 0.012$ & $0.809\pm 0.029$ & $0.809\pm 0.043$\\
\hline \multicolumn{5}{c}{$A$\qquad Scenario B}\\
\hline
nG & $0.812\pm 0.011$ & $0.812 \pm 0.011 $ & $0.812\pm 0.022$ & $0.812 \pm 0.033$\\
Rfit & $0.804^{+0.029}_{-0.014}$ & $0.804^{+0.029}_{-0.014}$ & $0.804^{+0.033}_{-0.025}$ & $0.804^{+0.038}_{-0.030}$\\
1-hypercube & $0.812\pm 0.004\pm 0.027$ & $0.812\pm 0.029$ & $0.812\pm 0.034$ & $0.812\pm 0.038$\\
adaptive hyperball & $0.812 \pm 0.004 \pm 0.010$ & $0.812\pm 0.012$ & $0.812\pm 0.027$ & $0.812\pm 0.042$
\end{tabular}}
\pgfuseimage{scenarioA_A}\pgfuseimage{scenarioB_A}

Scenario A\qquad\qquad \qquad\qquad \qquad\qquad\qquad\qquad \qquad  Scenario B

\vspace{0.4cm}

{\small \begin{tabular}{c|cccc}

Method & Fit result & 1 $\sigma$ & 2 $\sigma$ & 3 $\sigma$ \\
\hline  \multicolumn{5}{c}{$\lambda$\qquad Scenario A}\\
\hline
nG & $0.2254\pm 0.0007$ & $0.2254\pm 0.0007$ & $0.225\pm 0.0013$ & $0.2254\pm 0.0020$ \\
Rfit &  $0.2254 \pm 0.0010$& $0.2254 \pm 0.0010$ & $0.2254 \pm 0.0010$ & $0.2254 \pm 0.0010$\\
1-hypercube & $0.2254 \pm 0.0000\pm 0.0010$ & $0.2254\pm 0.0010$ & $0.2254\pm 0.0010$ & $0.2254\pm 0.0010$ \\
adaptive hyperball & $0.2254 \pm 0.0000 \pm 0.0007$ &  $0.2254\pm 0.0007$ & $0.2254\pm 0.0014$ & $0.2254\pm 0.0020$ \\
\hline  \multicolumn{5}{c}{$\lambda$\qquad Scenario B}\\
\hline
nG & $0.2252\pm 0.0007$ & $0.2252\pm 0.0007$ & $0.2252\pm 0.0013$ & $0.2252\pm 0.0020$  \\
Rfit & $0.2245^{+0.0011}_{-0.0001}$ &  $0.2245^{+0.0011}_{-0.0001}$ & $0.2245^{+0.0020}_{-0.0001}$ & $0.2245^{+0.0020}_{-0.0001}$\\
1-hypercube & $0.2252 \pm 0.0001 \pm 0.0011$ & $0.2252\pm 0.0011$ & $0.2252 \pm 0.0012$ & $0.2252 \pm 0.0013$\\
adaptive hyperball & $0.2252 \pm 0.0001 \pm 0.0007$ & $0.22525\pm 0.00070$ & $0.2252 \pm 0.0015$ & $0.2252 \pm 0.0022$\end{tabular}}
\pgfuseimage{scenarioA_lambda}\pgfuseimage{scenarioB_lambda}

Scenario A\qquad\qquad \qquad\qquad \qquad\qquad\qquad\qquad \qquad  Scenario B
\end{center}
\caption{Numerical results and $p$-values for the CKM parameters in $A$ and $\lambda$ for Scenarios A and B, depending on the method chosen. For each quantity, we provide the error budget, whenever possible, and the plots of the $p$-values for Scenarios A (left) and B (right).}\label{tab:scenarioAlambda}

\end{table}

\begin{table}[t]
\begin{center}
{\small \begin{tabular}{c|cccc}
Method & Fit result & 1 $\sigma$ & 2 $\sigma$ & 3 $\sigma$ \\
\hline  \multicolumn{5}{c}{$\bar\rho$\qquad Scenario A}\\
\hline
nG & $0.164\pm 0.012$ & $0.164\pm 0.012$ & $0.164\pm 0.025$ & $0.164\pm 0.037$ \\
Rfit & $0.164 \pm 0.032$ & $0.164 \pm 0.032$ & $0.164 \pm 0.039$ & $0.164 \pm 0.046$\\
1-hypercube & $0.164 \pm 0.007\pm 0.026$ & $0.164\pm 0.029$ & $0.164\pm 0.038$ & $0.164\pm 0.045$\\
adaptive hyperball & $0.164 \pm 0.007\pm 0.010$ & $0.164\pm 0.014$ & $0.164\pm 0.032$ & $0.164\pm 0.051$\\
\hline \multicolumn{5}{c}{$\bar\rho$\qquad Scenario B}\\
\hline
nG & $0.145\pm 0.009$ & $0.145\pm 0.009$ & $0.145\pm 0.018$ & $0.145\pm 0.027$\\
Rfit & $0.138 \pm 0.007$ & $0.138 \pm 0.007$ & $0.138^{+0.016}_{-0.013}$ & $0.138^{+0.028}_{-0.020}$\\
1-hypercube & $0.145\pm 0.007 \pm 0.011$ & $0.145\pm 0.015$ & $0.145\pm 0.024$ & $0.145\pm 0.031$ \\
adaptive hyperball & $0.145 \pm 0.007 \pm 0.005$ & $0.145\pm 0.009$ & $0.145\pm 0.023$ & $0.145\pm 0.036$
\end{tabular}}
\pgfuseimage{scenarioA_rhobar}\pgfuseimage{scenarioB_rhobar}

Scenario A\qquad\qquad \qquad\qquad \qquad\qquad\qquad\qquad \qquad  Scenario B

\vspace{0.4cm}

{\small \begin{tabular}{c|cccc} 
Method & Fit result & 1 $\sigma$ & 2 $\sigma$ & 3 $\sigma$\\
\hline \multicolumn{5}{c}{$\bar\eta$\qquad Scenario A}\\
\hline
nG & $0.353\pm 0.021$  & $0.353\pm 0.021$ & $0.353\pm 0.042$ & $0.353\pm 0.063$ \\
Rfit & $0.354^{+0.050}_{-0.049}$ & $0.354^{+0.050}_{-0.049}$ & $0.354^{+0.059}_{-0.058}$ & $0.354^{+0.068}_{-0.067}$\\
1-hypercube & $0.353 \pm 0.009 \pm 0.041$ & $0.353\pm 0.046$ & $0.353\pm 0.057$ & $0.353\pm 0.067$\\
adaptive hyperball & $0.353  \pm 0.009 \pm 0.019$ & $0.353 \pm 0.023$ & $0.353\pm 0.054$ & $0.353\pm 0.083$\\
\hline \multicolumn{5}{c}{$\bar\eta$\qquad Scenario B}\\
\hline
nG & $0.343\pm 0.008$ & $0.343\pm 0.008$ & $0.343\pm 0.016$ & $0.343\pm 0.023$\\
Rfit &  $0.342 \pm 0.008$  & $0.342 \pm 0.008$ & $0.342^{+0.016}_{-0.015}$ & $0.342^{+0.024}_{-0.022}$\\
1-hypercube & $0.343 \pm 0.007 \pm 0.007$ & $0.343\pm 0.011$ & $0.343\pm 0.019$ & $0.343\pm 0.027$\\
adaptive hyperball & $0.343 \pm 0.007 \pm 0.003$ & $0.343\pm 0.008$ & $0.343\pm 0.018$ & $0.343\pm 0.028$\\
\end{tabular}}
\pgfuseimage{scenarioA_etabar}\pgfuseimage{scenarioB_etabar}

Scenario A\qquad\qquad \qquad\qquad \qquad\qquad\qquad\qquad \qquad  Scenario B
\end{center}
\caption{Numerical results and $p$-values for the CKM parameters in $\bar\rho$ and $\bar\eta$ for Scenarios A and B, depending on the method chosen. For each quantity, we provide the error budget, whenever possible, and the plots of the $p$-values for Scenarios A (left) and B (right).}\label{tab:scenariorhoeta}
\end{table}

\clearpage

\subsection{Global fits}

In order to illustrate the impact of the treatment of theoretical uncertainties, we consider a global fit including mainly
observables that come with a theoretical uncertainty. The list of observables is given in Tab.~\ref{tab:inputssystdominated}. Their values are motivated by the CKMfitter inputs used in Summer 2014, but they are used only for purposes of illustration~\footnote{In particular, most of the inputs have several sources of theoretical uncertainties, which should be combined together linearly or in quadrature according to the model of theoretical uncertainties chosen. Since we just want to illustrate the difference between the various approaches at the level of the fit, we take as inputs the values obtained in a given framework (Rfit) without recomputing the averages and uncertainties for each approach.}.
We consider two fits: Scenario A involves only constraints dominated by theoretical uncertainties whereas Scenario B includes also constraints from the angles (statistically dominated).

As far as the CKM matrix elements are concerned the Standard Model is linear but it is not linear in all the other fundamental parameters of the Standard Model. For the illustrative purposes of this note, the first step thus consists in determining the minimum of the full (non-linear) $\chi^2$, and to linearise the Standard Model formulae for the various observables around this minimum (we choose the inputs of scenario B to determine this point): this define an exactly linear model, which at this stage should not be used for realistic phenomenology but is useful for the comparison of the methods presented here. One can use the results presented in the previous section in order to determine the $p$-value as a function of each of the parameters of interest. In the case of the nuisance-$\delta$ approach, we can describe this $p$-value using the same parameters as before, namely a central value, a statistical error and a theoretical error. 

We provide the results for the 4 CKM parameters in both scenarios in Tabs.~\ref{tab:scenarioAlambda} and \ref{tab:scenariorhoeta} (using the same linearised theory described above). We also indicate the profiles of the $p$-values. As before, we observe that the methods give similar results at the 2-3 $\sigma$ level, although the adaptive hyperball method tends to be more conservative than the others.

\begin{table}[t]
\begin{center}
\begin{tabular}{c|c}
Observable & Input\\
\hline
$|V_{ud}|$ & $0.97425\pm 0\pm 0.00022$\\
$|V_{ub}|$ & $(3.70\pm 0.12\pm 0.26)\times 10^{-3}$\\
$|V_{cb}|$ & $(41.00\pm 0.33\pm 0.74)\times 10^{-3}$ \\
$\Delta m_{d}$ & $(0.510\pm 0.003)$ ps$^{-1}$ \\
$\Delta m_{s}$ & $(17.757\pm 0.021)$ ps$^{-1}$ \\
$B_s/B_d$ & $1.023\pm 0.013\pm0.014$\\
$B_s$ & $1.320\pm 0.017\pm0.030$\\
$f_{B_s}/f_{B_d}$ & $1.205\pm 0.004\pm 0.007$\\
$f_{B_s}$ & $225.6\pm 1.1\pm 5.4$ MeV\\
$\eta_B$ & $0.5510\pm 0\pm 0.0022$\\
$\bar{m}_t$ & $165.95\pm 0.35 \pm 0.64$ GeV\\
  \hline
  $\alpha$ & $(87.8\pm 3.4)^\circ$  \\
  $\sin(2\beta)$ &  $0.682\pm 0.019$ \\
  $\gamma$ & $(72.8\pm 6.7)^\circ$
\end{tabular}
\caption{Inputs for the theory-dominated CKM fits, inspired by the data available in Summer 2014. Scenario A is restricted to the upper part of the table, whereas Scenario B includes all inputs}\label{tab:inputssystdominated}
\end{center}
\end{table}

%% file: singular.tex
\section{Singular covariance matrices}\label{app:singcov}

\subsection{Inversion of the covariance matrix}

In Sec.~\ref{sec:averageN}, we perform the average of $N$ measurements relying on a test statistic
involving the inverse of the statistical covariance and the theoretical correlation matrices. In the case where at least two observables are fully correlated, these matrices are singular and they cannot be inverted naively. One must thus determine a generalised inverse for these matrices. For definiteness, we consider the case where only statistical uncertainties are involved. The statistical test reads
\begin{equation}
   T=(X-\mu U)^T. \bar{W}. (X-\mu U)
\end{equation}
where $U$ is a vector containing $N$ times the unit value, $\bar{W}=C_s^+$ is a generalised inverse of the covariance matrix $C_s$ (identical to $C_s^{-1}$ if the matrix $C_s$ is not singular).

Minimising $T$ yields 
\begin{eqnarray}
\hat\mu&=&\frac{U^T.\bar{W}.X}{U^T.\bar{W}.U}=\sum_i w_iX_i\qquad w_i=\frac{(\bar{W}.U)_i}{U^T.\bar{W}.U}\nonumber\\
\sigma_\mu^2&=&w^T.C_s.w=\frac{U^T.\bar{W}.C_s.\bar{W}.U}{(U^T.\bar{W}.U)^2}
\end{eqnarray}

We have to choose a generalised inverse $C_s^{+}$. We cannot rely on arguments based on the case where $C_s$ is invertible (for instance taking a correlation $0<\rho<1$, followed by the limit $\rho\to 1$) since this limit is singular. We can start by constraining the structure of $C_s^{+}$ due to the particular structure of $C_s$. We have
\begin{equation}
C_s=\Sigma.\Gamma.\Sigma=\Sigma.R.D.R^T.\Sigma
\end{equation}
where $\Sigma$ is a diagonal matrix with uncertainties as entries $\{\sigma_1,\ldots \sigma_n\}$, $\Gamma$ the correlation matrix with entries between -1 and 1 (and diagonal entries equal to 1),
$R$ is an orthogonal matrix, and $D$ is a diagonal matrix with entries in decreasing order
\begin{equation}
d_1 \geq d_2 \geq \ldots \geq d_m > 0 = d_{m+1} = \ldots = d_n
\end{equation}
The entries of $D$ are positive since $C_s$ is assumed to be positive, with 
\begin{equation}
\sum_{i=1}^N d_i={\rm Tr}(D)={\rm Tr}(\Gamma)=n\Longrightarrow d_1\leq n
\end{equation}

A generalised inverse for $C_s$ can be expressed in terms of a generalised inverse for $D$, if we define
\begin{equation}
C_s^+=\Sigma^{-1}.R.D^+.R^T.\Sigma^{-1}
\end{equation}
Indeed a generalised inverse for $C_s$ obeys $C_sC_s^+C_s=C_s$, which is equivalent to the condition
\begin{equation}\label{eq:structD}
D.D^+.D=D \quad \Longrightarrow\quad  D^+=\left[
\begin{array}{c|c} 1/d & A\\
\hline
A^T & B
\end{array}
\right]
\end{equation}
where $d$ is the $m\times m$ diagonal matrix with entries $d_i$, $A$ is an $m \times (n-m)$ arbitrary matrix and $B$ is an $(n-m)\times (n-m)$ arbitrary matrix. $A$ and $B$ can only depend on $d_1,\ldots d_m$, and 
each choice of $A$ and $B$ correspond to an admissible generalised inverse.

Under these conditions, we find for the weights and the variance
\begin{eqnarray}
w_i&=&\frac{(\Sigma^{-1}. R.D^{+}.R^T.\Sigma^{-1}.U)_i}{U^T.C_s^{+}.U}\nonumber\\
\sigma_\mu^2&=&w^T.C_s.w=\frac{U^T.C_s^+.C_s.C_s^+.U}{(U^T.C_s^{+}.U)^2}
\end{eqnarray}

\subsection{Choice of a generalised inverse}\label{app:geninv}

The most common generalised inverse  is the Moore-Penrose pseudoinverse, obtained by adding three other conditions on $C_s^+$ on top of the definition of a generalised inverse. The condition $C_s^+C_sC_s^+=C_s^+$ (reflexive generalised inverse) would translate as
$D^+.D.D^+=D^+$ leading to the condition $B=A^T.d.A$ in Eq.~(\ref{eq:structD}), whereas the two other conditions for the Moore-Penrose inverse of $C_s^+$ do not translate easily on $D^+$. Unfortunately, we will see in explicit examples that this pseudoinverse gives more weight to measurements with a poor accuracy, and is thus not appropriate in our case.

An alluring alternative to obey Eq.~(\ref{eq:structD})
 consists in considering $A=0$ and $B=\lambda\times 1_{(n-m)\times (n-m)}$ proportional to the identity, with $\lambda$ a real number to be fixed. In this case, the weights read
\begin{equation}
w_i=\frac{1}{U^T C_s^+ U}\sum_{j=1}^n \frac{1}{\sigma_i\sigma_j} (RD^+R^T)_{ij}
\end{equation}
Let us assume that $\sigma_a$ becomes much smaller than the other $\sigma_i$, the weights are dominated by
\begin{equation}
w_i\sim \frac{1}{U^T C_s^+ U} \frac{1}{\sigma_i\sigma_a} (RD^+R^T)_{ia}
\end{equation}
Since the first (normalisation) factor is the same for all the inputs, the dominant weight will be $w_a$, under the condition that 
\begin{equation}
0\neq (RD^+R^T)_{aa}=\sum_{j=1}^n (R_{aj})^2 \frac{1}{d_j} + \lambda \sum_{j=n+1}^N (R_{aj})^2 
  = \lambda  + \sum_{j=1}^n (R_{aj})^2 \left(\frac{1}{d_j}-\lambda\right)
\end{equation}
which is a condition fulfilled for $0<\lambda\leq 1/d_1$. 
We see that the family of generalised inverses thus defined~\footnote{The definition of $C_s^+$ can be extended for an arbitrary matrix $C$ in the following way. $\Sigma$ is defined as the diagonal matrix  with entries $\{\sqrt{|C_{11}|},\ldots \sqrt{|C_{NN}|}\}$ (if a diagonal entry is 0, one defines $\Sigma$ with  1 in the corresponding entry). The matrix $\Gamma=\Sigma^{-1}.C.\Sigma^{-1} $ can be written according to a singular value decomposition $\Gamma=R.D.S$ with
two rotation matrices $R$ and $S$. Once the generalised inverse $D^+$ is defined, the corresponding generalised inverse of $C$ is defined as $C^+=\Sigma^{-1}.S^T.D^+.R^T.\Sigma^{-1}$.
}
 has the following properties
\begin{itemize}
\item they can be computed in a very simple way
\item for $0<\lambda\leq 1/d_1$, if a determination is much more precise than the others, it will dominate the average
\end{itemize}
For $\lambda=1/d_1$, we call $C_s^+$ the $\lambda$-inverse of $C_s$. For $\lambda=0$, we recover
the Moore-Penrose pseudoinverse for $D$, and call this generalised inverse the 0-inverse of $C_s$. As said earlier, one could also consider the possibility of taking the Moore-Penrose pseudoinverse of $C_s$ directly. We will illustrate these three possibilities with a few simple examples.

\subsection{Examples}

\subsubsection{Two measurements}

In the case of two uncorrelated measurements, there is no problem with inversion, and we get for all methods
\begin{eqnarray}
C_s^{-1}&=&\left(\begin{array}{ccc}
\frac{1}{\sigma_1^2} & 0 \\
0 & \frac{1}{\sigma_2^2}
\end{array}\right)
\qquad 
w=\frac{1}{\sigma_1^2+\sigma_2^2}\left(\begin{array}{c}
\sigma_2^2 \\
\sigma_1^2
\end{array}\right)\nonumber\\
\sigma_\mu^2&=&\frac{\sigma_1^2\sigma_2^2}{\sigma_1^2+\sigma_2^2}
\end{eqnarray}

For partially correlated measurements ($|\rho|<1$), the same inversion can be performed, leading to
\begin{eqnarray}
C_s^{-1}&=&\frac{1}{1-\rho^2}\left(\begin{array}{ccc}
\frac{1}{\sigma_1^2} & -\frac{\rho}{\sigma_1\sigma_2} \\
-\frac{\rho}{\sigma_1\sigma_2} & \frac{1}{\sigma_2^2}
\end{array}\right)\nonumber\\
w&=&\frac{1}{\sigma_1^2-2\sigma_1\sigma_2\rho+\sigma_2^2}\left(\begin{array}{c}
\sigma_2(\sigma_2-\rho \sigma_1)\\
\sigma_1(\sigma_1-\rho \sigma_2)
\end{array}\right)
\sim \left(\begin{array}{c}
1 \\
-\rho \sigma_1/\sigma_2
\end{array}\right)\end{eqnarray}
and the expression for the uncertainty
\begin{equation}
\sigma_\mu^2=\frac{\sigma_1^2\sigma_2^2(1-\rho^2)}{\sigma_1^2-2\rho\sigma_1\sigma_2+\sigma_2^2}\sim \sigma_1^2(1-\rho^2)
\end{equation}
In each case, we indicate the limit where $\sigma_1$ becomes much smaller than $\sigma_2$ with the $\sim$ symbol, i.e., one measurement is much more accurate than the other. A comment is in order with respect to the HFAG approach at this stage.
As noticed in Ref.~\cite{Amhis:2014hma}, the maximal uncertainty is $\min(\sigma_1^2,\sigma_2^2)$ and
corresponds to the correlation coefficient $\rho=\min(\sigma_1/\sigma_2,\sigma_2/\sigma_1)$ (it is not $\rho=1$).

In the case of two fully correlated measurements, we have
\begin{equation}
C_s=\left(\begin{array}{cc} \sigma_1^2 & \sigma_1\sigma_2 \\
\sigma_1\sigma_2 & \sigma_2^2 
\end{array}
\right)
\end{equation}
with $d_1=2$, $d_2=0$. The $\lambda$-inverse for $C_s$ yields
\begin{eqnarray}
C_s^+&=&\left(\begin{array}{ccc}
\frac{1}{2 \sigma_1^2} & 0 \\
0 & \frac{1}{2 \sigma_2^2}
\end{array}\right)\nonumber\\
w&=&\frac{1}{\sigma_1^2+\sigma_2^2}\left(\begin{array}{c}
\sigma_2^2 \\
\sigma_1^2
\end{array}\right)\sim \left(\begin{array}{c}
1 \\
\sigma_1^2/\sigma_2^2 
\end{array}\right)\nonumber\\
\sigma_\mu^2&=&\frac{\sigma_1^2\sigma_2^2[\sigma_1+\sigma_2]^2}{[\sigma_1^2+\sigma_2^2]^2}\sim \sigma_1^2
\end{eqnarray}
where we indicated the limit when $\sigma_1\to 0$.
The 0-inverse yields
\begin{eqnarray}
C_s^+&=&\left(\begin{array}{ccc}
\frac{1}{4 \sigma_1^2} & \frac{1}{4 \sigma_1\sigma_2} \\
\frac{1}{4 \sigma_1\sigma_2} & \frac{1}{4 \sigma_2^2} 
\end{array}\right)
\nonumber\\
w&=&\frac{1}{\sigma_1+\sigma_2}\left(\begin{array}{c}
\sigma_2 \\
\sigma_1
\end{array}\right)\sim \left(\begin{array}{c}
1\\
\sigma_1/\sigma_2
\end{array}\right)
\nonumber\\
\sigma_\mu^2&=&\frac{4\sigma_1^2\sigma_2^2}{[\sigma_1+\sigma_2]^2}\sim 4 \sigma_1^2
\end{eqnarray}
and the Moore-Penrose pseudoinverse yields
\begin{eqnarray}
C_s^+&=&\frac{1}{(\sigma_1^2+\sigma_2^2)^2}\left(\begin{array}{ccc}
\sigma_1^2 & \sigma_1\sigma_2 \\
\sigma_1\sigma_2 &\sigma_2^2 
\end{array}\right)\nonumber\\
w&=&\frac{1}{(\sigma_1+\sigma_2)}\left(\begin{array}{c}
\sigma_1\\ \sigma_2 \end{array}\right)\sim 
\left(\begin{array}{c}
\sigma_1/\sigma_2\\ 1\end{array}\right)\nonumber\\
 \sigma_\mu^2&=&\frac{(\sigma_1^2+\sigma_2^2)^2}{(\sigma_1+\sigma_2)^2}\sim \sigma_2^2
\end{eqnarray}

\subsubsection{$n$ fully correlated measurements}

We have a correlation matrix $\tilde{C}_s$ with unit entries everywhere.
This yields $d_1=n$, $d_{i>1}=0$. 
The $\lambda$-inverse yields
\begin{eqnarray}
C_s^+&=&\left(\begin{array}{ccc}
\frac{1}{n \sigma_1^2} & \cdots & 0\\
\vdots & \ddots & \vdots  \\
0 & \cdots & \frac{1}{n \sigma_n^2}
\end{array}\right)\nonumber\\
w&=&\frac{1}{\sum_i 1/\sigma_i^2}
\left(\begin{array}{c}
1/\sigma_1^2 \\
\vdots \\
1/\sigma_N^2 
\end{array}\right)\sim \left(\begin{array}{c}
1 \\
\sigma_1^2/\sigma_2^2\\
\vdots \\
\sigma_1^2/\sigma_n^2
\end{array}\right)\nonumber\\
\sigma_\mu^2&=&\frac{(\sum 1/\sigma_i)^2}{(\sum 1/\sigma_i^2)^2}
\sim \sigma_1^2
\end{eqnarray}
where we indicated the limit when $\sigma_1\to 0$.
The 0-inverse yields
\begin{eqnarray}
C_s^+&=&\left(\begin{array}{cccc}
\frac{1}{n^2 \sigma_1^2} & \frac{1}{n^2 \sigma_1\sigma_2} & \cdots & \frac{1}{n^2\sigma_1\sigma_n}\\
\vdots &  &  & \vdots \\
 \frac{1}{n^2\sigma_1\sigma_n} &  \frac{1}{n^2 \sigma_2\sigma_n} & \cdots &  \frac{1}{n^2 \sigma_n^2}
\end{array}\right)\nonumber\\
w&=&\frac{1}{\sum_i 1/\sigma_i}
\left(\begin{array}{c}
1/\sigma_1 \\
\vdots \\
1/\sigma_n
\end{array}\right)\sim \left(\begin{array}{c}
1 \\
\sigma_1/\sigma_2\\
\vdots \\
\sigma_1/\sigma_n
\end{array}\right)\nonumber\\
\qquad \sigma_\mu^2&=&\frac{n^2}{(\sum_i 1/\sigma_i)^2}
\sim n^2\sigma_1^2
\end{eqnarray}

The Moore-Penrose pseudoinverse yields
\begin{eqnarray}
C_s^+&=&\frac{1}{(\sum_i \sigma_i^2)^2}\left(\begin{array}{ccc}
\sigma_1^2 & \cdots & \sigma_1\sigma_n \\
\vdots & \ddots & \vdots \\
 \sigma_1\sigma_n &\cdots & \sigma_n^2 \\
\end{array}\right)
\nonumber\\
w&=&\frac{1}{\sum \sigma_i}\left(\begin{array}{c}
\sigma_1\\ \vdots\\ \sigma_n \end{array}\right)\sim 
\frac{1}{\sum_{i>1} \sigma_i}\left(\begin{array}{c}
\sigma_1\\ \vdots\\ \sigma_n \end{array}\right)\nonumber\\
 \sigma_\mu^2&=&\frac{(\sum_i \sigma_i^2)^2}{(\sum \sigma_i)^2}\sim 
 \frac{(\sum_{i>1} \sigma_i^2)^2}{(\sum_{i>1} \sigma_i)^2}
\end{eqnarray}

We can actually show that in this situation, the choice of the $\lambda$-inverse is optimal in the family of generalised inverses defined in App.~\ref{app:geninv}. Indeed, there is only one non-vanishing eigenvalue $d_1=n$, leading to
\begin{equation}
\sigma_\mu^2
 =\frac{(\sum 1/\sigma)^2/n^2}{\left[(\sum 1/\sigma)^2/n^2
    +\lambda \left[\sum 1/\sigma^2-(\sum 1/\sigma)^2/n\right]\right]^2}
\end{equation}
which is minimal for the maximal value $\lambda=1/d_1$, corresponding to the $\lambda$-inverse. 

\subsubsection{Two fully correlated measurements with an uncorrelated measurement }

Let us consider
\begin{equation}
C_s=\left(\begin{array}{ccc} \sigma_1^2 & \sigma_1\sigma_2 & 0\\
\sigma_1\sigma_2 & \sigma_2^2 & 0\\
0 & 0 & \sigma_3^2
\end{array}
\right)
\end{equation}
with $d_1=2$, $d_2=1$, $d_3=0$.

The $\lambda$-inverse for $C_s$ yields
\begin{eqnarray}
C^+&=&\left(\begin{array}{ccc}
\frac{1}{2 \sigma_1^2} & 0 & 0\\
0 & \frac{1}{2 \sigma_2^2} & 0 \\
0 & 0 & \frac{1}{\sigma_3^2}
\end{array}\right)\\
w&=&\frac{1}{2\sigma_1^2\sigma_2^2+\sigma_1^2\sigma_3^2+\sigma_2^2\sigma_3^2}\left(\begin{array}{c}
\sigma_2^2\sigma_3^2 \\
\sigma_1^2\sigma_3^2 \\
2\sigma_1^2\sigma_2^2
\end{array}\right)\sim \left(\begin{array}{c}
1 \\
\sigma_1^2/\sigma_2^2 \\
2\sigma_1^2/\sigma_3^2
\end{array}\right)\nonumber\\
\sigma_\mu^2&=&\frac{\sigma_1^2\sigma_2^2\sigma_3^2[2\sigma_1\sigma_2\sigma_3^2+4\sigma_1^2\sigma_2^2+\sigma_1^2\sigma_3^2+\sigma_2^2\sigma_3^2]}{[2\sigma_1^2\sigma_2^2+\sigma_1^2\sigma_3^2+\sigma_2^2\sigma_3^2]^2}\sim \sigma_1^2\nonumber
\end{eqnarray}
The 0-inverse for $C_s$ yields
\begin{eqnarray}
C_s^+&=&\left(\begin{array}{ccc}
\frac{1}{4 \sigma_1^2} & \frac{1}{4 \sigma_1\sigma_2} & 0\\
\frac{1}{4 \sigma_1\sigma_2} & \frac{1}{4 \sigma_2^2} & 0 \\
0 & 0 & \frac{1}{\sigma_3^2}
\end{array}\right)\nonumber\\
w&=&\frac{1}{4\sigma_1^2\sigma_2^2+\sigma_1^2\sigma_3^2+\sigma_2^2\sigma_3^2+2\sigma_1\sigma_2\sigma_3^2}\left(\begin{array}{c}
\sigma_2\sigma_3^2(\sigma_1+\sigma_2) \\
\sigma_1\sigma_3^2(\sigma_1+\sigma_2) \\
4\sigma_1^2\sigma_2^2
\end{array}\right)\sim \left(\begin{array}{c}
1\\
\sigma_1/\sigma_2\\
4\sigma_1^2/\sigma_3^2
\end{array}\right)\nonumber\\
\sigma_\mu^2&=&\frac{4\sigma_1^2\sigma_2^2\sigma_3^2}{4\sigma_1^2\sigma_2^2+\sigma_1^2\sigma_3^2+\sigma_2^2\sigma_3^2+2\sigma_1\sigma_2\sigma_3^2}\sim 4 \sigma_1^2
\end{eqnarray}
The Moore-Penrose pseudoinverse yields
\begin{eqnarray}
C_s^+&=&\left(\begin{array}{ccc}
\frac{\sigma_1^2}{(\sigma_1^2+\sigma_2^2)^2} & \frac{\sigma_1\sigma_2}{(\sigma_1^2+\sigma_2^2)^2} & 0 \\
 \frac{\sigma_1\sigma_2}{(\sigma_1^2+\sigma_2^2)^2} &\frac{\sigma_2^2}{(\sigma_1^2+\sigma_2^2)^2} & 0 \\
0 & 0 & \frac{1}{\sigma_3^2} \\
\end{array}\right)\nonumber\\
w&=&\frac{1}{(\sigma_1^2+\sigma_2^2)^2+(\sigma_1+\sigma_2)^2\sigma_3^2}\left(\begin{array}{c}
\sigma_1\sigma_3^2(\sigma_1+\sigma_2)\\ \sigma_2\sigma_3^2(\sigma_1+\sigma_2)\\ (\sigma_1^2+\sigma_2^2)^2 \end{array}\right) \sim 
\frac{1}{(\sigma_2^2+\sigma_3^3)}\left(\begin{array}{c}
\sigma_1\sigma_3^2/\sigma_2\\ \sigma_3^2\\ \sigma_2^2 \end{array}\right)\nonumber\\
 \sigma_\mu^2&=&\frac{(\sigma_1^2+\sigma_2^2)^2\sigma_3^2}{(\sigma_1^2+\sigma_2^2)^2+(\sigma_1+\sigma_2)^2\sigma_3^2}\sim \frac{\sigma_2^2 \sigma_3^2}{\sigma_2^2+\sigma_3^2}
\end{eqnarray}

\subsection{Choice of the inverse}

In the above examples, the $\lambda$-inverse yields interesting results for the generalised inverse in cases that are likely to be useful. In the limit where one measurement becomes very accurate, it dominates the average. In this situation, other generalised inverses of the same family, like the 0-inverse, yield results of the same order, but larger, for the combined uncertainty, whereas the Moore-Penrose pseudoinverse yields a combined uncertainty dominated by the least precise measurements.
For 100\% correlated uncertainties, the $\lambda$-inverse recovers Schmelling's proposal~\cite{Schmelling:1994pz} used by the Flavour Lattice Averaging Group~\cite{Aoki:2016frl}, and it does not run into the danger of underestimating the resulting uncertainty as  discussed by the Heavy Flavour Averaging Group~\cite{Amhis:2014hma}.

For these reasons, we choose the $\lambda$-inverse to compute both the inverse statistical covariance matrix and the inverse theoretical correlation matrix when these matrices are singular (the regular case being trivial).

%% file: correlation.tex
\section{Varying the biases in the presence of theoretical correlations}\label{app:theocorrelation}

\subsection{Range of variations for the biases}

Another issue consists in implementing correlations for the biases describing theoretical uncertainties. Some differences occur compared to statistical uncertainties, since different models are used in both cases (random variables versus biases). As described in Sec.~\ref{sec:averageN}, once the weights $w_i$ are determined, the theoretical uncertainty is given by $\delta_\mu=\sum_i w_i \Delta_{i\alpha} \tilde\delta_\alpha$, which requires one to determine the range of variation for the normalised biases $\tilde\delta_\alpha$.
We want to describe their variation starting from variations of uncorrelated variables. This can be achieved through a linear transformation by introducing
the Cholesky decomposition for the theoretical correlation matrix $C_t=P\cdot P^T$ with $P$ a lower triangular matrix with diagonal positive entries. We obtain
the expression for the theoretical uncertainty
\begin{equation}
\delta_\mu=\sum_i w_i \Delta_i \tilde\delta_i=\sum_{i,j} w_i \Delta_{i\alpha} P_{\alpha\beta} (P^{-1}\tilde\delta)_\beta
\end{equation}
where $(P^{-1}\tilde\delta)_j$ are uncorrelated biases varied in a hyperball, leading to
\begin{equation}\label{eq:hyperballDeltamu}
\Delta_\mu = \sqrt{\sum_\beta \left(\sum_{i,\alpha} w_i \Delta_{i\alpha} P_{\alpha\beta} \right)^2} \ \ (\mathrm{hyperball})
\end{equation}
There is an ambiguity in the definition of $P$ when $C_t$ is only semi-definite positive (which occurs when $C_t$ is singular due to 100\% correlations, and exhibits not only positive but also vanishing eigenvalues). We define then $P$ by computing $P(\epsilon)$ for the shifted matrix $C_t + \epsilon\times 1_{m\times m}$ and defining $P=\lim_{\epsilon\to 0^+}P(\epsilon)$. This limit is not singular, and it allows one to define the limit of two measurements fully correlated theoretically as a smooth limit of the general case with a partial correlation.

One should emphasise that in the case of a singular correlation matrix $C_t$ for theoretical uncertainties, we may have to treat this singularity at two different stages: first when we build the test statistic involving $\bar{W}$ (depending on the structure of the statistical and theoretical correlation matrices), second when we consider the domain of variation for the parameters $\tilde\delta$. We stress that we used different procedures in both cases ($\lambda$-inverse for $\bar{W}$, Cholesky decomposition for $\tilde\delta$), which involves some arbitrariness, but reproduces desirable properties for the combined uncertainties and domains of variation of the biases in this singular limit.

In the case of a hypercube, we may want to follow the same procedure and define
\begin{equation}\label{eq:hypercubecorrDeltamu}
\Delta_\mu = \sum_\beta \left| \sum_{i,\alpha} w_i \Delta_{i\alpha} P_{\alpha\beta}  \right| \qquad\qquad  (\mathrm{hypercube\ with\ correlations\ ?})
\end{equation}
The question mark indicates that this definition is only tentative, and will not actually be used. Indeed
as discussed in Sec.~\ref{sec:otheraverages} and illustrated in the following sections, this definition has the rather unpleasant feature that the ranges of variations depend on the order of the inputs used, and we have not been able to identify an alternative
choice for the range of variations that would avoid this problem, which does not occur in the hyperball case. These difficulties could be somehow expected from the properties of the hypercube case discussed in Sec.~\ref{sec:average2}. Indeed, in the case of two measurements, the hypercube corresponds to values of $\delta_1$ and $\delta_2$ left free to vary without relation among them (contrary to the hyperball case). It seems therefore difficult to introduce correlations in this case which was designed to avoid them initially. Our failure to introduce correlations in this case might be related to the fact that the hypercube is somehow designed to avoid 
such correlations from the start and cannot accommodate them easily.

We thus propose the alternative definition, ignoring theoretical correlations to determine the range of variations for the biases
\begin{equation}\label{eq:hypercubenocorrDeltamu}
\Delta_\mu = \sum_\alpha \left| \sum_i w_i \Delta_{i\alpha} \right| \ \ (\mathrm{hypercube\ no\ correlation})
\end{equation}

\subsection{Averaging measurements with theoretical correlations}

If we take two measurements $X_1\pm \sigma_1 \pm \Delta_1$ and $X_2\pm \sigma_2\pm \Delta_2$ with $\sigma_1$ and $\sigma_2$ uncorrelated, but $\Delta_1$ and $\Delta_2$ correlated with a correlation $\rho$, one gets the Cholesky decomposition
\begin{equation}
C_t=P.P^T \qquad P=\left(\begin{array}{cc}1 & 0\\ \rho & \sqrt{1-\rho^2}\end{array}\right)
\end{equation}
so that the variations for the two (normalised) biases $\tilde\delta_1$ and $\tilde\delta_2$ are given by
\begin{equation}\label{eq:biasesdomain2}
\tilde\delta_1=d_1 \qquad \tilde\delta_2=\rho d_1 + \sqrt{1-\rho^2} d_2
\end{equation}
where $d_1$ and $d_2$ are varied in a hyperball or a hypercube following Eqs.~(\ref{eq:hyperballDeltamu}) and (\ref{eq:hypercubecorrDeltamu}) respectively. Eq.~(\ref{eq:hypercubenocorrDeltamu}) would correspond to neglecting correlations and setting $\rho=0$ in Eq.~(\ref{eq:biasesdomain2}).

\begin{figure*}[t]
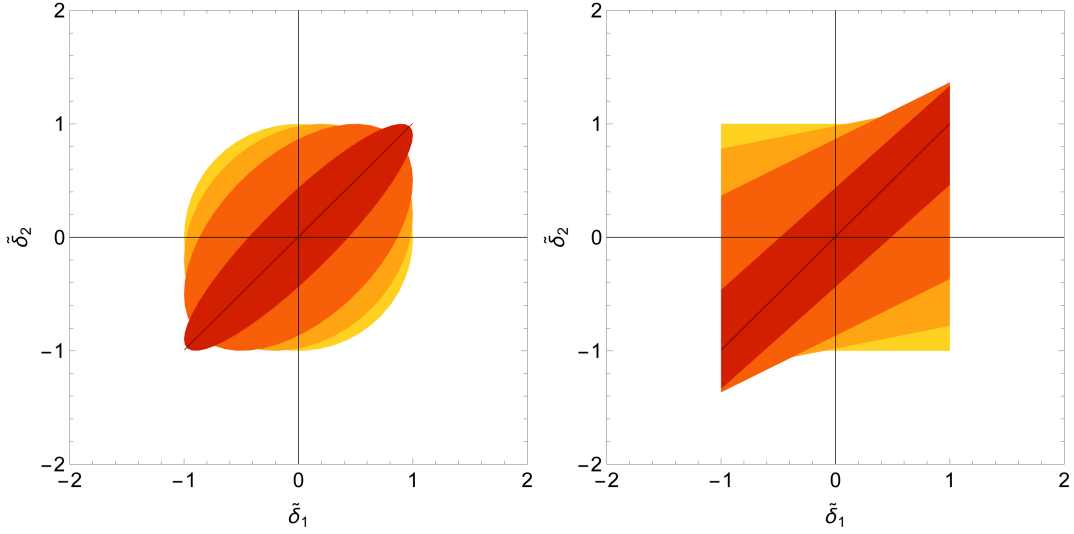

\begin{center}
\pgfuseimage{theocorr2ball}
\pgfuseimage{theocorr2cube}
\caption{Ranges of variation for $\tilde\delta_1$ and $\tilde\delta_2$ for $\rho=0,0.2,0.5,0.9,1$, going from light (yellow) to dark (red). The variation over a hyperball (left) or a hypercube (right) is considered.} \label{fig:theocorrrange}
\end{center}
\end{figure*}

In the case of a hypercube with correlations, $\tilde\delta_1,\tilde\delta_2$ are varied in a parallelogram with two sides parallel to the $\tilde\delta_2$ axis, whereas they are varied in a tilted ellipse in the hyperball case, as can be seen in Fig.~\ref{fig:theocorrrange}. In both cases, the limiting case where $\rho\to \pm1$ corresponds to $\tilde\delta_1$ and $\tilde\delta_2$ varied along a diagonal line, meeting our expectations for fully correlated theoretical uncertainties. We see that this treatment yields a symmetric domain for $\tilde\delta_1$ and $\tilde\delta_2$ in the hyperball case, but not in the hypercube case, which means that the two uncertainties are not treated in a symmetric way~\footnote{One could try to symmetrise the problem, but one would lose the connection with the Cholesky decomposition, with the unpleasant feature that all domains of variation would be identical and thus do not take into account correlations.}. As indicated before, Eq.~(\ref{eq:hypercubenocorrDeltamu}) 
corresponds to the hypercube with $\rho=0$, i.e., a square domain for $\tilde\delta_1$ and $\tilde\delta_2$.

\begin{figure*}[t]
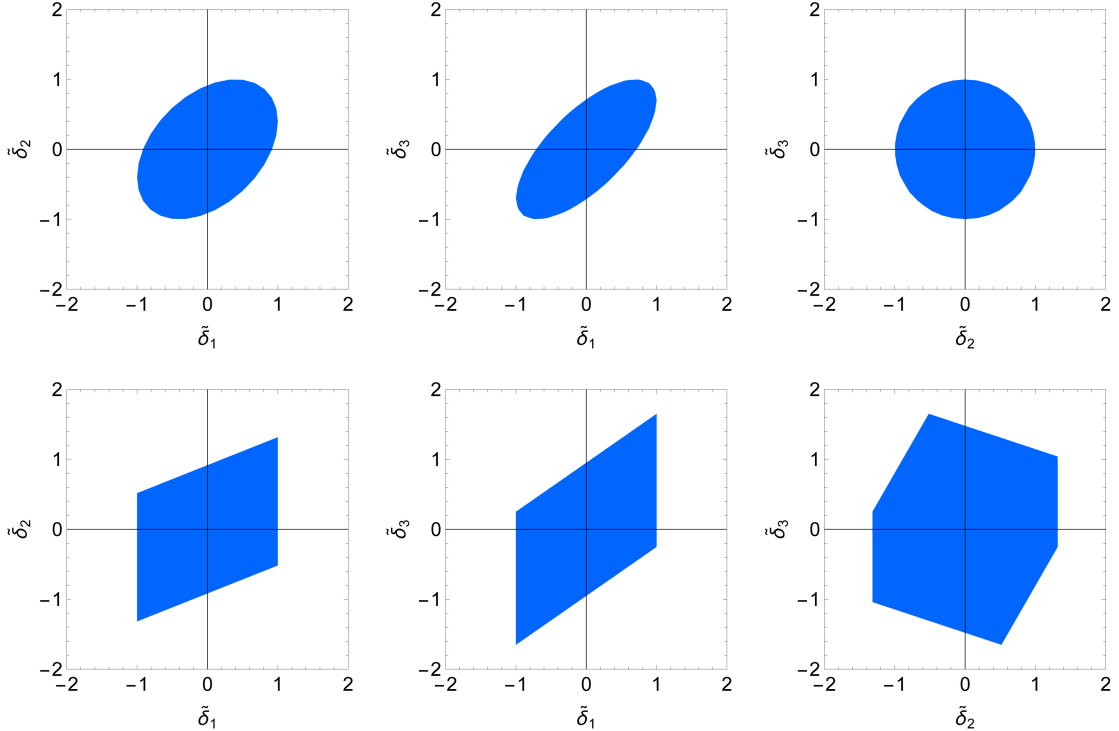

\begin{center}
\pgfuseimage{theocorr3ball}
\pgfuseimage{theocorr3cube}
\caption{Ranges of variation for $\tilde\delta_1,\tilde\delta_2,\tilde\delta_3$ with $\rho_{12}=0.4,\rho_{13}=0.7,\rho_{23}=0$. The variation over a hyperball (top) or a hypercube (bottom) with correlations is considered. Neglecting correlations would yield discs (top) and squares (bottom).}\label{fig:theocorr3}
\end{center}
\end{figure*}

One can easily extend the same procedure to a larger number of correlated theoretical uncertainties. As indicated above, the hyperball with correlations  yields domains of variations which are symmetric for any pair  $(\tilde\delta_k,\tilde\delta_l)$ whereas the hypercube with correlations does not. This means that the range of variation chosen for the biases will depend on the order of the inputs: a mere reshuffling of the inputs will yield different ranges of variations for the biases and (in general) different outcomes for averages and fits. In addition, we should emphasise that a total correlation $(C_t)_{k,l}=0$ between two biases does not  have the same impact for the domain of variation in the $(\tilde\delta_k,\tilde\delta_l)$ plane in both approaches: in the hyperball case, one obtains an undeformed disk, whereas the hypercube case yields a complicated convex polytope depending on the other elements of the correlation matrix (see Fig.~\ref{fig:theocorr3} in the case of three biases) (a symmetrisation of the Cholesky 
decomposition in the form $P+P^T$ or a  different choice of linear transformation would yield similar results).

These features lead us to neglect correlations in the hypercube range of variations, whereas we keep them when considering the hyperball case. We thus discard Eq.~(\ref{eq:hypercubecorrDeltamu}) and consider only Eqs.~(\ref{eq:hyperballDeltamu}) and (\ref{eq:hypercubenocorrDeltamu}) in our analyses.

%% file: reduction.tex
\section{Definition of the test statistic in $n$ dimensions}\label{app:reduction}

\subsection{Ambiguities in the definition of a 100\% theoretical correlation}

In Eq.~(\ref{eq:masterTstat}), one may be uncertain about the case where a theoretical uncertainty is fully correlated between two observables. Let us imagine that we have two quantities $X_1=X_{10}\pm \sigma_1 \pm \Delta_1$ and $X_2=X_{20}\pm \sigma_2 \pm \Delta_2$ with the two theoretical uncertainties being fully correlated. We can imagine describing the theoretical uncertainties either via $m=2$ parameters fully correlated through $\widetilde{C}_t$:
\begin{equation}
I: \qquad \Delta=\left[\begin{array}{cc} \Delta_1 & 0\\ 0 & \Delta_2\end{array}\right]
 \qquad \tilde{C}_t=\left[\begin{array}{cc} 1 & 1\\ 1 & 1\end{array}\right]\qquad 
\end{equation}
or as $m=1$ parameter intervening in the two quantities via $\Delta$
\begin{equation}
II: \qquad \Delta=\left[\begin{array}{c} \Delta_1\\ \Delta_2\end{array}\right]
 \qquad \tilde{C}_t=\left[1 \right]
\end{equation}
We can see in the above discussion that the only relevant combination of $\Delta$ and $\widetilde{C}_t$ is actually $\Delta P$, whether in the definition of $\bar{W}$ that involves $\Delta \widetilde{C}_t \Delta^T=(\Delta P) (\Delta P)^T$, or in the discussion of the theoretical uncertainty $\Delta_\mu$. We have
\begin{eqnarray}
I&:& P=\left[\begin{array}{cc} 1 & 0\\ 1 & 0\end{array}\right] \quad
  \Delta P=\left[\begin{array}{cc} \Delta_1 & 0\\ \Delta_2 & 0\end{array}\right]\nonumber\\
II&:&P=\left[1\right] \quad \Delta P=\left[\begin{array}{cc} \Delta_1 \\ \Delta_2\end{array}\right]
\end{eqnarray}
leading to  the same $\Delta \widetilde{C}_t \Delta^T$ and showing that only one uncorrelated bias parameter is needed in both cases, even though we started from a different number of bias parameters.
The discussion can be extended to an arbitrary number of fully correlated theoretical uncertainties. Obviously, for partial correlations, only $\widetilde{C}_t$ can be used with an unchanged number of bias parameters.

\subsection{Reducing the problem to one bias parameter per observable}

We can define a reduced version of the problem Eq.~(\ref{eq:masterTstat}), with only $n$ bias parameters rather than $m$. We have to determine an equivalent problem
\begin{equation}
T'(x,\tilde\delta')=[X-x-\Delta' \tilde\delta']^T W_s [X-x-\Delta' \tilde\delta']+\tilde\delta^T \bar{W}'_t \tilde\delta'
\end{equation}
where $\bar{W}'_t$ and $\Delta'$ are $n\times n$ matrices, and $\Delta'$ is diagonal. From what was discussed before, we see that we will obtain the same result for the weights $w^{(q)}$, the variances and the correlations, if we ensure that $\Delta P=\Delta' P'$. 

This can be achieved by defining $\Delta'$ and the correlation matrix $\widetilde{C}'_t$ using
\begin{equation}
\Delta \widetilde{C}_t \Delta^T = \Delta' \widetilde{C}'_t \Delta'
\end{equation}
$\widetilde{C}_t$ is positive semi-definite, which means that $\Delta \widetilde{C}_t \Delta^T$ will also be. The diagonal elements of a positive semi-definite matrix are positive, and therefore, one can define $\Delta'$ so that $\widetilde{C}'_t$ has 1 as a diagonal. 

It could occur that $\Delta \widetilde{C}_t \Delta^T$ has 0 on the diagonal for some $k^{th}$ entry. But since $\Delta \widetilde{C}_t \Delta^T$ is positive semi-definite, one can prove that the corresponding row and column then vanish, meaning that the corresponding bias parameter does not actually occur in the reduced problem. In such a case, one can define $\Delta'_k=0$ and $C'_t$ vanishing on the  $k^{th}$ row and column, and $C'_{t,kk}=1$ (this is the case for instance if there is no theoretical uncertainty for some of the observables).

Moreover, one can check that $\widetilde{C}'$ is indeed a correlation matrix by defining the scalar product $(x,y)=x^T \Delta \widetilde{C}_t \Delta^T y$. We can apply the Cauchy-Schwartz inequality to the basis vectors $u^{(i)}$ defined so that $u^{(i)}_j=\delta_{ij}$ (i.e., only one non-vanishing component):
\begin{eqnarray}
(u^{(i)},u^{(j)})^2&\leq& (u^{(i)},u^{(i)}) (u^{(j)},u^{(j)})\nonumber\\
(\Delta'_i)^2 (\Delta'_j)^2 (\widetilde{C}'_{t,ij})^2 &\leq& (\Delta'_i)^2 (\Delta'_j)^2 \widetilde{C}'_{t,ii}\widetilde{C}'_{t,jj} 
\end{eqnarray}
so that $|\widetilde{C}'_{t,ij}|\leq 1$ and $\widetilde{C}'_{t,ii}= 1$, with the appropriate structure of a correlation matrix.

Finally, the Cholesky decomposition of $\widetilde{C}'_t$ corresponds to $P'=(\Delta')^{-1}\Delta P$. Therefore, the determination of the theoretical uncertainties for $\Delta_\mu$ remains indeed the same with the new set of biases. 

We have thus reduced the 
problem of $n$ measurements and $m$ theoretical biases to the case with $n$ measurements, each of them having with a single bias parameter, with correlations among the biases. Without loss of generality we can consider that $\Delta$ is diagonal and $m=n$.

%% file: asymmetric.tex
\section{Asymmetric uncertainties}\label{app:asym}

In this article, statistical uncertainties are assumed to be strictly Gaussian and hence symmetric.  In practice, if asymmetric uncertainties are quoted, we symmetrise in the following manner
\begin{equation}\label{eq:asymunc}
X=\mu^{+\sigma_+}_{-\sigma_-} \to X=\left(\mu+\frac{\sigma_+-\sigma_-}{2} \right) \pm \left(\frac{\sigma_++\sigma_-}{2}\right)
\end{equation}
This is also the case for the theoretical uncertainties in the random-$\delta$ approach. 

In contrast, it is perfectly possible to have asymmetric theoretical uncertainties in the nuisance-$\delta$ or external-$\delta$ approaches described above. A theoretical uncertainty that is modeled by a bias parameter $\delta$ may be asymmetric: that is, the region in which $\delta$ is varied may depends on the sign of $\delta$, \textit{e.g.} $\delta\in [-\Delta_-,+\Delta_+]$ in one dimension ($\Delta_\pm\ge 0$).

In the case of a quadratic test statistic, we want to keep
 the stationarity property stemming from the symmetric quadratic shape, by using a test statistic  Eq.~(\ref{quadstat}) with $(\Delta_++\Delta_-)/2$ or Max$(\Delta_+,\Delta_-)$ in the definition, the second possibility being more conservative and our preferred choice in the following. As indicated in Sec.~\ref{sec:averageN}, this is independent
 of the range of variation  $\Omega$ chosen, which will be kept asymmetric, e.g., $[-\Delta_-,\Delta_+]$ in the fixed nuisance approach.

In the case of the Rfit approach~\cite{Hocker:2001xe,Charles:2004jd}, we can use the fact that the  well test statistic has a shape that is independent of the central value chosen, as long as the position of the flat bottom remains unchanged. One can thus shift the central value by an  arbitrary quantity if one remains at the bottom of the well. It is thus completely equivalent to take asymmetric theoretical ranges or to take  symmetric theoretical ranges following Eq.~(\ref{eq:asymunc}) where  $\sigma_\pm$ is replaced by $\Delta_\pm$.

%% file: ckmappendix.tex
\section{3-$\sigma$ intervals for CKM-related examples} \label{app:ckm3sigma}

\begin{figure}
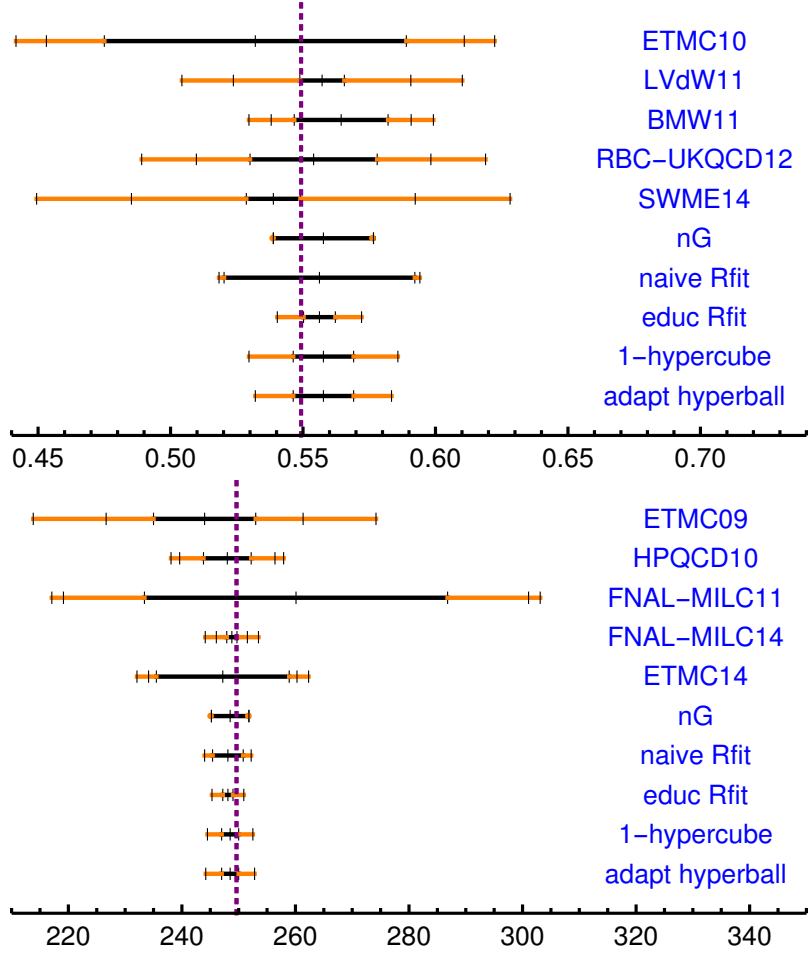

\begin{center}
\pgfuseimage{3sigma_Image_BK} \hspace{0.5cm} \pgfuseimage{3sigma_Image_Ds}
\end{center}
\caption{(Top) Inputs for $B_K^{\bar{\rm MS}}(2{\rm GeV})$ and the averages resulting from the different models considered here. (Bottom) Same for the lattice determinations of the $D_s$-meson decay constant (in MeV). The black range gives the statistical error. For each individual input, the solid yellow range indicates the 3 $\sigma$ interval according to the adaptive hyperball approach, whereas the interval corresponding to the   1-fixed hypercube approach is given by the vertical lines in the middle of the solid yellow intervals. For average according to the different approaches, the black range corresponds again to the 3 $\sigma$ statistical error, whereas the yellow range corresponds to the 3 $\sigma$ interval following the corresponding approach. The comparison between black and yellow ranges illustrates the relative importance of statistical and theoretical errors.
 Finally, for illustrative purposes, the vertical purple line gives the arithmetic average of the inputs (same weight for all central values).}
\label{fig:comparisonBKandDs3sigma}
\end{figure}

We collect here the intervals at 3 $\sigma$ for the various approaches applied to the CKM examples discussed in Sec.~\ref{sec:ckm}. Figs.~\ref{fig:comparisonBKandDs3sigma}, \ref{fig:comparisonalphaS3sigma} and \ref{fig:comparisonVubandVcb3sigma} are the 3-$\sigma$ equivalents of Figs.~\ref{fig:comparisonBKandDs}, \ref{fig:comparisonalphaS} and \ref{fig:comparisonVubandVcb} showing 1 $\sigma$ intervals. The comparison between the two series of plot shows how the intervals evolve with the confidence level. In particular,  the adaptive hyperball approach appears more (less) conservative than the 1-hypercube approach at high (low) significance. This change of hierarchy explains why we choose a different convention to plot the 1 $\sigma$ (dashed horizontal line) and 3 $\sigma$ (vertical lines in the middle of the solid intervals) intervals for the 1-hypercube approach in Figs.~\ref{fig:comparisonBKandDs}, \ref{fig:comparisonalphaS}, \ref{fig:comparisonVubandVcb} on one hand and Figs.~\ref{fig:comparisonBKandDs3sigma}, \ref{fig:comparisonalphaS3sigma}, \ref{fig:comparisonVubandVcb3sigma} on the other hand.

\begin{figure}
\begin{center}
\pgfuseimage{3sigma_Image_alphaSMZ}
\end{center}
\caption{Determinations of the strong coupling constant at $M_Z$ through $e^+e^-$ annihilation, and the averages resulting from the different models considered. The intervals are given at 3 $\sigma$. See Fig.~\ref{fig:comparisonBKandDs3sigma} for the legend.}
\label{fig:comparisonalphaS3sigma}
\end{figure}

\begin{figure}
\begin{center}
\pgfuseimage{3sigma_Image_Vub} 

 \pgfuseimage{3sigma_Image_Vcb}
\end{center}
\caption{(Top) Inclusive and exclusive inputs for the CKM matrix element $ \vert V_{ub} \vert $ (times $ 10^{3} $) and the averages resulting from the different models considered here. (Bottom) Same for the determinations of $ \vert V_{cb} \vert $ (times $ 10^{3} $) CKM matrix element. The intervals are given at 3 $\sigma$. See Fig.~\ref{fig:comparisonBKandDs3sigma} for the legend.}
\label{fig:comparisonVubandVcb3sigma}
\end{figure}